\documentclass[11pt,a4paper]{article}
\usepackage{jcappub}

\usepackage{amssymb,amsmath,amstext,amsfonts}
\usepackage{bbold}
\usepackage{bm}
\usepackage{graphicx}
\usepackage{xcolor}
\usepackage[normalem]{ulem}
\usepackage{overpic}
\usepackage{subcaption}
\usepackage{tikz}
\usepackage{enumitem}
\usepackage{comment}

\newcommand{\beq}{\begin{equation}}
\newcommand{\eeq}{\end{equation}}
\newcommand{\bal}{\begin{aligned}}
\newcommand{\eal}{\end{aligned}}

\newcommand{\kstar}{k_\star}

\newcommand{\Pzeta}{\mathcal{P}_{{\cal R}}}

\newcommand{\Pbar}{\overline{\mathcal{P}}_{{\cal R}}}

\newcommand{\omegalin}{\omega_{\textrm{lin}}}
\newcommand{\omegagwlin}{\omega_{\textrm{lin}}^{\textsc{gw}}}
\newcommand{\omegalog}{\omega_{\textrm{log}}}

\newcommand{\omegalogc}{\omega_{\textrm{log,c}}}
\newcommand{\omegalogx}{\omega_{\textrm{log,x}}}

\newcommand{\Alog}{A_{\textrm{log}}}
\newcommand{\thetalog}{\vartheta_{\textrm{log}}}

\newcommand{\OGW}{\Omega_\textrm{GW}}
\newcommand{\ObarGW}{\overline{\Omega}_\textrm{GW}}
\newcommand{\kref}{k_\textrm{ref}}
\newcommand{\Tw}{\mathcal{T}_w}

\newcommand{\C}{\Xi}

\title{Expansion history-dependent oscillations in the scalar-induced gravitational wave background}

\author[a]{Lukas~T.~Witkowski,}
\author[b]{Guillem Dom\`{e}nech,}
\author[a]{Jacopo Fumagalli}
\author[a]{and S\'{e}bastien Renaux-Petel}

\affiliation[a]{Institut d'Astrophysique de Paris, GReCO, UMR 7095 du CNRS et de Sorbonne Universit\'{e},\\ 98bis boulevard Arago, 75014 Paris, France}
\affiliation[b]{INFN Sezione di Padova, I-35131 Padova, Italy}

\emailAdd{lukas.witkowski@iap.fr}
\emailAdd{domenech@pd.infn.it}
\emailAdd{jacopo.fumagalli@iap.fr}
\emailAdd{renaux@iap.fr}

\abstract{Oscillations in the frequency profile of the scalar-induced contribution to the stochastic gravitational wave background are a characteristic signal for small-scale features during inflation. We investigate how this oscillatory frequency profile is affected by the expansion history of the post-inflationary universe. Our results are applicable as long as the equation of state of the universe can be taken as constant during the period in which the gravitational waves are produced, and we compute the spectrum of gravitational waves induced by both sharp and resonant features, associated with oscillations in $k$ and $\log(k)$, respectively. For resonant features, the frequencies of the oscillatory contributions to the gravitational wave spectrum are unaffected by the equation of state, but not their relative amplitudes, allowing one to extract information about both inflationary physics and the post-inflationary expansion history from the oscillatory pattern. For sharp features we find that the gravitational wave spectrum only exhibits prominent modulations as long as the propagation speed of density fluctuations is $c_s<1$, with a frequency larger by a factor $c_s^{-1}$ than that of the scalar power spectrum. We find that the stiffer the equation of state, the larger the relative amplitude of the oscillations. In particular, a relative amplitude significantly higher than $20 \%$ is not achievable for the `standard' case of radiation domination, and would be a smoking-gun signal of both nontrivial inflationary dynamics on small scales, and a post-inflationary universe not dominated by radiation.}

\begin{document}

\maketitle

\section{Introduction}
\label{sec:intro}
Despite the success of the standard model of cosmology, most of the early universe remains unexplored. Thus far, we are certain of the physics since roughly the time of neutrino decoupling; we know that since then, the early universe was dominated by a plasma of relativistic particles. Furthermore, the observation of primordial fluctuations provides strong evidence that a period of inflation took place roughly 120 \textit{e}-folds prior to today and that lasted for at least half of that. However, only the first few \textit{e}-folds have been probed by the observation of large scale primordial fluctuations \cite{Aghanim:2018eyx,Akrami:2018odb}. The physics of the remaining fraction of inflation, small scales fluctuations and the universe afterwards until neutrino decoupling are largely unknown. Encouragingly, extraordinary events that occurred within those periods might leave a gravitational wave signature that falls right inside the range of future gravitational wave detectors \cite{Caprini:2018mtu}, such as LISA. As gravitational waves travel barely scattered by matter, we will have in the next decades a unique window to probe the unexplored periods of the universe, otherwise inaccessible by conventional observations with electromagnetic waves. 

Promising sources of cosmic gravitational waves are (see Ref.~\cite{Caprini:2018mtu} for a review): first order phase transitions, cosmic strings, resonances during preheating, quantum fluctuations during inflation \cite{Guzzetti:2016mkm} and gravitational waves induced by primordial fluctuations \cite{Tomita,Matarrese:1992rp,Matarrese:1993zf}. The latter is the most promising candidate to explore the physics of the last stages of inflation and afterwards \cite{Ananda:2006af,Baumann:2007zm,Saito:2008jc,Saito:2009jt,Assadullahi:2009jc,Assadullahi:2009nf,Inomata:2018epa,Cai:2019amo,Liu:2020oqe,Hajkarim:2019nbx,Bhattacharya:2019bvk,Domenech:2019quo,Inomata:2019zqy,Inomata:2019ivs,Gow:2020bzo,Domenech:2020kqm,Fumagalli:2020nvq,Dalianis:2020cla,Abe:2020sqb,Braglia:2020eai,Braglia:2020taf,Atal:2021jyo,Fumagalli:2021cel} (see Refs.~\cite{Yuan:2021qgz,Domenech:2021ztg} for recent reviews). The intuitive physical picture is as follows. First, primordial fluctuations are set by inflation on scales larger than the cosmological horizon. After inflation ends, density fluctuations start to evolve once they are back inside the horizon. Then, the time evolution of such fluctuations induces space-time oscillations, which are the so-called induced gravitational waves. At first glance, the resulting induced gravitational wave spectrum depends mostly on the primordial spectrum of fluctuations. This is so if we extrapolate the standard model of cosmology and assume that radiation dominated the universe right after inflation. In reality though, we have scarce evidence of the content of the universe at the time of wave generation \cite{Allahverdi:2020bys} and we should factor it in. For instance, it has been shown that the induced gravitational wave spectrum could change significantly if the universe is not dominated by radiation \cite{Assadullahi:2009nf,Inomata:2019zqy,Inomata:2019ivs,Domenech:2019quo,Domenech:2020kqm,Dalianis:2020cla}. Thus, we should also take into account the dependence of the gravitational wave spectrum on the unknown expansion history of the early universe.

Future measurements of the primordial spectrum of induced gravitational waves have the potential of granting access to the physics of inflation that cannot be tested by Cosmic Microwave Background (CMB) data. In particular, inflation may depart from the single-field slow-roll paradigm at a later stage of inflation without affecting CMB predictions, but with distinct observational effects in GWs. Such departures, also referred to as primordial features, can be classified by their effect on the scalar power spectrum $\Pzeta(k)$, which takes the form of a sinusoidal modulation (see the reviews \cite{Chen:2010xka,Chluba:2015bqa,Slosar:2019gvt}). So-called \emph{sharp features} arise when inflation exhibits a sudden transition, as e.g.~caused by a step in the potential or a sharp turn in the inflationary trajectory, and are associated with an oscillation in $\Pzeta(k)$ that is periodic in $k$. So-called \emph{resonant features} refer to models where a background quantity oscillates with a frequency larger than the expansion rate $H$, with monodromy inflation the paradigmatic example, resulting in an oscillation in $\Pzeta(k)$ that is periodic in $\log(k)$. The existence of a feature implies that over a range of scales $k$ the scalar power spectrum adheres to the following templates:
\begin{align}
    \label{eq:P-of-k-sharp}
    \textrm{Sharp:} \quad & \Pzeta(k)=\Pbar(k) \Big[1 + A_{\textrm{lin}} \cos \big(\omega_{\textrm{lin}} k + \vartheta_{\textrm{lin}} \big) \Big] \, , \\
    \label{eq:P-of-k-resonant}
    \textrm{Resonant:} \quad & \Pzeta(k)=\Pbar(k) \Big[1 + \Alog \cos \big(\omegalog \log(k/k_\textrm{ref}) + \thetalog \big) \Big] \, ,
\end{align}
where the envelope $\Pbar(k)$ depends on the particular realisation of the feature.

While the defining property of a `feature' is the oscillatory behaviour of $\Pzeta$, feature models can also enhance the amplitude of the primordial spectrum by several orders of magnitude. For the templates in \eqref{eq:P-of-k-sharp} and \eqref{eq:P-of-k-resonant} this corresponds to an envelope $\Pbar(k)$ that is elevated compared to the power spectrum at CMB scales or exhibits a peak. Such an enhancement is of great interest as it may lead to the formation of a significant fraction of primordial black holes in the early universe, by the collapse of large primordial fluctuations \cite{Zeldovich:1967lct,Hawking:1971ei,Carr:1974nx} (see Refs.~\cite{Sasaki:2018dmp,Carr:2020gox,Carr:2020xqk,Green:2020jor} for recent reviews). Primordial black holes are gathering the attention of both the theoretical and the experimental communities as plausible explanation to several observations \cite{Sasaki:2018dmp}, such as some of the binary black holes observed by LIGO \cite{Inomata:2016rbd,Nakama:2016gzw,Ando:2017veq,Kohri:2018qtx,Franciolini:2021tla}. The enhancement is also advantageous for the observability of the induced GWs, as an increased amplitude of scalar fluctuations boosts the amplitude of the induced GW spectrum. Interestingly, as described in \cite{Fumagalli:2020nvq}, the enhancement of fluctuations and the properties of the oscillations in $\Pzeta$ are related: if a sharp feature is responsible for a significant enhancement of $\Pbar(k)$, large oscillations are unavoidable as their amplitude approaches $A_\textrm{lin} \rightarrow 1$.

What makes feature models particularly suitable for experimental detection is that the oscillations in $\Pzeta$ lead to corresponding modulations of the energy density fraction $\OGW(k)$, which are in principle observable in the upcoming generation of GW observatories. The spectrum of induced GWs due to sharp and resonant features has been analysed in \cite{Fumagalli:2020nvq, Fumagalli:2021cel} assuming that the relevant fluctuations re-entered during a period of radiation-domination. The existence of oscillations in $\OGW(k)$ has also been observed numerically in \cite{Braglia:2020taf,Dalianis:2021iig} for a family of feature models. The GW spectrum $\OGW(k)$ can be shown to follow the templates \cite{Fumagalli:2020nvq, Fumagalli:2021cel}:
\begin{align}
    \label{eq:sharp-template-intro} 
    \textrm{Sharp feature:} \quad & \Omega_{\textrm{GW}}(k) = \overline{\Omega}_{\textrm{GW}}(k) \Big[1+ \mathcal{A}_\textrm{lin} \cos \big(\omegagwlin k + \phi_\textrm{lin}  \big) \Big] \, , \\
    \label{eq:resonant-template-intro}
    \textrm{Resonant feature:} \quad & \Omega_{\textrm{GW}}(k) = \overline{\Omega}_{\textrm{GW}}(k) \Big[1+ \mathcal{A}_{\textrm{log},1} \cos \big(\omegalog \log (k/k_\textrm{ref}) + \phi_{\textrm{log},1} \big) \\
    \nonumber & \hphantom{\Omega_{\textrm{GW}}(f) = \overline{\Omega}_{\textrm{GW}}(k) \Big[1} + \mathcal{A}_{\textrm{log},2} \cos \big(2 \omegalog \log (k/k_\textrm{ref}) + \phi_{\textrm{log},2} \big) \Big] \, ,
\end{align}
For a sharp feature the frequencies in $\OGW(k)$ and $\Pzeta(k)$ are related as $\omegagwlin = \sqrt{3} \omegalin$. As the processing of scalar fluctuations into GWs is a non-linear effect, the amplitude of oscillations in $\OGW$ tends to be washed out, with $\mathcal{A}_\textrm{lin} \sim \mathcal{O}(10 \%)$ for an amplitude in $\Pzeta$ that is $A_\textrm{lin}=1$. In particular, one can show that for sharp features the amplitude cannot significantly exceed $\mathcal{A}_\textrm{lin} \approx 20 \%$ for GWs induced during a period of radiation domination. For a resonant feature in $\Pzeta$ the corresponding GW spectrum exhibits two oscillatory parts, one with the original frequency $\omegalog$ and one with $2 \omegalog$. For $\omegalog \lesssim 5$ the term with frequency $\omegalog$ dominates over the other, with an amplitude $\mathcal{A}_\textrm{log,1}$ that can be $\mathcal{O}(1)$ in the maximal case. Increasing $\omegalog$ both $\mathcal{A}_{\textrm{log},1,2}$ decrease, but the relative importance of the term with frequency $2 \omegalog$ grows, leading to a complicated oscillatory pattern in $\OGW$ for $\omegalog \gtrsim 5$. For large values of $\omegalog$ the oscillatory piece with frequency $2 \omegalog$ eventually dominates, but the amplitude of oscillation is smaller.\footnote{For resonant features with sufficiently broad envelope $\Pbar$, the dependence of the amplitudes $\mathcal{A}_{\textrm{log},1,2}$ and phases $\phi_{\textrm{log},1,2}$ on $\omegalog$ and $A_\textrm{log}$ is universal and can be computed exactly, while for narrower envelopes this is also affected by the envelope shape \cite{Fumagalli:2021cel}.}

The exciting prospect then is that by detecting an oscillation in $\OGW$ that has been caused by a primordial feature one can gain information about the scalar power spectrum, by measuring the properties of this oscillation and applying the `dictionary' between $\Pzeta$ and $\OGW$ from \cite{Fumagalli:2020nvq, Fumagalli:2021cel}. However, to be conclusive, such an analysis should also take into account the uncertainty regarding the thermal history of the universe after inflation. The reason is that a different equation of state $w$ during horizon re-entry of the relevant fluctuations will affect the relation between $\Pzeta$ and $\OGW$. For example, the factor $\sqrt{3}$ relating $\omegagwlin$ and $\omegalin$ for a sharp feature is specific to radiation-domination and is expected to change for a different post-inflationary equation of state. Thus, to allow for a trustworthy reconstruction of $\Pzeta$ from $\OGW$, one which also takes into account the ignorance regarding the thermal history after inflation, the `dictionary' between $\Pzeta$ and $\OGW$ has to be generalised beyond the case of radiation-domination.

This is the subject of this paper, where we analyse the induced gravitational wave spectrum due to primordial features during inflation, assuming a general equation of state parameter of the universe with $0<w<1$ and a general propagation speed for the fluctuations $c_s^2$, thus going beyond the `standard' case of radiation-domination, which corresponds to $c_s^2=w=1/3$. Our main results are as follows: For sharp features we find that oscillations are only significantly imprinted when $c_s^2<1$, in which case they still adhere to the template \eqref{eq:sharp-template-intro}. However, there is a degeneracy between the equation of state and the original frequency of the oscillations in $\Pzeta$ for a given value of $\omegagwlin$ in \eqref{eq:sharp-template-intro}. Interestingly, for $w>1/3$ we find that the amplitude of the oscillations in the induced gravitational wave spectrum can be larger than for $w=1/3$, surpassing $20 \%$ in certain examples, and can thus be used to break the degeneracy and deduce information regarding $\Pzeta$ and the equation of state. For resonant features, oscillations in the GW spectrum also occur for $c_s^2=1$, and can be described by the template \eqref{eq:resonant-template-intro}. The equation of state and the value of $c_s$ affect the amplitude of the oscillations in $\OGW$, but not the frequency, thus giving easier access to information about both inflation and the thermal history.  
This type of test of the expansion history of the universe is independent and complementary to the observation of the low frequency tail of the induced gravitational wave spectrum \cite{Domenech:2020kqm}. The observation of both the low frequency tail and the oscillations can further break the degeneracy and allow to extract the oscillations in the primordial spectrum. Thus, induced gravitational waves might hint at new physics during and after inflation.

This paper is organised as follows. In sec.~\ref{sec:GWreview} we collect expressions for computing the spectral density of GWs induced during a period with general equation of state parameter $w$ and scalar fluctuation propagation speed $c_s$, and clearly state the assumptions for their validity. In sec.~\ref{sec:feat-general} we use analytical tools to characterise the oscillations in $\OGW$ induced by either a sharp or resonant feature in $\Pzeta$. We then take a more detailed look at the spectrum of GWs induced by a sharp feature and by a resonant feature in secs.~\ref{sec:sharp} and \ref{sec:resonant}, respectively, using both numerical methods as well as the analytical tools from sec.~\ref{sec:feat-general} to study the GW spectrum for different values of $w$, focussing on the case of an adiabatic perfect fluid ($c_s^2=w$) and a canonical scalar field ($c_s^2=1$). We briefly discuss the results and consider avenues for further work in sec.~\ref{sec:conclusions}. Various formulae and an analysis of the tails of the GW spectrum can be found in the appendix. 

\section{Review: induced gravitational waves}
\label{sec:GWreview}

\begin{figure}
    \centering
    \includegraphics[width=0.8\columnwidth]{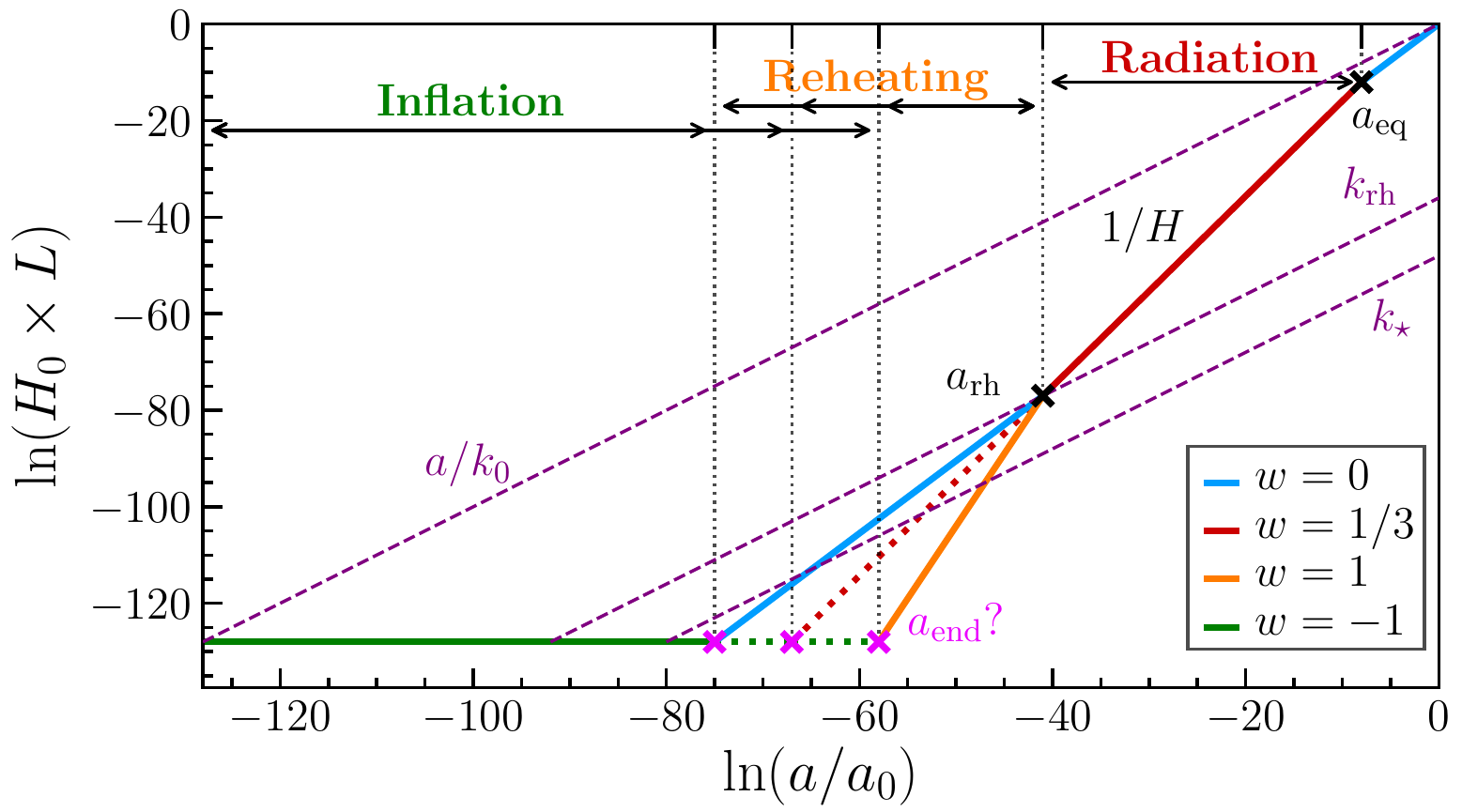}
    \caption{Illustration of the scenario considered for the early universe, where we show in logarithmic scale a physical scale $L$ normalised to the current Hubble parameter $H_0^{-1}$ versus the scale factor $a$ normalised to today. In solid lines we show the Hubble parameter $1/H$ and in purple dashed lines we show a physical scale $L=a/k$. We used the equations derived in appendix \ref{app:exphistory} for a reheating temperature $T_{\rm rh}=5\times 10^4\,{\rm GeV}$ and $H_{\rm inf}\approx 10^{14}\,{\rm GeV}$. See how, for a fixed reheating temperature, the total \textit{e}-folds of observable inflation depends on the equation of state parameter $w$ after inflation. As examples, we consider an early matter domination $w=0$ which could be due to the oscillations of heavy moduli fields, and a stiff fluid with $w=1$ which is typical of quintessential inflation scenarios \cite{Spokoiny:1993kt,Peebles:1998qn,Brax:2005uf,Hossain:2014xha}, see \cite{Allahverdi:2020bys} for other mechanisms. In general, there are several \textit{e}-folds between the end of inflation and the time where the reheating scale $k_{\rm rh}$ exited the Hubble radius. We consider a scalar power spectrum peaked at $k_\star$ and for illustration purposes we took $k_\star=4\times 10^5 \,k_{\rm rh}$. Induced GWs are generated when $k_\star$ re-enters the sound horizon. This figure shows that we have a large parameter space where induced GWs could be generated in a stage with $w\neq 1/3$. Note that the reheating temperature could be as low as $T_{\rm rh}>4\,{\rm MeV}$ \cite{Kawasaki:1999na,Kawasaki:2000en,Hannestad:2004px,Hasegawa:2019jsa}.}
    \label{fig:expansionhistory}
\end{figure}

The evolution of primordial fluctuations generate space-time oscillations which result in gravitational waves. In more mathematical terms, the equations of motion for tensor modes have a source term proportional to scalar modes squared, at second order in cosmological perturbation theory \cite{Tomita,Matarrese:1992rp,Matarrese:1993zf}. In this way, primordial fluctuations induce gravitational waves. In general, the spectral density of gravitational waves is given by
\begin{align}\label{eq:omegaGW}
    \Omega_{\textrm{GW}}(k,\tau) =\frac{1}{12}\frac{k^2}{a^2H^2}{\cal P}_{t}(k,\tau) \, ,
\end{align}
where ${\cal P}_{t}(k,\tau)$ is the dimensionless tensor mode power spectrum summed over the two polarizations and $\tau$ is conformal time. In the case of induced gravitational waves and assuming Gaussian primordial fluctuations, the induced tensor mode spectrum ${\cal P}_{t}(k,\tau)$ is an integral over the internal momentum $\vec{q}$ of two scalar power spectra, say ${\cal P}_{\cal R}(|\vec{q}|){\cal P}_{\cal R}(|\vec{k}-\vec{q}|)$, times a transfer function. The transfer function as well as the scale factor $a$ and expansion rate $H$ in Eq.~\eqref{eq:omegaGW} depend on the equation state parameter $w$ of the universe. Assuming a constant $w$ we have that
\begin{align}
a\propto \tau^{1+b}\quad,\quad aH=\frac{1+b}{\tau}\quad{\rm with}\quad b=\frac{1-3w}{1+3w}\,,
\end{align}
where we introduced $b$ for later convenience, with $b=0$ for a radiation dominated universe ($w=1/3$) and $b>0$ and $b<0$ respectively for a $w$ softer ($w<1/3$) and stiffer ($w>1/3$) than radiation. See figure \ref{fig:expansionhistory} for a detailed illustration of the scenario we are considering. On top of that, the transfer function also depends on the propagation speed of scalar fluctuations $c_s^2$. For instance, an adiabatic perfect fluid has $c_s^2=w$ and a canonical scalar field $c_s^2=1$. Through the transfer function, induced gravitational waves keep a record of the content of the universe.

Assuming that induced gravitational waves are generated during a $b\neq0$ epoch, their spectral density after the universe transitioned to the standard radiation-dominated epoch\footnote{In this paper we are interested in the shape of the induced GW spectrum. This is why we limit our study, for simplicity, to the spectral density of GWs right after the transition to radiation domination. For the reader interested in the spectral density of GWs evaluated today, we have that
\begin{align}\label{eq:spectraldensitytoday2}
\Omega_{\rm GW,0}h^2&=1.62\times 10^{-5}\left(\frac{\Omega_{r,0}h^2}{4.18\times 10^{-5}}\right)\left(\frac{g_*(T_{\rm rh})}{106.75}\right)\left(\frac{g_{*s}(T_{\rm rh})}{106.75}\right)^{-4/3}\Omega_{\rm GW}\,,
\end{align}
where $h=H_0/(100 \,{\rm km/s/Mpc})$, $\Omega_{r,0}$ is the density fraction of radiation today and $g_*$ and $g_{*s}$ are respectively the effective degrees of freedom in the energy density and entropy.} is given by \cite{Domenech:2019quo,Domenech:2020kqm,Domenech:2021ztg} 
\begin{align}\label{eq:OmegaGW-ts} 
\Omega_{\textrm{GW}}(k)&= \hphantom{\frac{1}{2}} {\left(\frac{k}{k_{\rm rh}}\right)^{-2b}} \int_1^{\infty} \textrm{d} s \int_{0}^{1} \textrm{d} d \, \Tw(d,s) \, \mathcal{P}_{\cal R} \bigg(\frac{k}{2}(s+d)\bigg) \mathcal{P}_{\cal R} \bigg(\frac{k}{2}(s-d)\bigg) \, ,
\end{align}
where 
\begin{align}\label{eq:Twcs}    
\Tw(d,s) = & \hphantom{\times} \mathcal{N}(b) {\bigg(\frac{(d^2-1)(s^2-1)}{d^2-s^2}\bigg)}^2 \, \frac{|1-y^2|^b }{(s^2-d^2)^2} \, \times \\
\nonumber & \times \Bigg\{\hphantom{+} \hphantom{\frac{4}{\pi^2}} \bigg[\mathsf{P}_{b}^{-b}(y) + \frac{2+b}{1+b} \mathsf{P}_{b+2}^{-b}(y) \bigg]^2 \Theta \big(s-c_s^{-1} \big) \\
\nonumber &\hphantom{\times \Bigg\{} +  \frac{4}{\pi^2} \bigg[\mathsf{Q}_{b}^{-b}(y) + \frac{2+b}{1+b} \mathsf{Q}_{b+2}^{-b}(y) \bigg]^2 \Theta \big(s-c_s^{-1} \big) \\
\nonumber &\hphantom{\times \Bigg\{} +  \frac{4}{\pi^2} \bigg[\mathcal{Q}_{b}^{-b}(-y) + 2 \frac{2+b}{1+b} \mathcal{Q}_{b+2}^{-b}(-y) \bigg]^2 \Theta \big(c_s^{-1}-s \big) \Bigg\} \, ,
\end{align}
and
\begin{align}
    \label{eq:ydef}
    y \equiv \frac{s^2+d^2-{2}{c_s^{-2}}}{s^2-d^2} \quad{,}\quad 
    \mathcal{N}(b)=\frac{1}{3}\left( \frac{4^{1+b}(b+2)}{\left(1+b\right)^{1+b}(2b+3)c_s^2} \, \Gamma^2 \Big[b+\tfrac{3}{2} \Big] \right)^2\, .
\end{align}
We have introduced $k_{\rm rh}$ to refer to the comoving wavenumber that entered the horizon at the transition, that is $k_{\rm rh}=a_{\rm rh}H_{\rm rh}$. The subscript ``rh'' stands for evaluation at the time of ``reheating''. For reference, we have defined
\begin{align}
 s\equiv\frac{|\vec{k}-\vec{q}|+q}{k}\quad{\rm and}\quad d\equiv\frac{|\vec{k}-\vec{q}|-q}{k}\,.
\end{align}
The functions $\mathsf{P}_{\nu}^{\mu}(x)$, $\mathsf{Q}_{\nu}^{\mu}(x)$ and $\mathcal{Q}_{\nu}^{\mu}(x)$ are respectively the Ferrer's function of the first and second kind and the associated Legendre function of the second kind, all provided in terms of hypergeometric functions in appendix \ref{app:Legendre}. Equation \eqref{eq:OmegaGW-ts} reduces to the results of Espinosa, Racco and Riotto \cite{Espinosa:2018eve} and Kohri and Terada \cite{Kohri:2018awv}  for $c_s^2=w=1/3$ ($b=0$). It should be noted that the expression \eqref{eq:OmegaGW-ts} is valid under the following assumptions:
\begin{enumerate}[label=\textit{(\roman*)}]
\item We focus on gravitational waves that have been generated much before the transition to radiation domination.
\item This transition is instantaneous.
\end{enumerate}
Condition $(i)$ entails that the primordial spectrum has a finite width and that all scales within that width enter the horizon well before the transition. As a result, most of the generated gravitational waves have wavelengths much smaller than the size of the cosmological horizon at the transition, that is $k\gg k_{\rm rh}$. If the primordial spectrum is peaked at $k=k_\star$ we thus consider that $\kstar\gg k_{\rm rh}$. Furthermore, we will be mainly interested in an adiabatic perfect fluid with $w>0$ ($b>1$). In this case, condition $(i)$ ensures that the low frequency tail of the induced gravitational wave spectrum always decreases with decreasing $k$. For $k<k_{\rm rh}$, $\Omega_{\rm GW}$ is at least suppressed by a factor $(k_{\rm rh}/\kstar)^2$ with respect to the amplitude of the peak in the gravitational wave spectrum \cite{Domenech:2020kqm,Domenech:2021ztg}. Thus, we will not be concerned with the low frequency tail of the gravitational wave spectrum. Note that such range of wavenumbers already covers the scales of interest, as the oscillatory features appear near the peak of the induced gravitational wave spectrum \cite{Fumagalli:2020nvq,Fumagalli:2021cel}. Lastly, condition $(ii)$ is required for analytical simplicity. Nevertheless, if the transition is not exactly instantaneous, we expect \eqref{eq:OmegaGW-ts} to be valid for modes that entered the horizon well before the transition. In other words, we only expect corrections for $k\sim k_{\rm rh}$ as tensor modes with $k\gg k_{\rm rh}$ are already propagating as free gravitational waves much before the transition. For more details on the induced gravitational wave spectrum and the transition we refer the reader to \cite{Domenech:2019quo,Domenech:2020kqm,Domenech:2021ztg}.

Before moving on to the next section, it is important to understand the general behaviour of the transfer function \eqref{eq:Twcs}. In the large and small momentum limit, that is for $s-d\gg1$ ($q\gg k$) and $s-d\ll1$ ($q\ll k$), the transfer function decays respectively as $q^{-4}$ and $q^2$ (see appendix \ref{app:IRUVtail}). This means that the transfer function peaks at some intermediate momentum. Interestingly, for $c_s^2<1$ and $w>0$ ($b<1$), the peak of the transfer function \eqref{eq:Twcs} is at $s=c_s^{-1}$ (for all $d$). This special point, which corresponds to $y\sim -1$, is where the associated Legendre functions may diverge. Such divergence is associated to the physical resonance that occurs when a harmonic oscillator has the same frequency as an external periodic force. For induced gravitational waves, if we consider that the primordial spectrum is highly peaked at $k_\star$, then we expect a resonant peak in the induced gravitational wave spectrum at $k=2c_s \kstar$ for $c_s<1$. For $c_s=1$ the resonance is kinematically forbidden and the gravitational wave spectrum peaks around $k\sim \kstar$ with an amplitude much smaller than for $c_s<1$. In more generality, if we regard the transfer function as the evolution of a single tensor mode with wavenumber $k$ per scalar momenta pair, $q$ and $|\vec{k}-\vec{q}|$, i.e.~${\cal T}_w\sim dh_k/dq/d|\vec{k}-\vec{q}|$, we also expect that the transfer function peaks at $k=c_s (|\vec{k}-\vec{q}|+q)$ for $c_s<1$. Thus, if the primordial spectrum consists of a series of sharp peaks or large oscillations, the presence of the resonance will play a crucial role in the way the induced GW spectrum captures such oscillations. We shall see this in more detail in the next section.

\section{Induced gravitational waves from primordial features: analytical considerations}
\label{sec:feat-general}
In this section we present two analytic approaches for extracting information about the spectrum of induced GWs due to a sharp or resonant feature during inflation. These methods have been previously employed in \cite{Fumagalli:2020nvq,Fumagalli:2021cel}, but here we generalise to the case of a  general equation of state parameter $w$.

\subsection{Resonance peak analysis}
\label{sec:resonance-peak-analysis}
As reviewed at the end of section \ref{sec:GWreview}, for $c_s<1$, a peak in $\Pzeta$ leads to a corresponding resonance peak in $\OGW$. This observation can be used to predict the peak-structure of $\OGW$ due to a primordial feature, by modelling the oscillation in $\Pzeta$ as a series of peaks and studying the corresponding resonance peaks. This has been employed in \cite{Fumagalli:2020nvq,Fumagalli:2021cel} for GWs sourced during a period of radiation domination, and here we extend to the case of general equation of state parameter $w$.

\begin{figure}[t]
\centering
\begin{overpic}[width=0.80\textwidth]{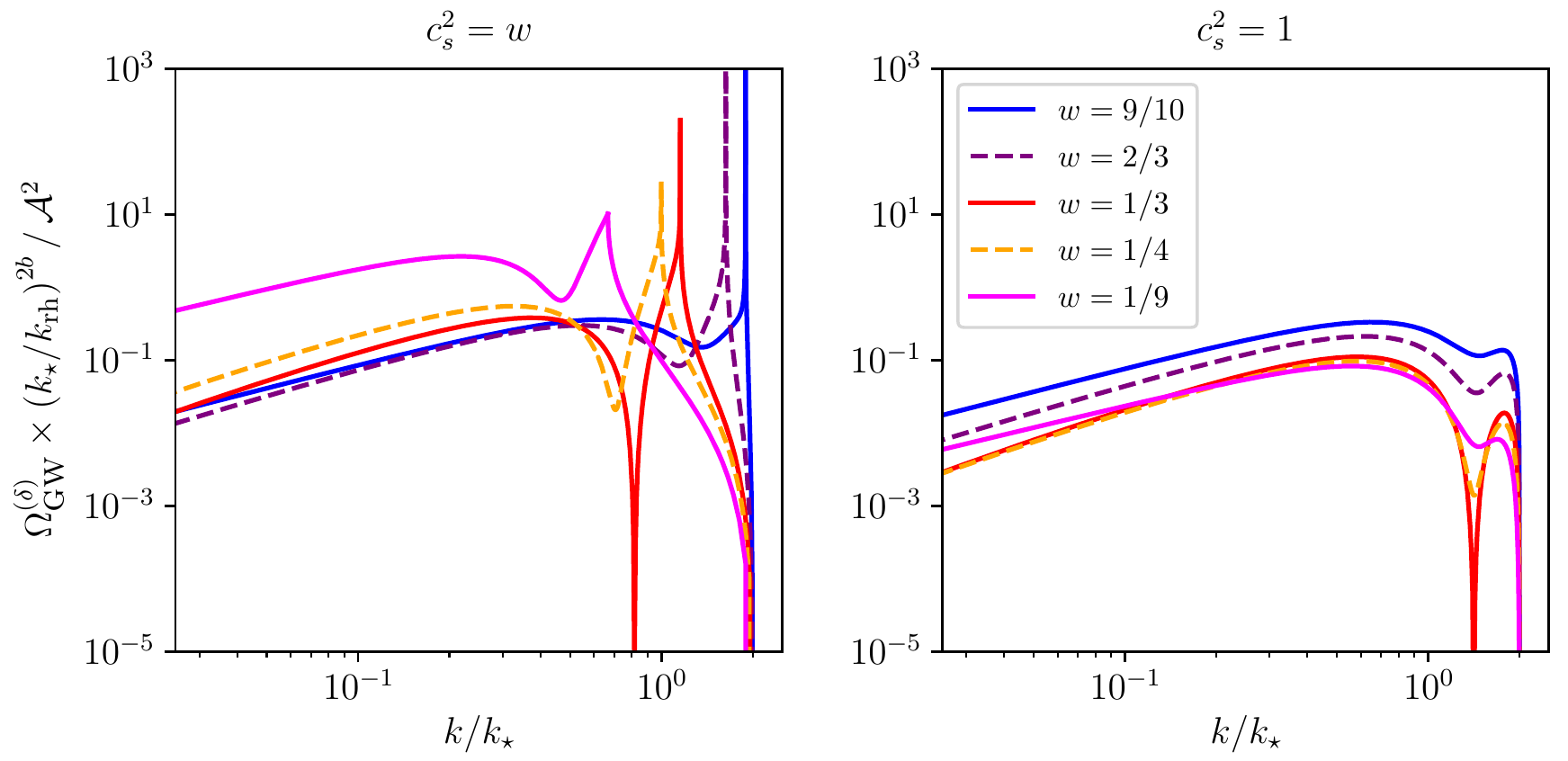}
\end{overpic}
\caption{GW spectrum for a $\delta$-peak scalar power spectrum for $c_s^2=w$ (left panel) and $c_s^2=1$ (right panel). The plot shows $\OGW^{(\delta)} (k_\star / k_\textrm{rh})^{2b} /\mathcal{A}^2$ as given in \eqref{eq:Omega-delta-w} vs.~$k/\kstar$ for several choices of $w<1$. For $c_s^2=w$ the GW spectrum exhibits a resonance peak at $k=2 \sqrt{w} \kstar$ that becomes narrower and spikier as $w$ is increased. For $c_s=1$ this resonance peak is absent altogether.}
\label{fig:OGW_delta_w_1}
\end{figure}

It will be convenient to start with the most extreme case of a peaked $\Pzeta$, which is a monochromatic scalar power spectrum $\Pzeta(k) = \mathcal{A} \, \delta \big(\log (k / k_\star) \big)$. Inserting into \eqref{eq:OmegaGW-ts} this gives:
\begin{align}
\label{eq:Omega-delta-w}
\OGW^{(\delta)}(k) = \mathcal{A}^2 \,  {\left( \frac{k}{k_\star} \right)}^{-2(1+b)} \, {\left(\frac{k_\star}{k_{\rm rh}}\right)^{-2b}} \, \Tw \bigg(0, \frac{2 k_\star}{k} \bigg) \, \Theta \bigg(2-\frac{k}{k_\star} \bigg) \, .
\end{align} 
One can check that for $c_s < 1$ this exhibits a peak at $k=2 c_s \kstar$, where the GW spectrum may even diverge. This can be e.g.~seen in the left panel of fig.~\ref{fig:OGW_delta_w_1} where we plot $\OGW^{(\delta)}(k)$ for the case of an adiabatic perfect fluid ($c_s^2=w$) for several values of $0<w<1$. In contrast, for $c_s=1$ this resonance peak is absent, as can be seen in the right panel of fig.~\ref{fig:OGW_delta_w_1}.

What is important is that these findings from the $\delta$-peak case also apply in the more realistic situation when the peak in $\Pzeta$ has finite width $\Delta k$, as long as the peak is sufficiently narrow, i.e.~$\Delta k / \kstar < 1$. Then the resonance peak at $k=2 c_s \kstar$ persists, but with a finite amplitude. The intuition from the $\delta$-peak case however breaks down if the peak in $\Pzeta$ is broad. In that case the resonance in $\OGW$ is washed out and the GW spectrum instead exhibits a broad peak in the vicinity $k \sim \kstar$.  

We now move on to scalar power spectra with multiple narrow peaks, thus generalising the findings from \cite{Cai:2019amo} to general $w$. To be specific, we denote the loci of the peaks in $\Pzeta$ by $k_{\star i}$ with $i$ running over the number of peaks present. In the GW spectrum these will give rise to a series of resonance peaks at $k= 2 c_s k_{\star i}$. In addition, there will be resonance peaks in $\OGW$ due to `interactions' between different peaks in $\Pzeta$. The reason is that two different peaks in $\Pzeta$ can coincide with one another in \eqref{eq:OmegaGW-ts} as the integral over $d$ is performed. For a particular value of $k$ this overlap happens for $s=c_s^{-1}$ in which case one gets resonant amplification and thus a peak in $\OGW$. One can make this quantitative by again modelling the various peaks in $\Pzeta$ as $\delta$-distributions. One finds that for a scalar power spectrum with a set of peaks at $k=k_{\star i}$, the GW spectrum will exhibit a corresponding series of resonance peaks at $k=k_{\textrm{max}, ij}$ given by 
\begin{align}
\label{eq:kmaxij-def}
 k_{\textrm{max}, ij} = c_s \big(k_{\star i} + k_{\star j} \big) \, , \quad \textrm{with} \quad  k_{\textrm{max}, ij} > \big| k_{\star i} - k_{\star j} \big| \, , \quad \textrm{for} \quad c_s < 1 \, ,
\end{align}
while for $c_s=1$ there are no pronounced resonance peaks.
The constraint $k_{\textrm{max}, ij} > | k_{\star i} - k_{\star j} |$ simply stems from momentum conservation, and technically arises from the fact that the overlap between two different peaks in \eqref{eq:OmegaGW-ts}, while requiring $d>0$, also has to happen within the domain of integration $d \in [0,1]$. In practice, this implies that resonance due to interactions between different peaks in $\Pzeta$ can only occur if the two peaks in question are not too far apart in $k$-space. The dependence on $c_s$ can be understood from the fact that for a sharp peak in the primordial spectrum, most of the induced GW generation occurs when the relevant scalar mode crosses the sound horizon at $c_sk_{\star i}=aH$ \cite{Domenech:2021ztg}.

Following \cite{Fumagalli:2020nvq,Fumagalli:2021cel}, we now employ \eqref{eq:kmaxij-def} to understand the peak-structure of $\OGW$ due to a primordial feature in the scalar power spectrum, i.e.~$\Pzeta$ as given in (\ref{eq:P-of-k-sharp}, \ref{eq:P-of-k-resonant}). The idea is that if the amplitude of oscillation is sufficiently large, i.e.~$A_\textrm{lin}, A_\textrm{log} \sim \mathcal{O}(1)$, the scalar power spectrum can be regarded as a series of individual peaks. For the resonance peak analysis to apply these peaks should be narrow, which is the case as long as the frequencies $\omegalin$ and $\omegalog$ are not too small. We consider the case of a sharp feature and of a resonant feature in turn. 

\vspace{0.1cm}

\subsubsection{Sharp feature} 
A sharp feature in $\Pzeta$ is characterised by an oscillation in $k$, see the template in \eqref{eq:P-of-k-sharp}, so that peaks in $\Pzeta$ appear periodically in $k$ and are separated from one another by the period $2 \pi / \omegalin$. Interestingly, the peaks due to a sharp feature are narrow by default: Consider a sharp feature active around some scale $\kstar$, i.e.~\eqref{eq:P-of-k-sharp} with an envelope $\Pbar(k)$ that is enhanced over scales $k \sim \kstar$. To have visible oscillations, this envelope needs to be broad enough to accommodate at least a few periods of oscillation. Taking the period as a measure for the width $\Delta k$ of individual peaks, it follows that $\Delta k / k_{\star i} \sim \Delta k / \kstar < 1$. Also, if the enhancement in the envelope $\Pbar$ is due to the feature (and not some other effect), the amplitude of oscillations is $A_\textrm{lin} \sim \mathcal{O}(1)$, which is a consequence of the sharp feature preparing an excited state with corresponding quantization constraints \cite{Fumagalli:2020nvq}. All this implies that the resonance peak analysis is applicable to effectively all sharp features of phenomenological relevance.

Applying \eqref{eq:kmaxij-def}, one finds that the constraint in \eqref{eq:kmaxij-def} is generically satisfied for most (if not necessarily all) combinations of peaks, in virtue of $\Delta k / k_{\star i} \sim \Delta k / \kstar < 1$. As a result, the resonance peaks in $\OGW$ are arranged periodically in $k$-space with frequency:
\begin{align}
\label{eq:omegagwlin-analytic}
    \omegagwlin = c_s^{-1} \, \omegalin \, ,
\end{align}
which generalises the result from \cite{Fumagalli:2020nvq}. That is, for a sharp feature $\OGW$ will exhibit a modulation in $k$ with frequency $\omegagwlin$. We stress that the $c_s$ dependence is due to the fact that for sharp peaks most of the GWs are induced at sound horizon crossing. The lower the value of $c_s$, the smaller the separation between the resonant peaks in the gravitational wave spectrum and the larger the frequency.

Note that the map \eqref{eq:omegagwlin-analytic} between the frequency $\omegagwlin$ in $\OGW$ and $\omegalin$ in $\Pzeta$ is only one-to-one if the value of $c_s$ is known. In general, permitting uncertainty regarding the value of $c_s$ when the GWs were induced, there is a degeneracy between $c_s$ and $\omegalin$ for a given value of $\omegagwlin$.   

Another important question concerns the amplitude of oscillations in $\OGW$. While the consideration of resonance peaks alone cannot give an exact quantitative result, there are still lessons to be learned. For example, for the radiation-domination case it was observed that even if the original amplitude of oscillations in $\Pzeta$ is $A_\textrm{lin}=1$, the amplitude of oscillations in $\OGW$ is reduced to $\mathcal{O}(10 \%)$, see \cite{Fumagalli:2020nvq}. The reason is that the oscillation in $\OGW$ arises from a superposition of resonance peaks, which have an inherent width and shape,\footnote{For example, for a monochromatic scalar power spectrum, the resonance peak in $\OGW$, even if it diverges at the `maximum', has finite width and is not monochromatic, see the left panel of fig.~\ref{fig:OGW_delta_w_1}.} leading to an averaging out of the modulation. 

Here we wish to highlight that the shape and width of the resonance peak depends on the equation of state during GW production, and hence the amplitude of oscillations can in principle encode information about $w$. To illustrate this point, consider the $\delta$-peak case, where the width of the resonance peak in $\OGW$, when approaching the singularity or maximum at $k=2 c_s \kstar$, appears to become narrower, and the peak thus `spikier' as $w$ is increased, see the left panel of fig.~\ref{fig:OGW_delta_w_1}. This is because near this resonant peak, the induced GW spectrum for $b<0$ ($w>1/3$) diverges as $\Omega_{\rm GW}\propto |1-2c_sk_\star/k|^{2b}$ \cite{Domenech:2019quo,Domenech:2021ztg}. Thus, as $w$ is increased and $b$ becomes more negative, the induced GW spectrum grows faster near the singularity. Hence, the steeper `spike' in the GW spectrum that can be seen in the left panel fig.~\ref{fig:OGW_delta_w_1}. The case $1>b>0$ ($0<w<1/3$) is different as no divergence occurs \cite{Domenech:2019quo,Domenech:2021ztg}. Nevertheless, the induced GW spectrum exhibits a peak which behaves as $\Omega_{\rm GW}\propto 1- |1-2c_sk_\star/k|^{b}$. This peak is much broader than for $b\leq0$, and broader and broader as $w$ decreases. 
Returning to the case of a sharp feature one may hence expect that overall, for larger values of $w$ there is less averaging out, as the overlapping resonance peaks are narrower each, resulting in a larger value of the amplitude of oscillation in $\OGW$.\footnote{This expectation is based on the observation that for a narrow peak in $\Pzeta$ the shape of the corresponding resonance peak in $\OGW$ can be well-approximated by taking the $\delta$-peak result and smoothing it with the typical width of the peak, \cite{Pi:2020otn}. This has also been successfully applied to a sharp feature in \cite{Fumagalli:2020nvq}.} We will later find in numerical examples that this expectation is indeed verified.\footnote{However, for $w=c_s^2 \rightarrow 1$ the resonance peak in the kernel $\Tw(d,s)$ becomes extremely narrow and is pushed to the edge of the integration domain $s \rightarrow 1$, so that the region containing the resonance peak only gives a subleading contribution to the integral in \eqref{eq:OmegaGW-ts}, while for smaller values of $w$ it gives the dominant part. As a result, the relative amplitude  of the oscillations will eventually decrease for $w=c_s^2 \rightarrow 1$ and the oscillations will not modulate the maximum of $\OGW(k)$ any more, but will appear principally on the UV tail. This is consistent with the fact that for $c_s^2=1$ we do not expect oscillations in $\OGW(k)$ at all.}

\subsubsection{Resonant feature}
\label{sec:resonance-peak-analysis-resonant}
For a resonant feature the scalar power spectrum is described by the template \eqref{eq:P-of-k-resonant}. Here we assume that $A_\textrm{log} \sim \mathcal{O}(1)$ and $\omegalog \gtrsim \pi$ so that individual peaks are sufficiently pronounced and narrow for the resonance peak analysis to apply.\footnote{The condition $\omegalog \gtrsim \pi$ comes from the requirement $\Delta k / k_{\star i} < 1$, where we identify $k_{\star i}$ with maxima of the $\cos$ in \eqref{eq:P-of-k-resonant} and $\Delta k$ with the half-period enclosing $k_{\star i}$ between two zeros of the $\cos$.}  

As the oscillation is periodic in $\log(k)$, the loci of neighbouring peaks in $k$-space are related to one another by a common factor $e^{2 \pi / \omegalog}$. Because of this, different peaks in $\Pzeta$ can easily become sufficiently separated so that the condition in \eqref{eq:kmaxij-def} for resonant interaction is not satisfied. As a result, for a resonant feature only a subset of peaks in $\Pzeta$ can typically interact resonantly, resulting in a peak-structure in $\OGW$ that is in general more complicated than in the sharp feature case \cite{Fumagalli:2021cel}.

Firstly, one always obtains one set of resonance peaks from self-interactions of peaks in $\Pzeta$, i.e.~$k_{\textrm{max}, ii} = 2 c_s k_{\star i}$. As neighbouring peaks in $\Pzeta$ are related to one another by the factor $e^{2 \pi / \omegalog}$, this is inherited by the resonance peaks from self-interactions.

Secondly, we turn to resonance peaks from interactions between peaks in $\Pzeta$ that are nearest neighbours. These can only arise if the neighbours are sufficiently close so that the condition in \eqref{eq:kmaxij-def} is satisfied. This can be re-expressed as a lower bound on the frequency $\omegalog$, i.e.
\begin{align}
\label{eq:omegalogc-def}
    \omegalog > \omegalogc \, , \quad \textrm{with} \quad \omegalogc \equiv 2 \pi / \log \bigg( \frac{1+c_s}{1-c_s} \bigg) \, ,
\end{align}
generalising the result from \cite{Fumagalli:2020nvq,Fumagalli:2021cel} beyond the case of radiation-domination. If \eqref{eq:omegalogc-def} is satisfied, the GW spectrum exhibits another series of resonance peaks, with the individual peaks again separated from one another by the factor $e^{2 \pi / \omegalog}$. These fall somewhere in-between the peaks from self-interaction, with the precise relation between these two series depending on the value of $\omegalog$. Note that $\omegalogc$ depends explicitly on $c_s$, i.e.~the existence or not of these resonance peaks is directly affected by the value of $c_s$. In particular, note that for $c_s \rightarrow 0$ one has that $\omegalogc \rightarrow \infty$, and hence for small $c_s$ interactions between different peaks can only occur for very large values of $\omegalog$. An immediate consequence is that for small $c_s$ the GW spectrum due to resonant features is mostly simple, with just one series of peaks from self-interactions, except for very large frequencies $\omegalog$.

We can also use $\omegalogc$ to write down conditions for the existence of resonance peaks in $\OGW$ from interactions between peaks in $\Pzeta$ beyond nearest neighbours. E.g.~resonant interactions between next-to-nearest neighbours occur if $\omegalog > 2 \omegalogc$, between next-to-next-to-nearest neighbours for $\omegalog > 3 \omegalogc$ etc. This suggests that for sufficiently large $\omegalog$ the peak-structure of $\OGW$ can become quite complicated. This is however not the case, as the positions of the various peak series arrange themselves with respect to one another to give rise to a simpler picture.\footnote{E.g.~for sufficiently large $\omegalog$ the various peak series collapse into a single series of peaks, with the neighbouring maxima related by a factor $e^{\pi / \omegalog}$, i.e.~consistent with an oscillation with frequency $2 \omegalog$. See \cite{Fumagalli:2021cel} for details.} Unfortunately, the analysis based on resonance peaks is not best-suited for unlocking this emerging structure. Thus, in the next section we will review another method for studying the effect of a resonant feature on $\OGW$. Unlike the resonance peak analysis here, this method will also be valid for any vale of $A_\textrm{log}$. 

\subsection{Analytic templates for resonant features}
\label{sec:analytic-templates}
For a resonant feature, i.e.~a scalar power spectrum adhering to the template \eqref{eq:P-of-k-resonant}, we can also make analytic progress in understanding expression \eqref{eq:OmegaGW-ts} directly. 

Inserting \eqref{eq:P-of-k-resonant} into \eqref{eq:OmegaGW-ts}, the integrand can be organised as follows: There will be one $k$-dependent factor that comes from the two instances of the envelope $\Pbar$. In addition, there will be $k$-dependent pieces from the sinusoidal oscillation. What has been shown in \cite{Fumagalli:2021cel} is that for a resonant feature the $k$-dependence of the sinusoidal terms can be separated from their dependence on the integration variables $(d,s)$, so that the oscillatory behaviour in $k$ can be factored out from the integral entirely. One can make further progress if the envelope $\Pbar$ is broad, i.e.~if $\Pbar$ near its maximum can be taken as effectively constant over a wide interval in $k$.\footnote{This requirement is compatible with the assumptions in \eqref{eq:OmegaGW-ts} as long as the smallest wavenumber of the non-vanishing $\Pbar$ enters the horizon before the transition to radiation domination.} Then, if we evaluate $\OGW$ for values of $k$ within this interval, the envelope will effectively only contribute a constant and can thus also be factored out from the integral. We will later comment on the case of non-broad envelopes. The result is a fully analytic expression for $\OGW$ as a function of $k$, with coefficients that can be computed numerically once and for all. The computational steps are as in \cite{Fumagalli:2021cel} giving: 
\begin{align}
    \label{eq:OGW-res-template-C} \OGW(k) = \, & A_0 \left(\frac{k}{k_{\rm rh}}\right)^{-2b} \Pbar^2 \, \times \\
    \nonumber & \times \Bigg\{ 1 + A_\textrm{log}^2 \C_0(\omegalog) + A_\textrm{log} \C_1(\omegalog) \cos \bigg[ \omegalog \log \bigg( \frac{k}{2 \kref} \bigg) + \theta_1(\omegalog) \bigg] \\
    \nonumber & \hphantom{\times \Bigg\{ 1 + A_\textrm{log}^2 \C_0(\omegalog)} + A_\textrm{log}^2 \C_2(\omegalog) \cos \bigg[2 \omegalog \log \bigg( \frac{k}{2 \kref} \bigg) + \theta_2(\omegalog) \bigg] \Bigg\} \, .
\end{align}
The coefficients $A_0$, $\C_{0,1,2}(\omegalog)$ and $\theta_{1,2}(\omegalog)$ are given by (ratios of) integrals of the type $\iint \textrm{d}d \textrm{d}s \, \Tw(d,s) \, f(d,s,\omegalog)$ with the explicit expressions given in appendix \ref{app:analytic-template}.\footnote{We can also bring \eqref{eq:OGW-res-template-C} into the form of the template in \eqref{eq:resonant-template-intro}, by also factoring out the constant piece $A_\textrm{log}^2 \C_0(\omegalog)$. The quantities $\ObarGW$, $\mathcal{A}_{\textrm{log},1}$ and $\mathcal{A}_{\textrm{log},2}$ in \eqref{eq:resonant-template-intro} are thus given by
$$
    \ObarGW(k) = A_0 \left(\frac{k}{k_{\rm rh}}\right)^{-2b} \Pbar^2 \big(1 + A_\textrm{log}^2 \C_0(\omegalog) \big) , \quad \mathcal{A}_{\textrm{log},1} = \frac{A_\textrm{log} \C_1(\omegalog)}{1 + A_\textrm{log}^2 \C_0(\omegalog)} , \quad \mathcal{A}_{\textrm{log},2} = \frac{A_\textrm{log}^2 \C_2(\omegalog)}{1 + A_\textrm{log}^2 \C_0(\omegalog)} .
$$}

The main observation from \eqref{eq:OGW-res-template-C} is that the modulations in $\OGW$ can be written as a superposition of two oscillatory pieces, one with the original frequency of oscillation $\omegalog$ from $\Pzeta$, and one with twice that frequency. The amplitudes and phases of these two pieces are set by the $\omegalog$-dependent coefficients $\C_{1,2}(\omegalog)$ and $\theta_{1,2}(\omegalog)$, and in the case of the amplitude also by $A_\textrm{log}$.

The new result from this work is that all the coefficients $A_0$, $\C_{0,1,2}(\omegalog)$ and $\theta_{1,2}(\omegalog)$ also depend on the equation of state of the universe and the value of $c_s$ (through $\Tw$). In fact, these coefficients are the only pieces that contain information about the equation of state and $c_s$ (together with the overall factor $(k / k_{\textrm{rh}})^{-2b}$). That is, different values for $w$ and $c_s$ do not affect the oscillatory parts of $\OGW(k)$ per se, but only their amplitudes and phases, and also the overall amplitude of the GW spectrum. This is different from the sharp feature case, where the frequency of oscillation $\omegagwlin$ in $\OGW$ depends on both the frequency $\omegalin$ in the scalar power and the value of $c_s$, see eq.~\eqref{eq:omegagwlin-analytic}. In contrast, for the resonant feature case the frequency $\omegalog$ appearing in $\OGW$ is the same as that in $\Pzeta$ irrespective of the values of $w$ or $c_s$.

For the case of radiation-domination it was observed in \cite{Fumagalli:2021cel} that for small values of $\omegalog$ the oscillatory piece with frequency $\omegalog$ dominates over that with $2 \omegalog$. As the frequency is increased, the relative amplitude of the term with frequency $2 \omegalog$ grows until for sufficiently large $\omegalog$ this piece dominates. Thus, for small and large values of $\omegalog$ the GW spectrum $\OGW$ effectively exhibits a single modulation with frequency $\omegalog$ or $2 \omegalog$, respectively, while for intermediate values of $\omegalog$ both oscillatory pieces contribute, resulting in a complicated oscillatory pattern of $\OGW$.\footnote{This is consistent with the results from the resonance peak analysis in sec.~\ref{sec:resonance-peak-analysis} as it should.} Given the dependence of the amplitudes and phases on the equation of state, these observations are expected to change with $w$ (and $c_s$). In particular, as we will find later, the threshold values between the regimes just described get shifted with $w$, resulting in a different oscillatory pattern in $\OGW$ for the same $\omegalog$ but different $w$.

We close by commenting on the situation when the envelope $\Pbar(k)$ is not broad. While the envelope-dependent terms cannot be separated from the integration over $(d,s)$, the sinusoidal pieces can still be factored out as before. As a result, the GW spectrum can still formally be written as in \eqref{eq:OGW-res-template-C}, but the coefficients $A_0$, $\C_{0,1,2}$ and $\theta_{1,2}$ are now also $k$-dependent through $\Pbar$. For $A_0$ and $\C_{0,1,2}$ this implies that both the overall GW spectrum and also the two oscillatory terms come with a $k$-dependent envelope. The $k$-dependence of $\theta_{1,2}$ in principle changes the behaviour of the oscillation. However, as the variation in $k$ of the envelope $\Pbar$ has to be slower than that of the oscillation by definition, one expects that the $k$-dependence of $\theta_{1,2}$ is sufficiently mild as to not change the oscillatory pieces substantially. This is confirmed by numerical tests in \cite{Fumagalli:2021cel}, where it was observed that the oscillatory part of $\OGW$ is still well-approximated by two sinusoidal modulations with frequencies $\omegalog$ and $2 \omegalog$, respectively, for various choices of non-broad envelopes.     

\section{Induced gravitational waves from sharp features}
\label{sec:sharp}
In this section we compute the spectrum of induced GWs for several examples of a sharp feature in the scalar power spectrum, focussing in particular on the effect of the expansion history on the resulting GW spectrum. To be specific, we assume that during production of the GWs the universe is either described by an adiabatic perfect fluid ($c_s^2=w$) or a canonical scalar field $(c_s^2=1)$ and evaluate the GW spectrum for various values of the equation of state parameter $w$.

\begin{figure}[t]
\centering
\begin{overpic}[width=0.80\textwidth]{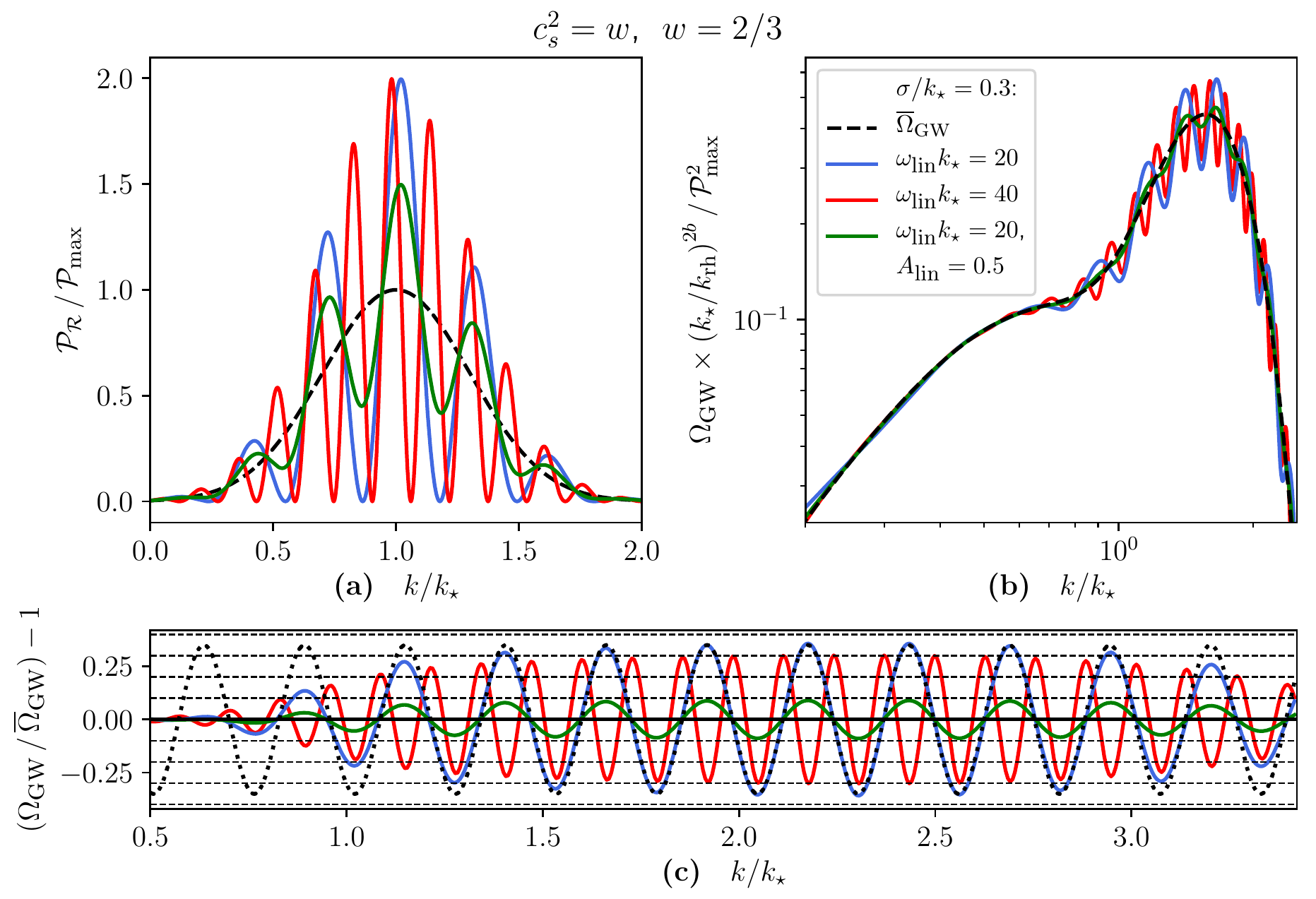}
\end{overpic}
\caption{$\Pzeta$, $\OGW$ and $(\OGW / \ObarGW-1)$ vs.~$k/k_\star$ for GWs induced due to a sharp feature during a phase with $c_s^2=w$ and $w=2/3$. The scalar power spectrum is given in \eqref{eq:P-sharp-turn-Gauss} with model parameters recorded in the legend. The black dashed lines in panels (a) and (b) are the envelope $\Pbar(k)$ of the scalar power spectrum and the corresponding GW spectrum $\ObarGW(k)$, respectively. The dotted line in panel (c) is sinusoidal with frequency $\omegagwlin = w^{-1/2} \omegalin$ for $\omegalin \kstar=20$, which we find to closely match the oscillations of the blue curve across the principal peak (and even the UV tail) of $\OGW$.}
\label{fig:P_O_ratios_w_sig0p3_om20_om40_w2o3}
\end{figure}

For a sharp feature the scalar power spectrum can be written as in \eqref{eq:P-of-k-sharp}, i.e.~it is given by a smooth envelope $\Pbar(k)$ modulated by sinusoidal oscillations that are periodic in $k$. The shape of the envelope $\Pbar(k)$ depends on the precise realisation of the sharp feature and is thus model-dependent. However, the subject of primary interest are the oscillations in $\Pzeta$ and their imprint on $\OGW$, which will be largely independent of the exact details of the envelope. Thus, to be specific, here we choose a Gaussian peak for the envelope, so that the scalar power spectrum used for numerical examples is given by
\begin{align}
\label{eq:P-sharp-turn-Gauss}
    \Pzeta(k) = \mathcal{P}_\textrm{max} \, e^{-\frac{(k - k_\star)^2}{2 \sigma^2}} \Big[1 + A_\textrm{lin} \sin \big( \omegalin k \big) \Big] \, .
\end{align}
Near the maximum this was also shown to be a good approximation to the exact scalar power spectrum for a sharp feature realised by a strong sharp turn in the inflationary trajectory with constant turn rate \cite{Fumagalli:2020nvq}.\footnote{Also see section 2.2 in \cite{Fumagalli:2021cel} for a dictionary between the quantities in \eqref{eq:P-sharp-turn-Gauss} and the parameters describing the turn.} A sharp feature only affects a finite range of scales and hence the expression \eqref{eq:P-sharp-turn-Gauss} should strictly speaking only be used for values of $k$ in the vicinity of $\kstar$. Note that \eqref{eq:P-sharp-turn-Gauss} exhibits oscillations on the UV tail for $k \gg \kstar$, which may not be the case for a realistic sharp feature. The presence of this oscillatory UV tail however does not pose any problems, as it does not impact the GW spectrum across the most enhanced scales, but only affects the UV tail where the induced GW spectrum decays exponentially fast. Interestingly, the oscillatory UV tail of $\Pzeta$ leads to corresponding oscillations on the UV tail of the induced GW spectrum, as we show in appendix \ref{app:IRUVtail}. A similar issue exists in the IR, which we address by cutting off expression \eqref{eq:P-sharp-turn-Gauss} at some value $k_\textsc{ir}$. We checked that the induced GW spectrum across its peak is not affected by the choice of $k_\textsc{ir}$ and that it exhibits the expected IR behaviour.

\begin{figure}[t]
\centering
\begin{overpic}[width=0.80\textwidth]{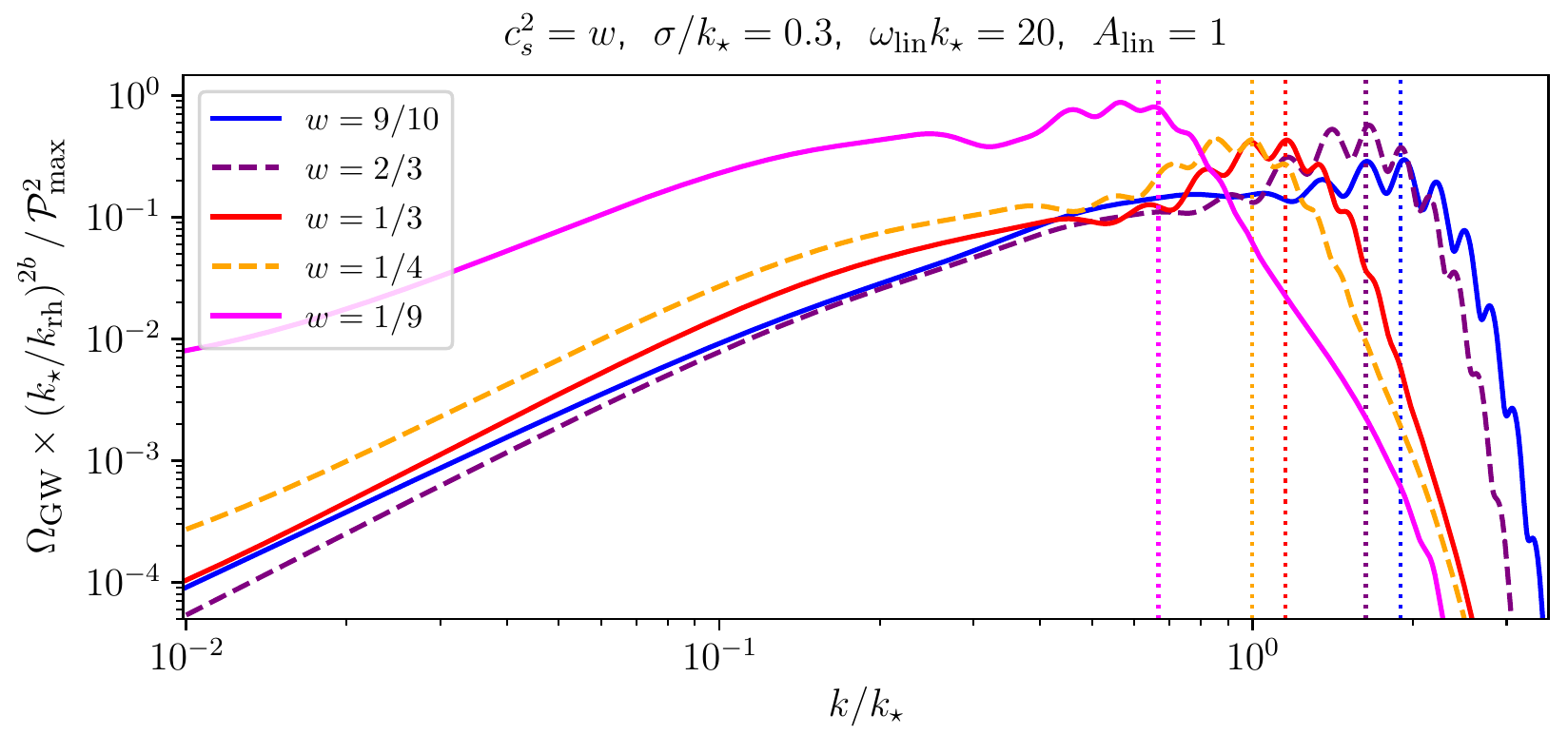}
\end{overpic}
\caption{$\OGW$ vs.~$k/k_\star$ for GWs due to a sharp feature in $\Pzeta$ as given by \eqref{eq:P-sharp-turn-Gauss} with $(\sigma/\kstar, \, \omegalin \kstar, \, A_\textrm{lin})$ $=$ $(0.3, \, 20, \, 1)$, induced during a phase with $c_s^2=w$ for various values of $w<1$ (see legend). In all cases the GW spectrum exhibits a principal peak modulated by $\mathcal{O}(10 \%)$ oscillations (that also affect the UV tail for larger values of $w$), but the overall shape of the GW spectrum and the exact frequency and amplitude of the oscillations depend on $w$. The dotted vertical lines denote $k=2 w^{1/2} \kstar$, i.e.~the expected position of the principal in $\OGW$ from resonance considerations. Note that in this figure the induced GW spectrum for $w=1/9$ has the largest amplitude because of the rescaling $(k_\star/k_{\rm rh})^{2b}$. Without this factor, the spectrum with $w=9/10$ has the largest amplitude due to the faster redshifting of the background compared to the energy density of gravitational waves.}
\label{fig:Omega_w_s0p3_w20}
\end{figure}

\subsection{Adiabatic perfect fluid ($c_s^2=w$)}
\label{sec:sharp-apf}
Here we insert \eqref{eq:P-sharp-turn-Gauss} for various choices of model parameters into \eqref{eq:OmegaGW-ts} and compute $\OGW$ numerically. To be specific, for all examples we choose $\sigma / \kstar =0.3$, so that the envelope $\Pbar$ is moderately narrow, consistent with the expectation for a sharp feature. To examine the effect of the oscillations on $\OGW$ we consider different combinations of $\omegalin$ and $A_\textrm{lin}$ and in particular $(\sigma/\kstar, \, \omegalin \kstar, \, A_\textrm{lin})$ $=$ $(0.3, \, 20, \, 1)$, $(0.3, \, 40, \, 1)$, $(0.3, \, 20, \, 0.5)$. In panel (a) of fig.~\ref{fig:P_O_ratios_w_sig0p3_om20_om40_w2o3} we plot $\Pzeta(k)$ for these three parameter choices. Note that $\omegalin \kstar=20$ lies close to the lower end of frequencies that still lead to visible modulations of the most enhanced scales for this choice of envelope, i.e.~for this choice $\Pzeta$ exhibits (only) three peaks with $\Pzeta > \mathcal{P}_\textrm{max}$.

In panel (b) of fig.~\ref{fig:P_O_ratios_w_sig0p3_om20_om40_w2o3} we show the corresponding results for $\OGW$ for these three examples and for an equation-of-state parameter $w=2/3$. For all three examples the GW spectrum exhibits a smooth IR tail, a principal peak modulated by oscillations, and a steep UV tail that still exhibits modulations. The behaviour on the IR and UV tails can also be understood analytically, see appendix \ref{app:IRUVtail} for details. This structure is already familiar from the radiation-domination-case, see \cite{Fumagalli:2020nvq}, and here we find that this also persists for a different equation of state. More precisely, the three spectra can be described as a modulation over an otherwise smooth background $\ObarGW$, shown as the black dashed line, and which corresponds to the GW spectrum for the envelope $\Pbar$. 

To better visualise the oscillatory component of $\OGW$, in panel (c) of fig.~\ref{fig:P_O_ratios_w_sig0p3_om20_om40_w2o3} we plot the ratio $\OGW / \ObarGW$. The oscillation is sinusoidal in $k$, with an amplitude that is approximately constant over the scales of the principal peak, and with frequency $\omegagwlin = w^{-1/2} \omegalin$ as predicted by the resonance peak analysis in sec.~\ref{sec:resonance-peak-analysis}. For example, the dotted curve in panel (c) is a sinusoid with frequency $\omegagwlin$, which we find to closely match the blue curve over the principal peak for $\omegalin \kstar =20$. Also note that the amplitude of oscillation for the red curve ($\omegalin \kstar=40$) is slightly smaller than that of the blue curve ($\omegalin \kstar=20$), with value $0.29$ vs.~$0.33$, even though in both cases the amplitude of oscillation in $\Pzeta$ is $A_\textrm{lin}=1$. This effect has already been observed for GWs sourced during radiation-domination, and can be explained using the resonance peak analysis of sec.~\ref{sec:resonance-peak-analysis}: According to this, the oscillation in $\OGW$ arises from the superposition of individual resonance peaks, which have an inherent width each. For larger $\omegalin$ more resonance peaks overlap over the same interval in $k$, leading to an averaging out between peaks and hence  a reduced amplitude.

\begin{figure}[t]
\centering
\begin{overpic}[width=0.80\textwidth]{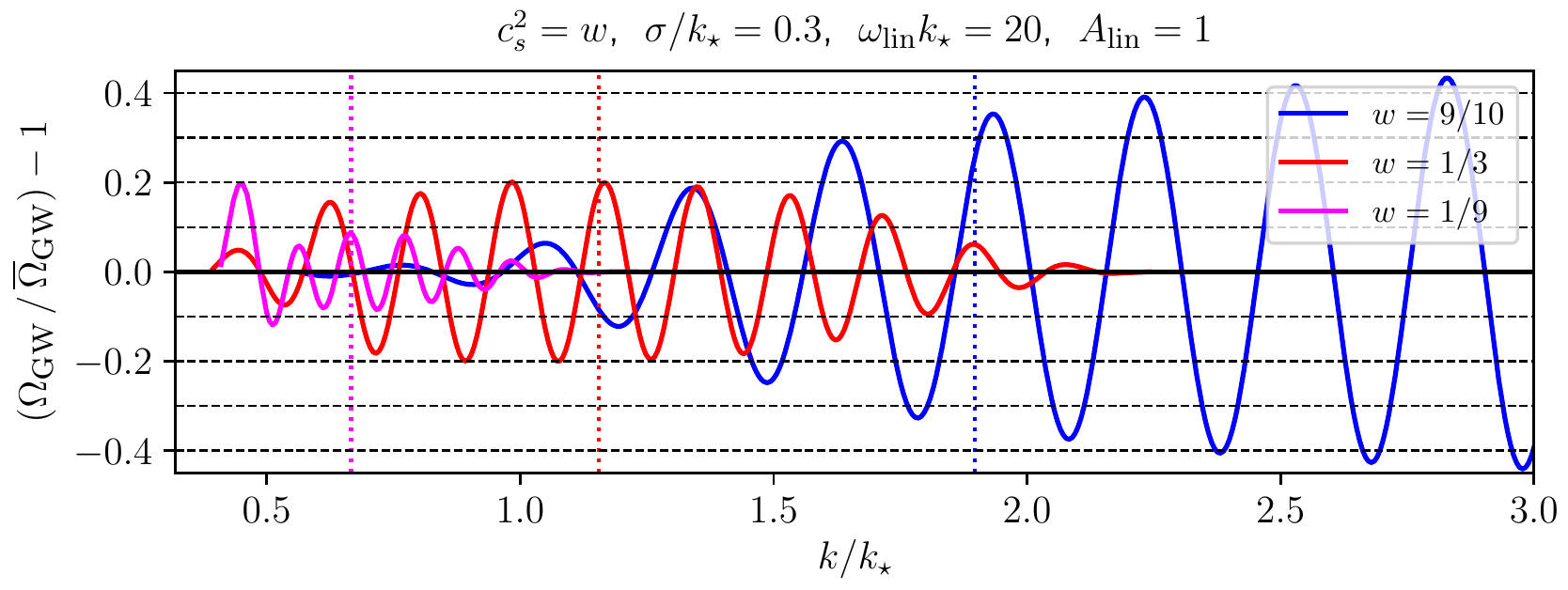}
\end{overpic}
\caption{$(\OGW / \ObarGW -1)$ vs.~$k/\kstar$, i.e.~the oscillatory component of $\OGW$, given by the ratio of the full GW spectrum $\OGW$ over the smooth background spectrum $\ObarGW$ for the examples in fig.~\ref{fig:Omega_w_s0p3_w20} with $w=1/9, \, 1/3, \, 9/10$. The dotted vertical lines denote $k=2 w^{1/2} \kstar$, i.e.~the expected position of the principal peak in $\OGW$ from resonance considerations. We observe that the amplitude of oscillations near the respective principal peak grows as $w$ is increased. Also note the large amplitude of oscillations in the UV for $w>1/3$.}
\label{fig:Ratios_w_s0p3_w20}
\end{figure}

We now examine how different values of the equation of state parameter during GW production affect the resulting GW spectrum for the same primordial feature. To this end in fig.~\ref{fig:Omega_w_s0p3_w20} we plot $\OGW$ for the example with $(\sigma/\kstar, \, \omegalin \kstar, \, A_\textrm{lin})$ $=$ $(0.3, \, 20, \, 1)$ for various values of $w<1$, hence respecting the positivity energy conditions of general
relativity. The main observation is that in all cases the GW spectrum shows the characteristic signature of a sharp feature, i.e.~a principal peak modulated by $\mathcal{O}(10 \%)$ oscillation. Hence a sharp feature during inflation can potentially be detected by its oscillatory contribution to the SGWB for a wide range of values for $w$ at the time of GW production. In other words, oscillations in the SGWB are a generic prediction of a sharp feature during inflation, even when accounting for the uncertainty regarding the thermal history of the universe after inflation. 

Quantitative details of the GW spectra in fig.~\ref{fig:Omega_w_s0p3_w20} do however depend on $w$. For example, the location of the principal peak in $\OGW$ varies with $w$. This can be understood as follows. The overall shape of $\OGW$ (i.e.~ignoring the modulations) is given by the GW spectrum $\ObarGW$ for the envelope $\Pbar$. As here the envelope is moderately narrow, $\ObarGW$ will exhibit a resonance peak at $k = 2 w^{1/2} \kstar$, see e.g.~sec.~\ref{sec:resonance-peak-analysis}. This is indicated by the dotted vertical lines in fig.~\ref{fig:Omega_w_s0p3_w20}, which we observe to adequately predict the position of the principal peak for every choice of $w$.  

To assess the effect of $w$ on the oscillations, in fig.~\ref{fig:Ratios_w_s0p3_w20} we plot the ratio $\OGW/ \ObarGW$ for the examples in fig.~\ref{fig:Omega_w_s0p3_w20}, limited to the choices $w=1/9, \, 1/3, \, 9/10$ for better visibility, but which is sufficient to illustrate the main points. Firstly, as one can check explicitly, the frequency of the oscillation is consistent with the prediction $\omegagwlin = w^{-1/2} \omegalin$ from the resonance peak analysis, see eq.~\eqref{eq:omegagwlin-analytic}. What is particularly important for a potential future detection of these oscillations is the amplitude of the modulations near the maximum of $\OGW$. In fig.~\ref{fig:Omega_w_s0p3_w20} we denote the approximate position of the maximum by a dotted vertical line.  Labelling the amplitude $\mathcal{A}_\textrm{lin}$, this can be read from fig.~\ref{fig:Ratios_w_s0p3_w20} and panel (c) of fig.~\ref{fig:P_O_ratios_w_sig0p3_om20_om40_w2o3} to find:\footnote{We also display the values of $\mathcal{A}_\textrm{lin}$ for $w=1/4, \, 8/10, \, 0.95$, which were obtained by the same method.}
\begin{center}
\begin{tabular}{| l || c | c | c | c | c | c | c |} \hline
  $w$ & 1/9 & 1/4 & 1/3 & 2/3 & 8/10 & 9/10 & 0.95 \\ \hline
  $\mathcal{A}_\textrm{lin}$ & 0.09 & 0.16 & 0.20 & 0.33 & 0.37 & 0.35 & 0.27 \\ \hline
\end{tabular}
\end{center}
The main observation is that $\mathcal{A}_\textrm{lin}$ increases with $w$ before this trend slightly reverses for $w \geq 9/10$. Nevertheless, for $w=0.95$ the amplitude $\mathcal{A}_\textrm{lin}$ is still larger than for $w=1/3$.  For even larger values of $w$ approaching $w \rightarrow 1$ the amplitude however quickly drops to effectively zero.\footnote{Here this is tied to the fact that we have chosen $w=c_s^2$ so that by letting $w \rightarrow 1$ we are also sending $c_s \rightarrow 1$. For $c_s=1$ resonant amplification of GWs is not allowed kinematically, while it is crucial for having visible oscillations across the principal peak in $\OGW(k)$, see sec.~\ref{sec:resonance-peak-analysis} and the following sec.~\ref{sec:sharp-csf}. The drop-off for $w \rightarrow 1$ is thus consistent with resonant amplification becoming increasingly constrained kinematically.} In fact, we expect the values in the table above to be close to the maximally attainable figures for a sharp feature for each choice of $w$. As observed before in the context of fig.~\ref{fig:P_O_ratios_w_sig0p3_om20_om40_w2o3}, the amplitude of oscillations and  $\omegalin$ are inversely related, so to further increase the amplitude one would need to decrease $\omegalin$. But this is barely possible, as the value $\omegalin \kstar=20$ employed here is close to the minimal value at which the period of oscillation becomes comparable to the width of the peak in $\Pbar$, and oscillations would cease to exist. This suggests that oscillations in $\OGW$ due to a sharp feature cannot significantly exceed $\mathcal{A}_\textrm{lin} \approx 20 \%$ if the post-inflationary period is radiation-dominated. In turn, a detection of an oscillation due to a sharp feature with $\mathcal{A}_\textrm{lin} > 20 \%$ would hint at the post-inflationary universe not dominated by radiation but described by a stiffer $w$. 

Returning to fig.~\ref{fig:Ratios_w_s0p3_w20}, also note that for $w \leq 1/3$ the amplitude of oscillations typically decreases along the UV tail immediately after the principal peak. In contrast, for $w> 1/3$ the amplitude of oscillations does not decrease significantly in the UV, or even keeps growing. The presence of these oscillations in the UV is due to the fact that $\Pzeta$ in \eqref{eq:P-sharp-turn-Gauss} exhibits modulations along its UV tail, which in turn induce oscillations in the UV in $\OGW$, see appendix \ref{app:IRUVtail}. For $w > 1/3$ this UV behaviour already sets in at moderately large values of $k /\kstar$, while for $w \leq 1/3$ this only occurs for larger values of $k /\kstar$ and hence is not visible in fig.~\ref{fig:Ratios_w_s0p3_w20}. In any case, as the overall amplitude of $\OGW$ drops (exponentially) fast in the UV, it is questionable whether the oscillations in the UV could ever be detected.\footnote{Also note that for a realistic sharp feature, the oscillations in $\Pzeta$ and hence in $\OGW$ are expected to eventually cease to exist for $k \gg \kstar$, so that the induced GW spectrum simply decays exponentially deep in the UV.}

\begin{figure}[t]
\centering
\begin{overpic}[width=0.80\textwidth]{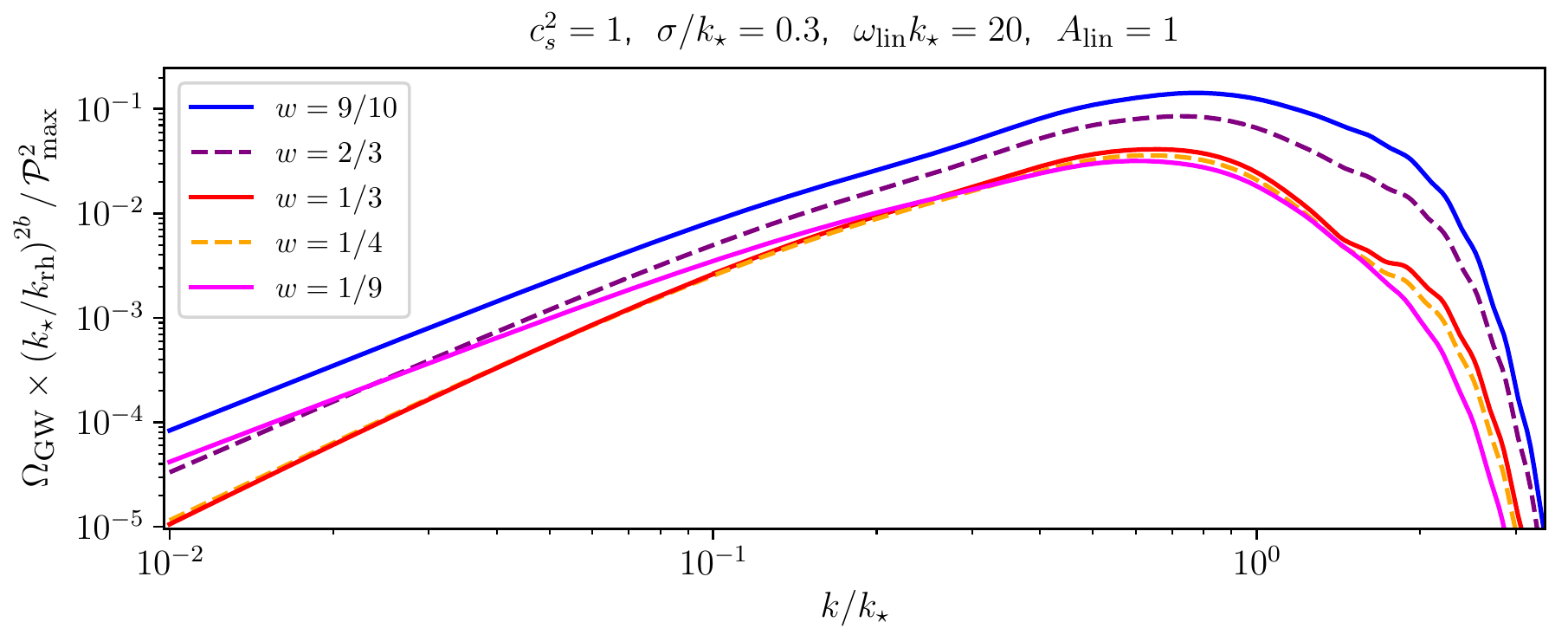}
\end{overpic}
\caption{$\OGW$ vs.~$k/k_\star$ for GWs due to a sharp feature in $\Pzeta$ as given by \eqref{eq:P-sharp-turn-Gauss} with $(\sigma/\kstar, \, \omegalin \kstar, \, A_\textrm{lin})$ $=$ $(0.3, \, 20, \, 1)$, induced during a phase with $c_s^2=1$ for various values of $w<1$ (see legend). The GW spectra do not exhibit any visible oscillations across the most enhanced scales, and at most show some mild modulations on the UV tail. This is consistent with analytical expectations: For $c_s=1$ resonant amplification of GWs is kinematically forbidden, leading to broadly-peaked GW spectra and hence a smoothing out of oscillations.}
\label{fig:Omega_1_s0p3_w20}
\end{figure}

\subsection{Canonical scalar field ($c_s^2=1$)}
\label{sec:sharp-csf}
For $c_s^2=1$ the spectrum of induced GWs does not exhibit any resonance peaks as resonant amplification of GWs is kinematically forbidden. Consequently, even for a monochromatic scalar power spectrum, the resulting GW spectrum is generically broad without sharp peaks, see the right panel of fig.~\ref{fig:OGW_delta_w_1}. The expectation thus is that for $c_s^2=1$ the GW spectrum due to a sharp feature will not exhibit any modulations, as this oscillation would have to have its origin in a superposition of resonance peaks, see sec.~\ref{sec:resonance-peak-analysis}. This is indeed what is observed in practice. In fig.~\ref{fig:Omega_1_s0p3_w20} we plot $\OGW$ for the example with $(\sigma/\kstar, \, \omegalin \kstar, \, A_\textrm{lin})$ $=$ $(0.3, \, 20, \, 1)$ for various values of $w<1$. For all cases the GW spectrum exhibits a broad peak in the vicinity of $k \sim \kstar$ that does not have any visible modulations. The only hint of an oscillation are mild wiggles along the UV tail. Note that the amplitude of $\OGW$ at the maximum is smaller by roughly one order of magnitude compared to the case with $c_s^2=w$ except for $w=9/10$, cf.~fig.~\ref{fig:Omega_w_s0p3_w20}.

This result also shows that the presence of oscillations in $\Pzeta$ does not automatically imply the existence of modulations in the induced GW spectrum, whose appearance crucially depends on the properties of the post-inflationary universe.

\section{Induced gravitational waves from resonant features}
\label{sec:resonant}
In this section we analyse the spectrum of induced GWs due to a resonant feature, again considering the case of a universe described by an adiabatic perfect fluid ($c_s^2=w$) and a canonical scalar field ($c_s^2=1$) in turn.

\begin{figure}[t]
\centering
\begin{overpic}[width=0.80\textwidth]{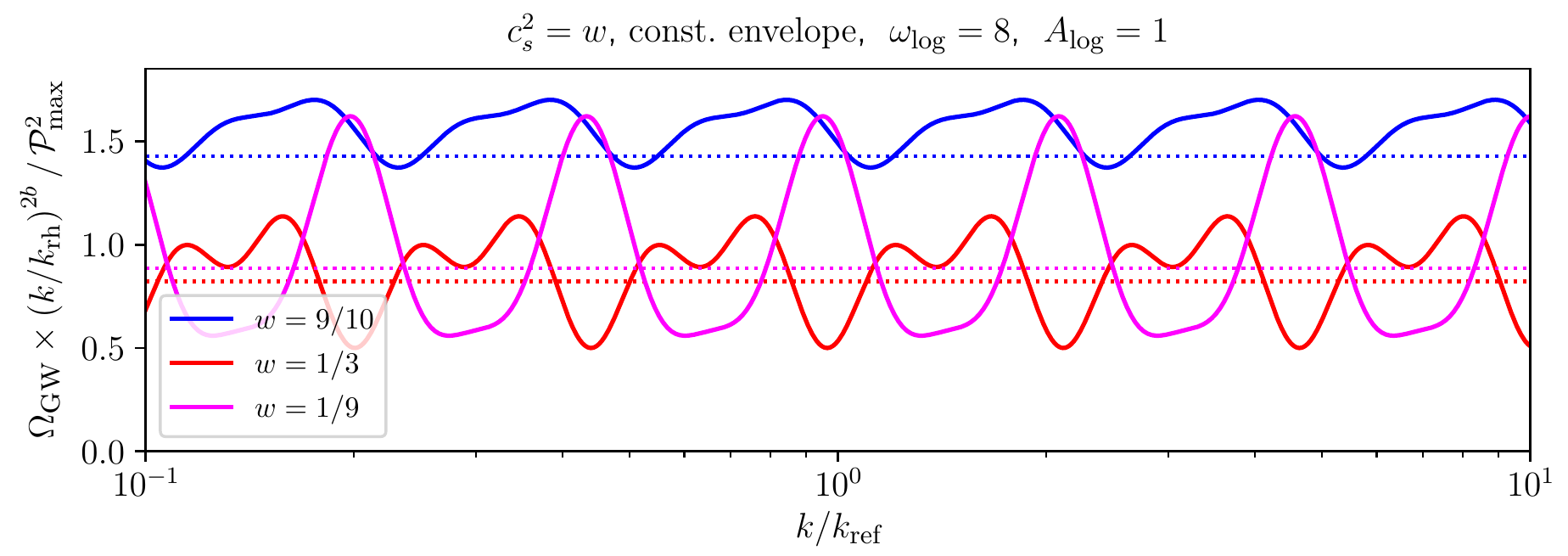}
\end{overpic}
\caption{$\OGW$ vs.~$k / \kref$ for a resonant feature in $\Pzeta$ as given in \eqref{eq:P-of-k-resonant} for constant envelope $\Pbar=\mathcal{P}_\textrm{max}$, frequency $\omegalog=8$ and amplitude $A_\textrm{log}=1$ for the choices $w=1/9, \, 1/3, \, 9/10$. The horizontal dotted lines correspond to the value of $A_0$ as it appears in \eqref{eq:OGW-res-template-C}, i.e.~the contribution to $\OGW$ from just the envelope of $\Pzeta$. The value $\omegalog=8$ was chosen as this lies below the criticial frequency $\omegalogc$ defined in \eqref{eq:omegalogc-def} for $w=1/9$, but above it for $w=1/3$ and $w=9/10$, leading to a qualitatively different oscillatory pattern in $\OGW$ for $w=1/9$ vs.~$w=1/3 , \, 9/10$. For better visibility of the oscillatory behaviour we multiplied $\OGW$ by the factor $(k / k_\textrm{rh})^{2b}$.}
\label{fig:Omega_w_res_top_om8_ws}
\end{figure}

\subsection{Adiabatic perfect fluid ($c_s^2=w$)}
\label{sec:resonant-apf}
As described in sec.~\ref{sec:analytic-templates}, for a resonant feature \eqref{eq:P-of-k-resonant} with a constant envelope $\Pbar$, the spectrum of induced GWs as a function of $k$ can be computed analytically, with the result for $\OGW(k)$ given in \eqref{eq:OGW-res-template-C}. For non-constant, but sufficiently broad envelopes $\Pbar(k)$ this is expected to still give a good approximation to $\OGW(k)$ over the most enhanced scales. As we will later see, many findings for the constant-envelope-case will also apply to or can be generalised for non-constant and even non-broad envelopes. Thus, we find it convenient to start with the case with $\Pbar=\mathcal{P}_\textrm{max}=\textrm{const.}$, where the analytic understanding is the most developed.

We are mainly interested in the oscillatory behaviour of $\OGW(k)$ and its dependence on the equation-of-state parameter $w$. 
In fig.~\ref{fig:Omega_w_res_top_om8_ws} we plot $\OGW(k)$ computed via \eqref{eq:OmegaGW-ts} for a resonant feature with constant envelope $\Pbar=\mathcal{P}_\textrm{max}$, frequency $\omegalog=8$ and amplitude $A_\textrm{log}=1$ for the choices $w=1/9, \, 1/3, \, 9/10$, i.e.~for radiation domination and for one example each of a softer and a stiffer equation of state. For better visibility of the oscillatory behaviour, in fig.~\ref{fig:Omega_w_res_top_om8_ws} we removed the $k$-dependent prefactor from $\OGW(k)$ by multiplying with $(k/k_\textrm{rh})^{2b}$. We make the following observations: 
\begin{itemize}
\item For $w=1/3$ the oscillation in $\OGW$ is given by a superposition of an oscillatory piece with frequency $\omegalog$ and one with frequency $2 \omegalog$, consistent with the analytic result in \eqref{eq:OGW-res-template-C}. Here the two oscillatory contributions have a comparable amplitude, resulting in a non-trivial oscillatory pattern with a double-peak structure. This superposition is also visible for $w=9/10$, albeit here the two oscillatory pieces combine to produce a series of irregularly shaped single peaks. For $w=1/9$, however, we have a single series of peaks with frequency $\omegalog$. This can be understood using the resonance-peak-analysis in sec.~\ref{sec:resonance-peak-analysis}, which states that there is a qualitative difference in $\OGW$ depending on whether $\omegalog$ is larger or smaller than the critical frequency $\omegalogc$: for $\omegalog < \omegalogc$ the resonance-peak-analysis predicts $\OGW$ to just exhibit a single series of peaks with frequency $\omegalog$, whereas for $\omegalog > \omegalogc$ one expects additional peaks in $\OGW$. The critical frequency $\omegalogc$ is given in \eqref{eq:omegalogc-def} and for the case of an adiabatic perfect fluid depends on $w$, with its value for various choices of $w$ recorded in the table below:
\begin{center}
\begin{tabular}{| l || c | c | c | c | c |} \hline
  $w$ & 1/9 & 1/4 & 1/3 & 2/3 & 0.9 \\ \hline
  $\omegalogc$ & 9.06 & 5.72 & 4.77 & 2.74 & 1.73 \\ \hline
\end{tabular}
\end{center}
Indeed, the value $\omegalog=8$ falls below the critical value for $w=1/9$, and the observed peak-structure of $\OGW$ in fig.~\ref{fig:Omega_w_res_top_om8_ws} is as predicted by the resonance-peak-analysis.
\item Another important observation concerns the relative amplitude of the oscillation, which differs considerably between the three examples in fig.~\ref{fig:Omega_w_res_top_om8_ws}. In decreasing order, the relative amplitude of oscillation, estimated as $\mathcal{A}=({\textrm{max}}-{\textrm{min}}) / ({\textrm{max}}+{\textrm{min}})$, is given by $\mathcal{A} \sim 0.5$ for $w=1/9$, $\mathcal{A} \sim 0.4$ for $w=1/3$ and $\mathcal{A} \sim 0.1$ for $w=9/10$. Interestingly, here the example with the softer (stiffer) equation of state has a larger (smaller) amplitude compared to the radiation case, while for a sharp feature this hierarchy was inverted, see e.g.~fig.~\ref{fig:Ratios_w_s0p3_w20}. We will return to this observation shortly, once we also consider other values of $\omegalog$.
\item While the relative amplitude of oscillation in fig.~\ref{fig:Omega_w_res_top_om8_ws} is larger for a softer equation of state, the overall level of $\OGW$ exhibits a different hierarchy, with the example with $w=9/10$ having the highest overall amplitude on average, followed by the one with $w=1/9$ and $w=1/3$, albeit with only a factor of $\sim 2$ between the largest and the smallest case.\footnote{Also note that the oscillation is not centred about the value $A_0$, denoted by the dotted line in fig.~\ref{fig:Omega_w_res_top_om8_ws}, corresponding to the GW spectrum for a constant scalar power spectrum in absence of oscillations. The reason is that the oscillations in $\Pzeta$ for a resonant feature also contribute a constant piece to $\OGW$, see the term $A_\textrm{log}^2 \C_0(\omegalog)$ in \eqref{eq:OGW-res-template-C}.} Regarding the detectability of the oscillations, there are hence two countervailing effects at play. While a smaller relative amplitude will make oscillations harder to detect, an overall larger value of $\OGW$ will make it easier to reconstruct the signal and hence the oscillations, and vice versa. We leave it for further work to determine the relative importance of these effects.    
\end{itemize}

\begin{figure}[t]
\centering
\begin{overpic}[width=0.90\textwidth]{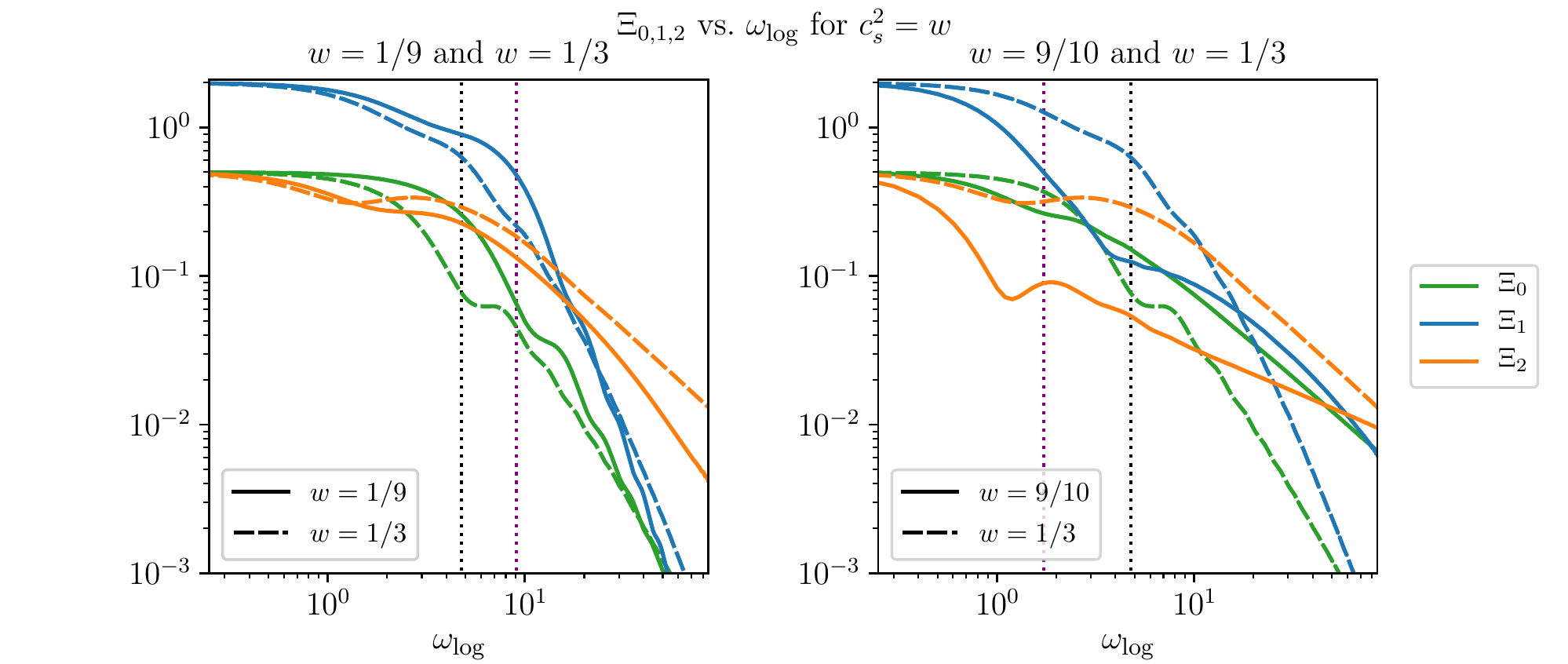}
\end{overpic}
\caption{$\C_{0,1,2}$ as appear in \eqref{eq:OGW-res-template-C} vs.~$\omegalog$ for $c_s^2=w$. In the left panel we display results for $w=1/9$ and $w=1/3$, and in the right for $w=9/10$ and $w=1/3$. The black dotted line indicates $\omegalogc$ as defined in \eqref{eq:omegalogc-def} for $w=1/3$, while the purple dotted line denotes $\omegalogc$ for $w=1/9$ or $w=9/10$, respectively.}
\label{fig:Xi012_w_w1o9_w1o3_w9o10}
\end{figure}

Fig.~\ref{fig:Omega_w_res_top_om8_ws} shows results for the parameter choice $\omegalog=8$ and $A_\textrm{log}=1$. Using the analytic results from sec.~\ref{sec:analytic-templates} we can study the oscillatory part of $\OGW$ for any $\omegalog$ and $A_\textrm{log}$, by computing the coefficients $\C_1(\omegalog)$ and $\C_2(\omegalog)$ that together with $A_\textrm{log}$ control the amplitude of the two oscillatory pieces in \eqref{eq:OGW-res-template-C}. Hence, in fig.~\ref{fig:Xi012_w_w1o9_w1o3_w9o10} we plot $\C_{1,2}(\omegalog)$ and for completeness also $\C_0(\omegalog)$ (which is the contribution to the constant part of $\OGW$ in \eqref{eq:OGW-res-template-C} due to the oscillation in $\Pzeta$). In the left panel we show the results for $w=1/9$ and $w=1/3$, while in the right panel we display the curves for $w=9/10$ and $w=1/3$. We make the following observations:
\begin{itemize}
\item Note that for all choices of $w$ the coefficients $\C_{0,1,2}$ fall with increasing $\omegalog$. That is, the amplitude of oscillation in $\OGW$ decreases with an increasing frequency, an effect that was also observed for sharp features.
\item For all choices of $w$ the coefficient $\C_1$ is larger than $\C_2$ for small $\omegalog$, while for large $\omegalog$ the coefficient $\C_2$ eventually dominates. There is then an intermediate regime in $\omegalog$ where both coefficients are roughly comparable. This implies that, independently of $w$, for small $\omegalog$ the oscillatory part of $\OGW$ is dominated by the oscillation with frequency $\omegalog$ (consistent with the expectation from the resonance peak analysis). For large $\omegalog$ the modulation in $\OGW$ is dominated by the oscillation with frequency $2\omegalog$, and in the intermediate regime we have a superposition of both oscillatory parts. 
The precise limits of these various regimes however depend on the value of $w$ and differ between the examples shown. In particular, both for $w=1/9$ and $w=9/10$ the value of $\omegalog$ where $\C_2$ becomes larger than $\C_1$ is higher than for $w=1/3$. See appendix \ref{app:analytic-template-numerical} for more details on this.
\item The oscillatory behaviour of $\OGW$ further depends on the value of $A_\textrm{log}$, which enters differently into the amplitudes of the two oscillatory pieces in \eqref{eq:OGW-res-template-C}. In particular, for smaller $A_\textrm{log}$ the oscillatory term with frequency $2 \omegalog$ is suppressed more than the one with frequency $\omegalog$, so that the latter dominates over a larger range of values of $\omegalog$.
\item Focussing further on the differences between the curves for different $w$, the drop-off of $\C_{0,1,2}$ at large $\omegalog > 10$ is slower for larger $w$, i.e.~$\C_{0,1,2}$ decrease at a smaller rate for larger $w$.  
However, for $\omegalog \lesssim 10$ this hierarchy is inverted. In this regime the coefficient $\C_{1}$ dominates over the others and it is larger for smaller values of $w$. Hence the largest amplitude of oscillation is achieved for small $w$ in this regime, as we have seen in fig.~\ref{fig:Omega_w_res_top_om8_ws} for $\omegalog=8$.
\item Finally, the values of $\C_{0,1,2}$ for $\omegalog \rightarrow 0$ are consistent with the result for $\OGW$ for a constant scalar power spectrum $\Pzeta = \mathcal{P}_{\textrm{max}}(1+A_\textrm{log})$ without oscillations, i.e.~$\OGW=(k / k_\textrm{rh})^{-2b} \, A_0 \mathcal{P}_\textrm{max}^2 (1+A_\textrm{log})^2$.
\end{itemize}
The key results for $c_s^2=w$ and a constant envelope $\Pbar$ can then be summarised as follows. For sufficiently large $\omegalog$ the oscillation in $\OGW$ is dominated by the term with frequency $2 \omegalog$. The amplitude in this regime is typically largest for a stiffer equation of state (larger $w$), but it is small in absolute terms ($\lesssim 0.1$). Oscillations with a large amplitude ($> 0.1$) occur for $\omegalog \lesssim 10$ where the modulation with frequency $\omegalog$ dominates. In this case the amplitude is largest for a softer equation of state (smaller $w$). 

\begin{figure}[t]
\centering
\begin{overpic}[width=0.80\textwidth]{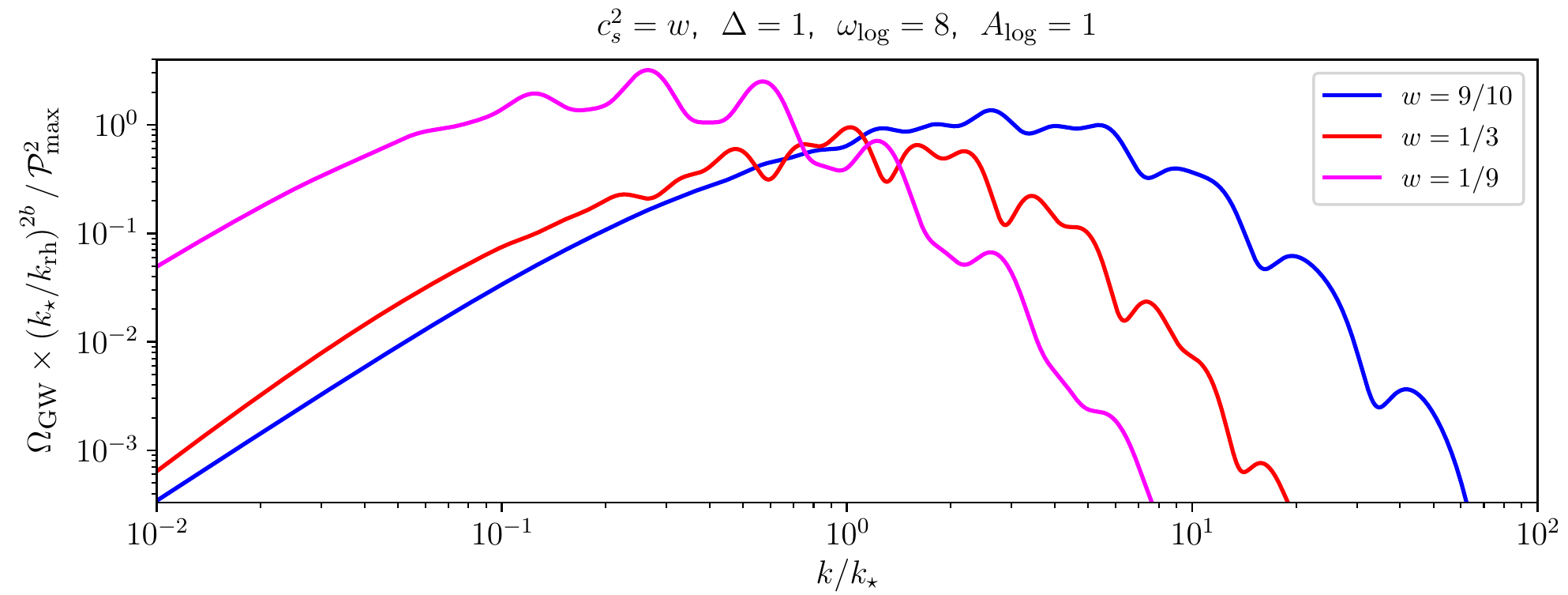}
\end{overpic}
\caption{$\OGW$ for GWs induced during a phase with $c_s^2=w$ vs.~$k / \kstar$ for a resonant feature in $\Pzeta$ as given in \eqref{eq:P-res-LN} with $\Delta=1$, frequency $\omegalog=8$ and amplitude $A_\textrm{log}=1$ for the choices $w=1/9, \, 1/3, \, 9/10$.}
\label{fig:Omega_w_res_LN1_om8_ws}
\end{figure}

An important question is how these results are modified when one considers the more realistic case of a non-constant envelope. As we will see, the behaviour of the oscillation with $\omegalog$ and $w$ follows the same trends as in the constant-envelope-case even though the precise quantitative results will differ. To be specific, consider for example an envelope given by a lognormal peak, so that the scalar power spectrum is given by
\begin{align}
\label{eq:P-res-LN}
    \Pzeta(k) = \mathcal{P}_\textrm{max} \, e^{-\frac{1}{2 \Delta^2} \big( \ln (k / \kstar)\big)^2} \bigg[ 1+ \Alog \cos \bigg( \omegalog \log \frac{k}{\kstar} + \frac{\pi}{4} \bigg) \bigg] \, ,
\end{align}
where we have set $\kref=\kstar$ and included a phase $\pi / 4$ to avoid the tuned situation where the maximum of the envelope coincides with an extremum of the $\cos$. In fig.~\ref{fig:Omega_w_res_LN1_om8_ws} we then plot $\OGW(k)$ for the same parameter choices $\omegalog=8$ and $A_\textrm{log}=1$ as in fig.~\ref{fig:Omega_w_res_top_om8_ws}, but now for a moderately narrow lognormal envelope with $\Delta=1$. Compared to fig.~\ref{fig:Omega_w_res_top_om8_ws} we now again include the prefactor $\sim k^{-2b}$. The main observation is that the oscillatory behaviour of $\OGW$ is qualitatively similar to the constant-envelope case: For $w=1/3$ we observe the double-peak-structure due to the superposition of the two oscillatory parts with frequencies $\omegalog$ and $2\omegalog$. This superposition is again less pronounced for $w=9/10$. For $w=1/9$ we once more have just one series of peaks with frequency $\omegalog$. As before, the amplitude of oscillation is largest for $w=1/9$ and smallest for $w=9/10$. The fact that here the result for $w=1/9$ has the largest overall amplitude of $\OGW$ is due to the factor $k^{-2b}$.\footnote{The maximum of $\OGW$, ignoring the maxima due to the oscillation, is well-approximated by the locus $k \simeq 2 w^{1/2} \kstar$ of the resonance peak of the maximum in the envelope $\Pbar$ at $k=\kstar$.} 

\begin{figure}[t]
\centering
\begin{overpic}[width=0.90\textwidth]{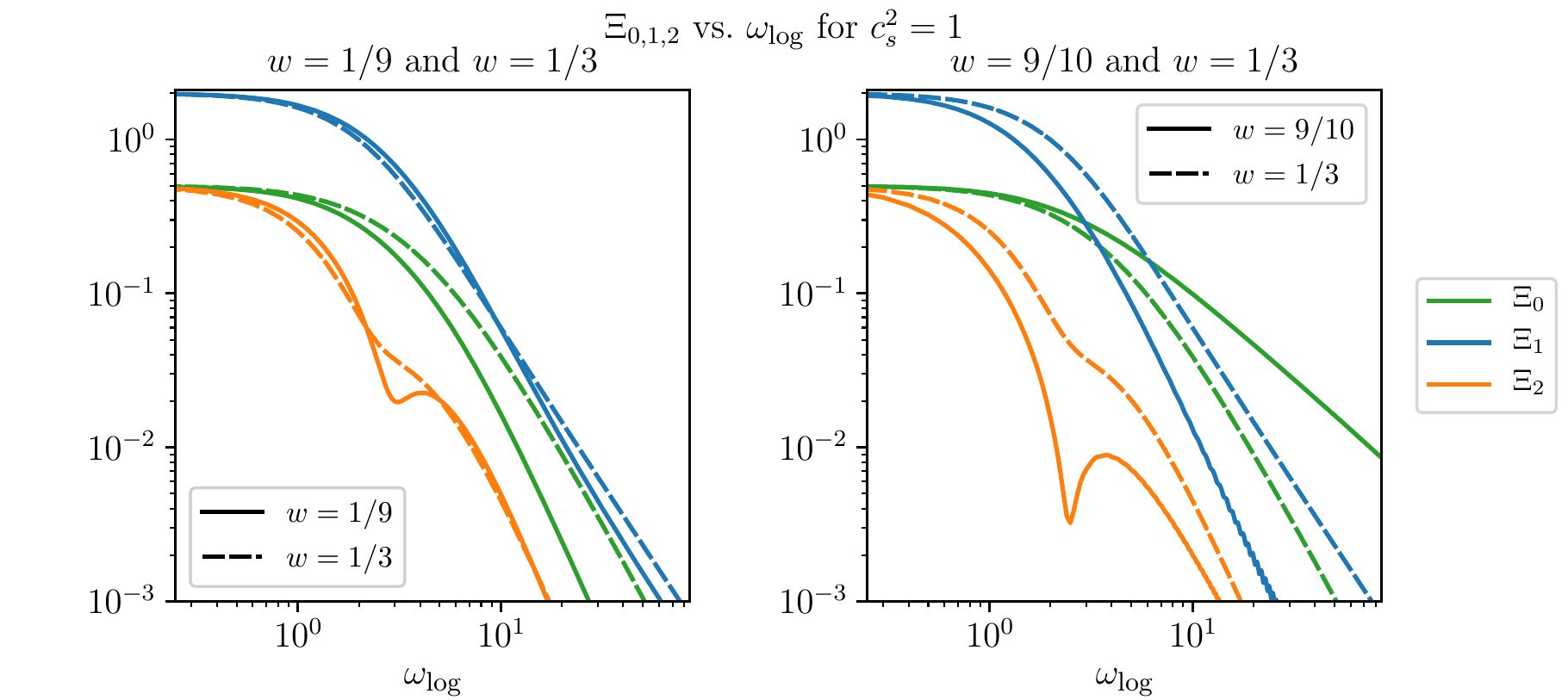}
\end{overpic}
\caption{$\C_{0,1,2}$ as appear in \eqref{eq:OGW-res-template-C} vs.~$\omegalog$ for $c_s^2=1$. In the left panel we display results for $w=1/9$ and $w=1/3$, and in the right for $w=9/10$ and $w=1/3$. As follows from the resonance peak analysis in sec.~\ref{sec:resonance-peak-analysis-resonant}, to have a significant oscillatory piece with frequency $2 \omegalog$ it is instrumental that different peaks due to the oscillation in $\Pzeta$ can interact resonantly. As resonant amplification is not possible for $c_s^2=1$, the coefficient $\C_2$ of the oscillatory piece with frequency $2 \omegalog$ can never become dominant in this case.}
\label{fig:Xi012_1_w1o9_w1o3_w9o10}
\end{figure}

For envelopes with even narrower peaks and for a large frequency $\omegalog$ we can also predict the oscillatory behaviour in $\OGW$ analytically. Consider an envelope $\Pbar$ with a narrow peak of width $\Delta k$ located at $k=\kstar$ so that $\Delta k / \kstar \ll 1$. To have at least one oscillation in $\Pzeta$ across this peak one then requires $\omegalog > 2 \pi \kstar / \Delta k \gg 1$. Across this narrow peak these oscillations will however be difficult to distinguish from the oscillations due to a sharp feature with frequency $\omegalin = \omegalog / \kstar$, i.e.
\begin{align}
    \Pzeta \supset \cos \bigg(\omegalog \log \frac{\kstar + \delta k}{\kref} \bigg) = \cos \bigg( \frac{\omegalog}{\kstar} \delta k + \ldots \bigg) = \cos (\omegalin \, \delta k + \ldots) \, ,
\end{align}
and $\delta k < \Delta k$.
We can thus predict the result for $\OGW$ for this parametric regime by matching to our findings for a sharp feature from sec.~\ref{sec:sharp}. These state that a narrow peak in $\Pzeta$ at $k=\kstar$ modulated by oscillations with frequency $\omegalin$ will produce a GW spectrum with a principal peak at $k=2 w^{1/2} \kstar$ modulated by oscillations with frequency $\omegagwlin= w^{-1/2} \omegalin$. This behaviour can be matched by the oscillatory contribution with frequency $2 \omegalog$ in the resonant feature template \eqref{eq:resonant-template-intro}, but not the one with frequency $\omegalog$. Expanding the relevant term in \eqref{eq:resonant-template-intro} around $k=2 w^{1/2} \kstar$ one finds:
\begin{align}
 \OGW \supset   \cos \bigg(2 \omegalog \log \frac{2 w^{1/2} \kstar + \delta k}{\kref} \bigg) = \cos \bigg( \frac{\omegalog}{w^{1/2} \kstar} \delta k + \ldots \bigg) = \cos (\omegagwlin \, \delta k + \ldots) \, .
\end{align}
Thus, consistency between the resonant and sharp feature results implies that for sufficiently large $\omegalog$ the oscillatory contribution with frequency $2 \omegalog$ has to dominate over that with frequency $\omegalog$ in \eqref{eq:resonant-template-intro}. This is as for the case of a constant envelope, see fig.~\ref{fig:Xi012_w_w1o9_w1o3_w9o10}, but here we find that this behaviour persists even for a narrowly peaked envelope $\Pbar$.

\subsection{Canonical scalar field ($c_s^2=1$)}
\label{sec:resonant-csf}
The absence of resonant amplification for $c_s^2=1$ results in a featureless integration kernel $\Tw(d,s)$ without any significant peaks. For a sharp feature this led to a spectrum of induced GWs without any significant oscillations, which were smoothed by the integration in \eqref{eq:OmegaGW-ts}. For a resonant feature we expect a more differentiated picture depending on the value of $\omegalog$. 

To illustrate this we compute the coefficients $\C_{1,2}(\omegalog)$ in \eqref{eq:OGW-res-template-C} that control the amplitude of oscillations in $\OGW$ for a constant envelope $\Pbar$. In fig.~\ref{fig:Xi012_1_w1o9_w1o3_w9o10} we plot $\C_{1,2}(\omegalog)$ and for completeness also $\C_0(\omegalog)$ for the choices $w=1/9$ and $w=1/3$ (left panel), and for $w=9/10$ and $w=1/3$ (right panel). We make the following observations. For all choices of $w$ the coefficients $\C_{1,2}$ fall rapidly with increasing $\omegalog$, implying that oscillations quickly become  undetectable for large $\omegalog$. This is similar to what we have seen for a sharp feature. In particular, we have that $\C_1 \lesssim 0.1$ for $\omegalog \gtrsim 10$ with $\C_2$ even more suppressed. In contrast, for $\omegalog \rightarrow 0$ the coefficients $\C_{0,1,2}$ take the finite values (0.5, 2, 0.5), consistent with the result for $\OGW$ due to a constant scalar power spectrum without oscillations. By continuity, there is a window $1 \lesssim \omegalog \lesssim 10$ where it appears that $\OGW$ exhibits large oscillations with $\C_1 \gtrsim 0.1$. This regime can be understood as follows. Interpreting the oscillation in $\Pzeta$ as a series of individual peaks, each such peak in $\Pzeta$ will in principle produce a broad peak in $\OGW$, just like a $\delta$-peak in $\Pzeta$ induces a broad peak in $\OGW$, see the right panel of fig.~\ref{fig:OGW_delta_w_1}. If the frequency $\omegalog$ is sufficiently large, the broad peaks would closely overlap, signalling that the oscillation is smoothed out just like in the sharp feature case. But if $\omegalog$ is sufficiently small, the individual peaks in $\OGW$ do not overlap significantly and one expects a periodic structure of peaks with frequency $\omegalog$.\footnote{For a sharp feature this regime is absent as the peaks in $\Pzeta$, arranged periodically in $k$ by default, can never be sufficiently separated to avoid significant overlap of the corresponding peaks in $\OGW$.} Comparing between the three choices for $w$, the value of $\C_1$ in this regime is comparable for $w=1/3$ and $w=1/9$, but smaller for $w=9/10$. Hence the amplitude is larger for a softer equation of state, roughly similar to what one observes for $c_s^2=w$ and $\omegalog \lesssim 10$, cf.~fig.\ref{fig:Xi012_w_w1o9_w1o3_w9o10}. 

\begin{figure}[t]
\centering
\begin{overpic}[width=0.80\textwidth]{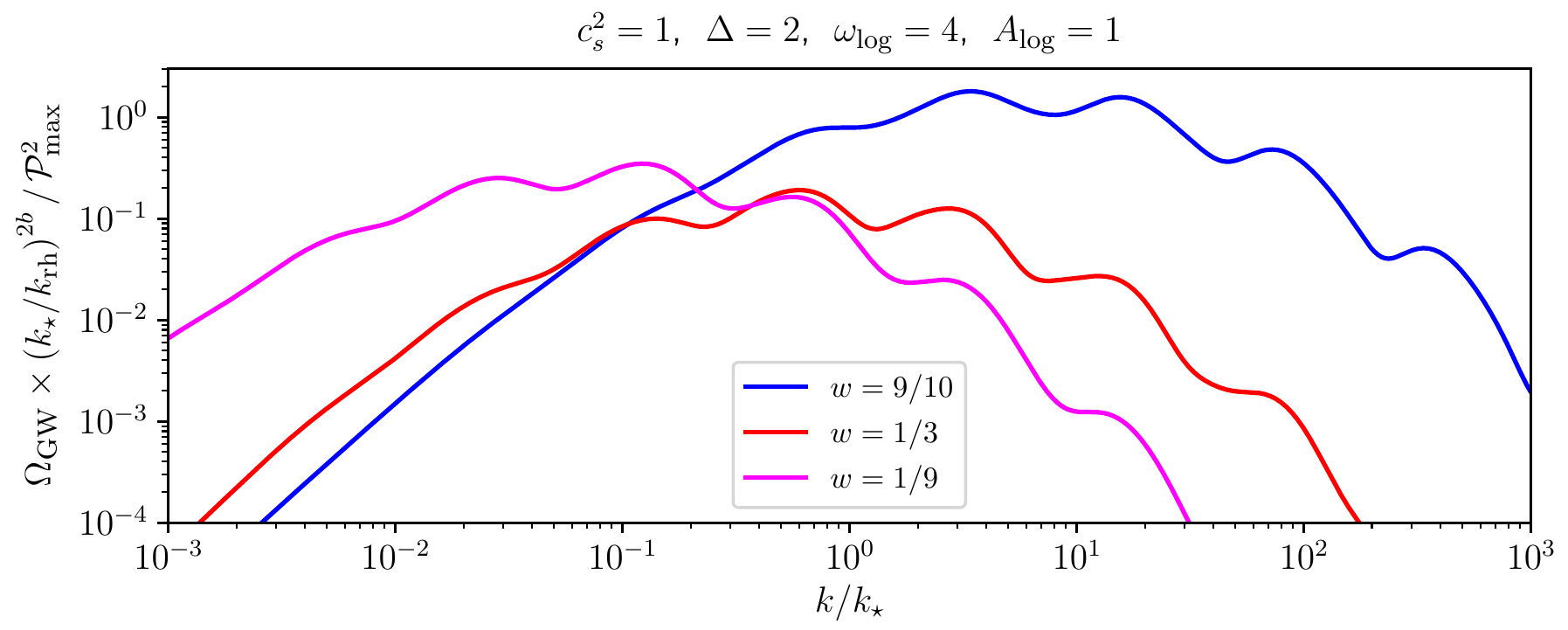}
\end{overpic}
\caption{$\OGW$ for GWs induced during a phase with $c_s^2=1$ vs.~$k / \kstar$ for a resonant feature in $\Pzeta$ as given in \eqref{eq:P-res-LN} with $\Delta=2$, frequency $\omegalog=4$ and amplitude $A_\textrm{log}=1$ for the choices $w=1/9, \, 1/3, \, 9/10$. The relative amplitude of oscillation in $\OGW$ close to the maximum is $\sim 40 \%$ for the three examples.}
\label{fig:Omega_1_res_LN2_om4_ws}
\end{figure}

To demonstrate this and also go beyond the constant-envelope-case, in fig.~\ref{fig:Omega_1_res_LN2_om4_ws} we plot $\OGW$ for a scalar power spectrum \eqref{eq:P-res-LN} with lognormal envelope for the model parameters $\Delta=2$, $\omegalog=4$, $A_\textrm{log}=1$ and for the choices $w=1/9$, $1/3$, $9/10$. As expected from the constant-envelope-results, this exhibits oscillations with a relative amplitude $\gtrsim 0.1$ about an otherwise smooth background.

Note that for $c_s^2=1$ the modulation in $\OGW$ due to a resonant feature is always dominated by the oscillatory piece in \eqref{eq:resonant-template-intro} with frequency $\omegalog$. For example, in the constant-envelope-case $\C_2$ is always suppressed compared to $\C_1$, see fig.~\ref{fig:Xi012_1_w1o9_w1o3_w9o10}. Hence one difference to the adiabatic fluid case is that the double-peak structure, e.g.~observed for $c_s^2=w$ with $w=1/3$ and $\omegalog=8$, cannot occur for $c_s^2=1$.

\section{Conclusions}
\label{sec:conclusions}
In this work we analysed how the expansion history of the universe after inflation affects the spectrum of induced GWs due to a sharp \eqref{eq:P-of-k-sharp} or resonant feature \eqref{eq:P-of-k-resonant} in the scalar power spectrum. To be specific, we considered a universe whose energy density in the post-inflationary era can be described by an adiabatic perfect fluid ($c_s^2=w$) or a canonical scalar field ($c_s^2=1$) and studied how the spectrum of GW induced during that era is affected by the equation of state, parameterised by $w$. We also assume that the value of $w$ remains constant over the limited period when the GWs are induced.

We find that for $c_s^2=w$ and any value $0<w<1$ the GW energy density fraction $\OGW(k)$ exhibits the characteristic oscillation for both sharp and resonant features \cite{Fumagalli:2020nvq,Fumagalli:2021cel} that can be described by the templates in \eqref{eq:sharp-template-intro} and \eqref{eq:resonant-template-intro}, respectively. For $c_s^2=1$ only a resonant feature produces modulations in $\OGW$ while the GW spectrum for a sharp feature is effectively smooth in this case. 

For a sharp feature and $c_s^2=w$ the frequency of the modulation in $\OGW$ is $\omegagwlin = w^{-1/2} \omegalin$, where $\omegalin$ is the frequency of oscillation in $\Pzeta$, showing a degeneracy between $w$ and $\omegalin$ for a given $\omegagwlin$. For a resonant feature and $c_s^2=w$ the modulation in $\OGW$ is a superposition of two oscillatory pieces, one with the original frequency $\omegalog$ of the oscillations in $\Pzeta$, and one with frequency $2\omegalog$, with the amplitudes of these two pieces dependent on $w$. For a resonant feature and $c_s^2=1$ there is just the oscillatory part with frequency $\omegalog$ with a $w$-dependent amplitude. Thus, information about the thermal history of the universe is encoded differently for a sharp vs.~a resonant feature. In all cases the amplitude of oscillation decreases with increasing frequency $\omegalin$ or $\omegalog$, so that for a sufficiently large value the oscillations become effectively undetectable. However, the rate of change of the amplitudes, and hence the threshold value of $\omegalin$ or $\omegalog$ where the oscillations become too small, is different for different $w$.

The main motivation for this work was to determine what information can be extracted about the early universe from a possible future detection of an oscillation in $\OGW$ due to a primordial feature. In particular, is it possible to both reconstruct from $\OGW$ the properties of the feature in $\Pzeta$ \emph{and} obtain information about the thermal history of the universe? For a sharp feature we show that a measurement of the frequency $\omegagwlin$ alone would be insufficient to uniquely determine $\omegalin$ and $w$. However, we also observe that this degeneracy can be broken when other observables like the amplitude of the oscillations in $\OGW$ are also taken into account. For example, for GWs induced during radiation-domination ($c_s^2=w=1/3$) the maximally attainable amplitude of oscillations for a sharp feature is $\sim 20 \%$, but we find that this upper bound is generally larger for a stiffer equation of state ($w>1/3$) and smaller for a softer equation of state ($w<1/3$). E.g.~for $w=9/10$ the amplitude of oscillation was observed to be $0.35$, see fig.~\ref{fig:Ratios_w_s0p3_w20}. Hence, an oscillation of sharp-feature type \eqref{eq:sharp-template-intro} with amplitude $\mathcal{A}_\textrm{lin} \gg 20 \%$ would be a clear indicator of GWs induced during an era with $w>1/3$.

For a resonant feature the situation is even more favourable, as the frequencies $\omegalog$ and $2\omegalog$ of the oscillation in $\OGW$ are unaffected by $w$. The absolute values of the corresponding amplitudes and their relative weight can then give information about $w$ and even help distinguish between the cases of an adiabatic perfect fluid and a canonical scalar field. Resonant features with $\omegalog \lesssim 10$ have the best prospect of detection, as the amplitude of oscillation in $\OGW$ is typically $> 10 \%$. In this regime the largest values of the amplitude of oscillation for a given value of $\omegalog$ is realised for a softer equation of state, see e.g.~fig.~\ref{fig:Xi012_1_w1o9_w1o3_w9o10}. For $\omegalog \gg 10$ this hierarchy is not realised any more, and for sufficiently large $\omegalog$ the amplitude of oscillation is largest for a stiffer equation of state, albeit the absolute value will be small. Finally, for a universe described by a canonical scalar field we find that the oscillatory term in \eqref{eq:resonant-template-intro} with frequency $2 \omegalog$ is always suppressed compared to the one with frequency $\omegalog$, while in the adiabatic fluid case the term with frequency $2 \omegalog$ can be comparable to or even dominate over the other, which in principle allows one to experimentally distinguish between these two descriptions.

There are several avenues for further work. Here we only focussed on GWs due to primordial features induced \emph{after} inflation. However, the physics responsible for sharp and resonant features will also produce gravitational waves during inflation, which on general grounds are also expected to exhibit an oscillation in their contribution to the stochastic gravitational wave background. For a complete picture of signals in $\OGW$ due to primordial features it would be important to also compute this contribution.\footnote{For the sharp feature case this will appear in \cite{primordial-GWs}.}  

It should be noted that here we computed the GW spectrum using \eqref{eq:OmegaGW-ts}, that is assuming Gaussian primordial fluctuations, hence ignoring any contributions from possible primordial non-Gaussianity that may be induced by the mechanism enhancing the primordial power spectrum. In general, non-Gaussianity may also impact the GW spectrum \cite{Unal:2018yaa,Atal:2021jyo,Adshead:2021hnm}. Since we are considering a scalar power spectrum that is sharply but not too sharply peaked at $\kstar$, that is the dimensionless width of the peak in \eqref{eq:P-sharp-turn-Gauss} and \eqref{eq:P-res-LN} is $\mathcal{O}(0.1)-\mathcal{O}(1)$, we expect that for a local non-Gaussianity parameter satisfying ${\cal P}_{\rm max}F^2_{NL}\gtrsim \mathcal{O}(0.1)$ there will be an additional visible bump at around $k\sim 3 \kstar$ \cite{Unal:2018yaa,Adshead:2021hnm}. Applied to our case, this has the interesting effect of extending the visibility of the oscillations up to $k\sim 3k_\star$. For ${\cal P}_{\rm max}F^2_{NL}>1$ the non-Gaussian contribution might dominate but it is unclear how this regime could be achieved in an inflationary model under perturbative control \cite{Atal:2021jyo}. We leave the study of the impact of local non-Gaussianity on the induced GW spectrum due to features for future work.

An exciting outcome of our work is that by detecting an oscillation in $\OGW$ and measuring its properties like the frequency and the amplitudes of the oscillatory contribution, one can in principle extract information about both the primordial scalar power spectrum and the expansion history of the universe. In simple cases this matching can be straightforward, but in practice this will have to be automatised, also taking into account the effect of the model-dependent envelope of $\Pzeta$, which determines the overall shape of the GW spectrum, together with $w$. One key result of this work is that such an endeavor is feasible in principle, as degeneracies between different parameters appear to be resolvable by measuring sufficiently many properties of $\OGW$. One way of implementing such a reconstruction technique in practice would then be to generate a large data bank of GW spectra for a wide range of primordial features and equations of state to be used as a training data for a machine-learning algorithm. We leave this exciting possibility for future work. Altogether, this further elevates oscillations in the stochastic gravitational wave background as an important target for future detection efforts. 

\begin{acknowledgments}
We are grateful to Sadra Jazayeri, Lucas Pinol and Denis Werth for interesting discussions. G.D.~as a Fellini fellow was supported by the European Union’s Horizon 2020 research and innovation programme under the Marie Sk{\l}odowska-Curie grant 
agreement No 754496. J.F, S.RP, and L.T.W are supported by the European Research Council under the European Union's Horizon 2020 research and innovation programme (grant agreement No 758792, project GEODESI).
\end{acknowledgments}

\appendix

\section{Possible expansion histories of the universe\label{app:exphistory}}

In this appendix we show that there is a large enough parameter space where the induced GW signal falls inside the frequency range of current and future GW detectors and where the primordial feature originates from inflation. To illustrate this, we consider the comoving scale equal to the size of the cosmological horizon at the time of reheating for a given reheating temperature $T_{\rm rh}$, which is given by \begin{align}\label{eq:kreh}
k_{\rm rh}=1.2\times 10^{12}\,{\rm Mpc}^{-1}\,\left(\frac{T_{\rm rh}}{5\times 10^4\,{\rm GeV}}\right)\left(\frac{g_*(T_{\rm rh})}{106.75}\right)^{1/2}\left(\frac{g_{*s}(T_{\rm rh})}{106.75}\right)^{-1/3}\,,
\end{align}
where $g_*$ and $g_{*s}$ are respectively the effective degrees of freedom in the energy density and entropy.
We choose a pivot reheating temperature of $T_{\rm rh}=5\times 10^4\,{\rm GeV}$ because its associated GW frequency is roughly at the LISA peak sensitivity, namely at
 \begin{align}\label{eq:freh}
f_{\rm rh}=1.8\times 10^{-3}\,{\rm Hz}\,\left(\frac{T_{\rm rh}}{5\times 10^4\,{\rm GeV}}\right)\left(\frac{g_*(T_{\rm rh})}{106.75}\right)^{1/2}\left(\frac{g_{*s}(T_{\rm rh})}{106.75}\right)^{-1/3}\,.
\end{align}
Note that we have another relevant scale in our set up, which is the position of the peak in the primordial spectrum at $k_\star$. The analysis in this work is valid for $k_\star\gg k_{\rm rh}$ and, therefore, the peak of the induced GW spectrum appears at higher frequencies. For the present choice of $f_{\rm rh}$ the peak could show up in LISA, DECIGO and ET.

Now, we estimate how many \textit{e}-folds prior to the end of inflation the mode $k_\star$ was generated, assuming that before the transition to radiation domination there was a period with an arbitrary constant equation of state parameter $w$. In what follows, we defined the number of \textit{e}-folds prior to today, that is $N=\ln a$ where $N_0=\ln a_0=0$. Denoting $N_\star$ as the time when the mode $k_\star$ exited the Hubble radius during inflation, i.e. when $aH=k_\star$, we have that
\begin{align}\label{eq:nstar}
N_\star-N_{\rm end}&\approx \ln\left(\frac{k_\star}{k_{\rm rh}}\right)\nonumber\\&+\frac{1+3w}{3(1+w)}\left[-51.9 + \frac{1}{2}\left(\frac{g_*(T_{\rm rh})}{106.75}\right) + 2 \ln\left(\frac{T_{\rm rh}}{5\times 10^4\,{\rm GeV}}\right) - \frac{1}{2}\ln\left(\frac{\epsilon_{\rm inf}}{10^{-2}}\right)\right]\,.
\end{align}
In equation \eqref{eq:nstar} we used that ${\cal P}_{\cal R}=H_{\rm inf}^2/(8\pi^2M_{\rm pl}^2\epsilon_{\rm inf})=2.09\times 10^{-9}$ from Planck's results \cite{Aghanim:2018eyx}, where $H_{\rm inf}$ is the Hubble parameter during inflation, which we took as constant for simplicity, and $\epsilon_{\rm inf}=-\dot H/H^2$ is the first slow-roll parameter during inflation. For example, if we consider that after inflation there was a period with $w\approx 0$ ($w\approx 1$), a mode $k_\star=10^2 k_{\rm rh}$ was generated $13$ ($30$) \textit{e}-folds before the end of inflation. Thus, there is plenty of room to produce such kind of features during inflation for modes which enter the Hubble radius before the end of the reheating phase. We illustrate this in figure \ref{fig:expansionhistory}. By decreasing the reheating temperature or/and having a high energy scale for inflation, the possible parameter space enlarges. For instance, the current constraints on the reheating temperature from BBN are as low as $T_{\rm rh}> 4\,{\rm MeV}$ \cite{Kawasaki:1999na,Kawasaki:2000en,Hannestad:2004px,Hasegawa:2019jsa}.

The total number of \textit{e}-folds from the end of inflation until today is completed with
\begin{align}
N_{\rm end}-N_{\rm rh}\approx \frac{2}{3(1+w)}\left[-51.9 + \frac{1}{2}\left(\frac{g_*(T_{\rm rh})}{106.75}\right) + 2 \ln\left(\frac{T_{\rm rh}}{5\times 10^4\,{\rm GeV}}\right) - \frac{1}{2}\ln\left(\frac{\epsilon_{\rm inf}}{10^{-2}}\right)\right] \,,
\end{align}
and
\begin{align}
N_{\rm rh}-N_{0}\approx -41-\frac{1}{3}\left(\frac{g_{*s}(T_{\rm rh})}{106.75}\right) - \ln\left(\frac{T_{\rm rh}}{5\times 10^4\,{\rm GeV}}\right)\,.
\end{align}

\section{Low and high frequency tails of the induced gravitational wave spectrum  \label{app:IRUVtail}}

In this appendix we show that there are no oscillations on the low frequency tail of the induced gravitational wave spectrum. We will focus on a radiation dominated universe for simplicity and we comment later on the straightforward generalisation to other values of the equation of state parameter. We also consider a general primordial spectrum of sharp feature type which is given by\footnote{In contrast to the main text, here we define $\omegalin$ to be dimensionless.} 
\begin{align}
{\cal P}_{\cal R}(\kappa)={\cal P}^{p}_{\cal R}(\kappa)\left(1+A\sin(\omega_{\rm lin}\kappa)\right),\quad{\rm with}\quad \kappa=\frac{k}{k_\star}\,,
\end{align}
where ${\cal P}^{\rm p}_{\cal R}(\kappa)$ is a fairly sharp peaked spectrum around $\kappa=k/k_\star=1$, e.g.~a log-normal peak.

\subsection{Low frequency tail}
For the low frequency limit, i.e.~$k\ll k_\star$ and hence $\kappa\ll1$, we may use Eq.~\eqref{eq:OmegaGW-ts} for $w=1/3$ ($b=0$), which reads
\begin{align}\label{eq:OmegaGW-tsappendix}
\Omega_{\textrm{GW}}(k)&= \int_1^{\infty} \textrm{d} s \int_{0}^{1} \textrm{d} d \, {\cal T}_{b=0}(d,s) \, \mathcal{P}_{\cal R} \left(\frac{\kappa}{2}(s+d)\right) \mathcal{P}_{\cal R} \left(\frac{\kappa}{2}(s-d)\right) \, .
\end{align}
The transfer function $\Tw(d,s)$ is given in Eq.~\eqref{eq:Twcs}. Since we are interested in the limit $\kappa\ll1$ and the variable $d$ is limited to $0<d<1$, the peak of the primordial spectrum inside the integral \eqref{eq:OmegaGW-tsappendix} will be at $\kappa s/2=1$, that is at $s\gg1$. In that case, we can safely drop the dependence in $d$ by setting it to $d=0$ and we arrive at
\begin{align}\label{eq:OmegaGW-tsappendix2}
\Omega_{\textrm{GW}}(k\ll k_\star)&\approx \int_1^{\infty} \textrm{d} s  \, {\cal T}_{b=0}(d=0,s\gg1) \, \mathcal{P}^2_{\cal R} \left(\frac{s\kappa}{2}\right)  \, .
\end{align}
We used the value $d=0$ since the integrand vanishes at $d=1$ due to momentum conservation. However, $d$ may take any value as long as $d\neq1$. We choose $d=0$ for simplicity. The transfer function in that limit is approximately given by
\begin{align}
 {\cal T}_{b=0}(d=0,s\gg1)\approx \frac{1}{4}{\cal N}(b=0)s^{-4}{y^2}\ln\left|\frac{1-y}{1+y}\right|\,.
\end{align}
In the large $s$ limit, we have from \eqref{eq:ydef} that
\begin{align}
y\sim 1-\frac{2}{c_s^{2}s^2}\,.
\end{align}
Only picking up the largest contribution in the Taylor expansion for $s\gg1$ we find
\begin{align}\label{eq:OmegaGW-tsappendix3}
\Omega_{\textrm{GW}}(k\ll k_p)&\approx {\cal N}(b=0)\int_1^{\infty} \textrm{d} s  \,s^{-4}\ln^2s \, \mathcal{P}_{\cal R}^2 \left(\frac{s\kappa}{2}\right) \, .
\end{align}
Changing the integration variable to $\tilde s=\kappa  s/2$ we arrive at
\begin{align}\label{eq:OmegaGW-tsappendix4}
\Omega_{\textrm{GW}}(k\ll k_p)&\approx \frac{1}{8}{\cal N}(b=0)\kappa^3\ln^2\kappa\int_{\kappa/2}^{\infty} \textrm{d} \tilde s  \,\tilde s^{-4} \, \mathcal{P}_{\cal R}^2 (\tilde s) \, ,
\end{align}
where we neglected terms $\sim \ln\tilde s$ since the peak of the integrand is at $\tilde s=1$ where the logarithm vanishes. Ultimately we have shown that in the low frequency regime
\begin{align}\label{eq:OmegaGW-tsappendix5}
\Omega_{\textrm{GW}}(k\ll k_\star)&\propto \left(\frac{k}{k_\star}\right)^3\ln^2\left(\frac{k}{k_\star}\right) \, ,
\end{align}
up to a constant factor which can be determined by performing the definite integral in Eq.~\eqref{eq:OmegaGW-tsappendix4}. Thus, we see that there are no oscillations in the low frequency tail. This is the standard low frequency behaviour for localized sources in a radiation dominated universe. One can recover the low frequency behaviour derived in \cite{Domenech:2020kqm} by using the large $s$ expansions of the transfer functions which can be found in detail in \cite{Domenech:2021ztg}. The main point is that such Taylor expansion for $s\gg1$ always yield a power-law and for $b\in \mathbb{Z}$ a logarithmic correction. This does not change the main conclusion that the low frequency tail does not present oscillations.

\subsection{High frequency tail}

To study the high frequency tail it is more convenient to work with
\begin{align}
s=u+v\quad,\quad d=u-v\,.
\end{align}
In terms of these new variables the integral \eqref{eq:OmegaGW-ts} now reads 
\begin{align}
\Omega_{\rm GW}(k)=\int_0^\infty dv\int_{|1-v|}^{1+v} du \,{\cal T}_{b=0}(u,v){\cal P}_{\cal R}(u\kappa){\cal P}_{\cal R}(v\kappa)\,.
\end{align}
As explained in \cite{Atal:2021jyo}, the relevant contribution to the integral in the high frequency regime, i.e.~$\kappa\gg1$, and for a peaked primordial spectrum occurs at $v\ll1$ and $u\sim 1$. There is a copy of this contribution at $u\ll1$ and $v\sim 1$, so that we have to multiply our results by $2$. Focusing on the $v\ll1$ case and expanding the $u$ integral around $v\sim 0$ we have
\begin{align}
\Omega_{\rm GW}(k\gg k_\star)=2{\cal P}_{\cal R}(\kappa)\int_0^\infty dv \,v \,{\cal T}_{b=0}(u=1,v\ll1){\cal P}_{\cal R}(v\kappa)\,.
\end{align}
The exact behaviour of ${\cal T}_{b=0}(u=1,v\ll1)$ depends on the value of $c_s^2$ as for $u\sim 1$ and $v\sim0$ we have that
\begin{align}
y\sim \frac{1-c_s^{-2}}{2v}+\frac{v}{2}\,.
\end{align}
For $c_s^2\neq 1$ we see that $y\to-\infty$. For $c_s^2=1$ we know that by momentum conservation $y\leq-1$ and in the exact limit $u\to 1$ we have that $y\sim v/2$. The only difference though will be a different power of $v$ for the expansion of the transfer function, which as we shall see does not matter in the present case. For simplicity, let us continue with the $c_s^2\neq 1$ case. Then, using the formulas in \cite{Domenech:2021ztg}, we find that
\begin{align}
{\cal T}_{b=0}(u=1,v\ll1)\approx {\cal N}(b=0)\frac{1}{\pi \Gamma^2\left[b+3/2\right]}y^{-2}\approx{\cal N}(b=0)\frac{4}{\pi\Gamma^2\left[b+3/2\right]}\frac{v^2}{(c_s^{-2}-1)^2 }\,.
\end{align}
The spectral density in the high frequency tail is then given by
\begin{align}\label{eq:omegaUV}
\Omega_{\rm GW}(k\gg k_\star)\approx{\cal N}(b=0)\frac{8}{\pi\Gamma^2\left[b+3/2\right]}\frac{1}{(c_s^{-2}-1)^2 }{\cal P}_{\cal R}(\kappa)\int_0^\infty dv \,v^3{\cal P}_{\cal R}(v\kappa)\,.
\end{align}
With the change of variables $\tilde v=v\kappa$ we arrive at
\begin{align}
\Omega_{\rm GW}(k\gg k_\star,c_s^2\neq 1)\propto \kappa^{-4}{\cal P}_{\cal R}(\kappa)\,,
\end{align}
where we used that the definite integral in Eq.~\eqref{eq:omegaUV} gives a numerical factor.
For $c_s^2=1$ we find that ${\cal T}_{b=0}(u=1,v\ll1)\approx {\rm constant}$ and therefore
\begin{align}
\Omega_{\rm GW}(k\gg k_\star,c_s^2= 1)\propto \kappa^{-2}{\cal P}_{\cal R}(\kappa)\,.
\end{align}
In both cases, we see that if ${\cal P}_{\cal R}(\kappa)$ exhibits oscillations in the high momenta region $k\gg k_\star$ then these oscillations are captured in the high frequency tail of the induced gravitational wave spectrum. This explains why we see oscillations in the UV when computing the residuals of the induced gravitational wave spectrum between the envelope and the full primordial spectrum, see e.g.~fig.~\ref{fig:Ratios_w_s0p3_w20}. However, note that in explicit models of e.g.~a sharp feature one does not expect the oscillations in the primordial spectrum to go much beyond $k>2k_\star$. This means that although the oscillations in the curvature power spectrum get captured in the UV tail of the induced GW spectrum, in realistic situations the oscillations should have a cut-off for $k\gg 2k_\star$. In any case, since the induced GW spectrum from a peaked curvature power spectrum decays at least as ${\cal P}_{\cal R}(\kappa)$, there will be in general a large suppression for $k>2k_\star$. 

\section{Analytic templates for resonant features}
\label{app:analytic-template}
Here we collect essential formulae for the analytic template \eqref{eq:OGW-res-template-C} for $\OGW(k)$ in case of a resonant feature with a sufficiently broad envelope. The analysis leading to these results is as in \cite{Fumagalli:2021cel}, except that the integration kernel in the integral expression for $\OGW(k)$ is here given by \eqref{eq:Twcs} and thus depends on $w$. Following \cite{Fumagalli:2021cel}, the parameters $\C_{0,1,2}$ and $\theta_{1,2}$ appearing in \eqref{eq:OGW-res-template-C} can be written as
\begin{align}
\label{eq:C012def}
    \C_0 (\omegalog) \equiv \frac{C_2 (\omegalog)}{A_0} , \quad \C_1 \equiv \frac{\big(A_1(\omegalog)^2+B_1(\omegalog)^2\big)^{1/2}}{A_0} , \quad \C_2 \equiv \frac{\big(A_2(\omegalog)^2+B_2(\omegalog)^2\big)^{1/2}}{A_0} ,
\end{align}
and
\begin{align}
\label{eq:tan1_and_tan2_def}
\tan \big( \theta_1(\omegalog) \big)= \frac{B_1(\omegalog)}{A_1(\omegalog)} \, , \qquad \tan \big( \theta_2(\omegalog) \big)= \frac{B_2(\omegalog)}{A_2(\omegalog)} \, , 
\end{align}
with the functions $A_{0,1,2}$, $B_{1,2}$ and $C_2$ given by
\begin{align}\label{eq:A0}
    A_0 &= \int_0^{1} \textrm{d} d \int_{1}^\infty \textrm{d} s \, \Tw (d,s) ,
\end{align}
\begin{align}\label{eq:A1}
    A_1(\omegalog) &= \int_0^{1} \textrm{d} d \int_{1}^\infty \textrm{d} s \, \Tw (d,s) \, \Big[ \cos \big( \omegalog \log(s+d) \big) + \cos \big( \omegalog \log(s-d) \big) \Big],\\
    \label{eq:B1}
    B_1(\omegalog) &= \int_0^{1} \textrm{d} d \int_{1}^\infty \textrm{d} s \, \Tw (d,s) \, \Big[\sin \big( \omegalog \log(s+d) \big) + \sin \big( \omegalog \log(s-d) \big) \Big],
\end{align}

\begin{figure}[t]
\centering
\begin{overpic}[width=0.90\textwidth]{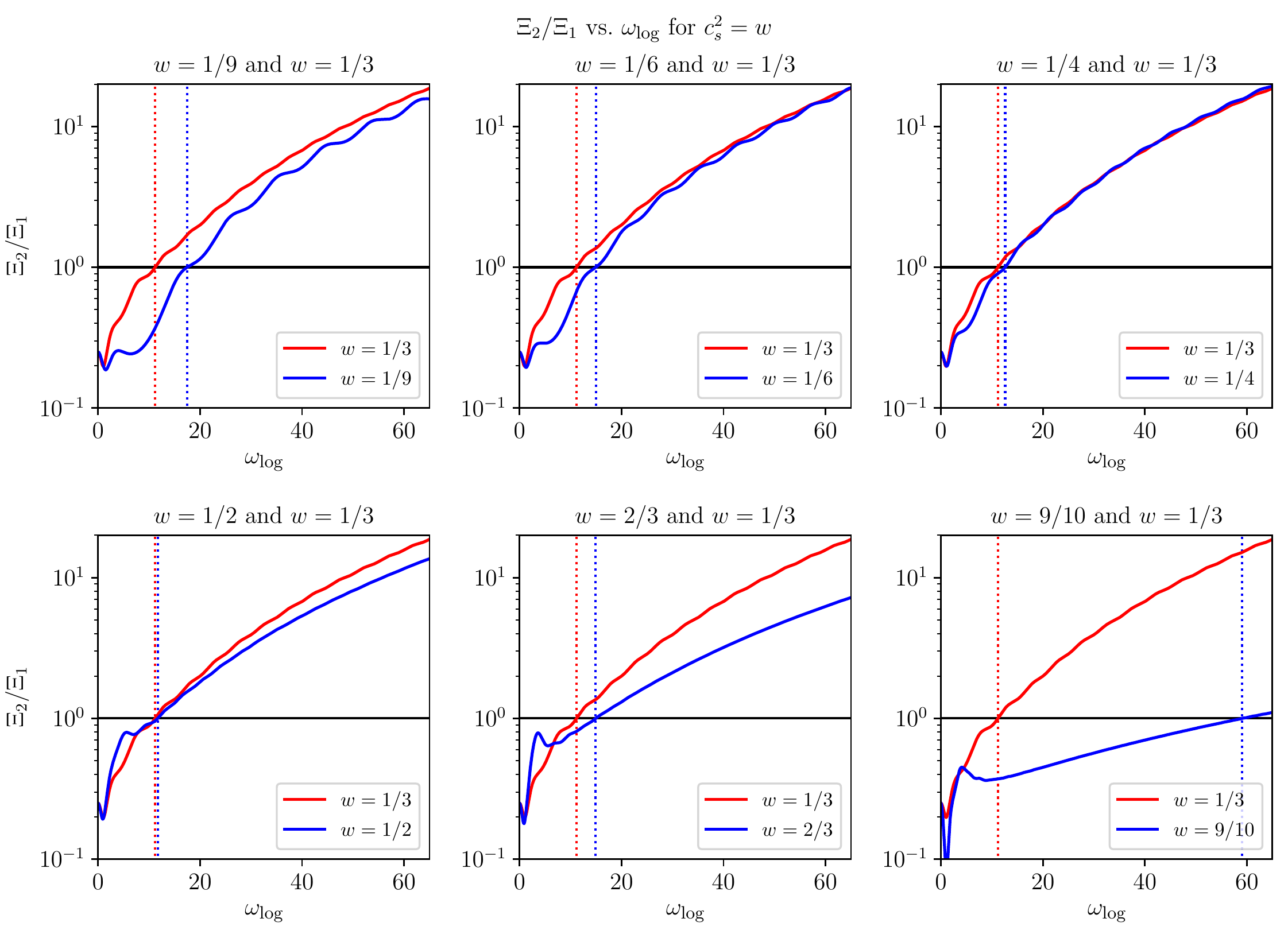}
\end{overpic}
\caption{$\C_{2} / \C_{1}$ vs.~$\omegalog$ with $\C_{1,2}$ defined in \eqref{eq:C012def} for $c_s^2=w$. The red lines in all panels correspond to the result for $w=1/3$, while the blue lines in the six panels are for $w=1/9, \, 1/6, \, 1/4, \, 1/2, \, 2/3, \, 9/10$, respectively. The dotted vertical lines denote the cross-over frequency $\omegalogx$ for which $\C_1(\omegalogx)=\C_2(\omegalogx)$. Note that $\omegalogx$ increases as $w$ is decreased from $w=1/3$ or increased from $w=1/3$.}
\label{fig:Xi2_over_Xi1_w_grid_ws}
\end{figure}

\begin{align}\label{eq:A2}
    A_2(\omegalog) &= \frac{1}{2} \int_0^{1} \textrm{d} d \int_{1}^\infty \textrm{d} s \, \Tw (d,s) \, \cos\big(\omegalog \log (s^2-d^2)\big),\\
    \label{eq:B2}
    B_2(\omegalog) &= \frac{1}{2} \int_0^{1} \textrm{d} d \int_{1}^\infty \textrm{d} s \, \Tw (d,s) \, \sin\big(\omegalog \log (s^2-d^2)\big),\\
    \label{eq:C2}
    C_2(\omegalog) &= \frac{1}{2} \int_0^{1} \textrm{d} d \int_{1}^\infty \textrm{d} s \, \Tw (d,s) \, \cos\bigg(\omegalog \log \frac{s+d}{s-d}\bigg) .
\end{align}

\subsection{Numerical coefficients: adiabatic perfect fluid ($c_s^2=w$)}
\label{app:analytic-template-numerical}
Numerical results for the coefficients $\C_{0,1,2}(\omegalog)$ for both $c_s^2=w$ and $c_s^2=1$ are displayed in figs.~\ref{fig:Xi012_w_w1o9_w1o3_w9o10} and \ref{fig:Xi012_1_w1o9_w1o3_w9o10}, respectively, for the choices $w=1/9, \, 1/3, \, 9/10$. For the case of an adiabatic perfect fluid ($c_s^2=w$) one finds that $\C_{1}$ is larger than $\C_2$ for small values of $\omegalog$, while for large values of $\omegalog$ eventually $\C_2$ dominates. The value $\omegalogx$ where the cross-over happens, i.e.~$\C_1(\omegalogx)=\C_2(\omegalogx)$, depends on the value of $w$, see fig.~\ref{fig:Xi012_w_w1o9_w1o3_w9o10}.\footnote{For a canonical scalar field ($c_s^2=1$) the coefficient $\C_2$ is always suppressed  compared to $\C_1$, see fig.~\ref{fig:Xi012_1_w1o9_w1o3_w9o10}.} To better illustrate this, in fig.~\ref{fig:Xi2_over_Xi1_w_grid_ws} we display the ratio $\C_{2} / \C_{1}$ vs.~$\omegalog$ for various values of $w<1$. 

As a reference value, note that for radiation, $w=1/3$, one finds $\omegalogx=11.2$. Then, as can be seen in fig.~\ref{fig:Xi2_over_Xi1_w_grid_ws}, decreasing $w$ from $w=1/3$ the value of $\omegalogx$ increases, i.e.~the oscillatory piece with frequency $2 \omegalog$ in \eqref{eq:OGW-res-template-C} only begins to dominate for larger values of $\omegalog$. This can be understood from the resonance peak analysis in sec.~\ref{sec:resonance-peak-analysis-resonant}. To have the oscillatory piece with frequency $2 \omegalog$, it is instrumental that different peaks due to the oscillation in $\Pzeta$ can interact resonantly, which is only possible for $\omegalog > \omegalogc$. As $\omegalogc \rightarrow \infty$ for $c_s \rightarrow 0$, it follows that $\omegalogx$ must increase as $w$ is decreased for $c_s^2=w$.

Increasing $w$ from $w=1/3$ the value of $\omegalogx$ also increases. This behaviour is expected from continuity with the result for $c_s^2=1$. For the examples shown here we have set $c_s^2=w$ and hence letting $w \rightarrow 1$ we are also sending $c_s^2 \rightarrow 1$. For $c_s^2=1$ the coefficient $\C_2$ never dominates, see e.g.~fig.~\ref{fig:Xi012_1_w1o9_w1o3_w9o10}, and hence for $c_s^2=w \rightarrow 1$ we expect $\omegalogx \rightarrow \infty$ by continuity. At a deeper level, the fact that $\C_2$ never dominates for $c_s^2=1$ is due to resonant amplification of GWs being kinematically forbidden in this case, see the comment in the caption of fig.~\ref{fig:Xi012_1_w1o9_w1o3_w9o10}. The fact that $\omegalogx \rightarrow \infty$ for $c_s^2=w \rightarrow 1$ can thus be understood as a consequence of resonant amplification being pushed to the edge of the kinematically allowed regime.

\section{Legendre functions on the cut and associated Legendre functions}
\label{app:Legendre}
For completeness, here we record the definitions for $\mathsf{P}_{\nu}^{\mu}(x)$, $\mathsf{Q}_{\nu}^{\mu}(x)$ and $\mathcal{Q}_{\nu}^{\mu}(x)$ as given in \cite{NIST:DLMF}: 
\begin{align}
    \mathsf{P}_{\nu}^{\mu}(x) &= {\bigg(\frac{1+x}{1-x} \bigg)}^{\mu/2} \frac{1}{\Gamma[1-\mu]} F \big(\nu +1, -\nu; 1-\mu ; \tfrac{1}{2}(1-x) \big) \, , \\
    \mathsf{Q}_{\nu}^{\mu}(x) &= \frac{\pi}{2 \sin(\pi \mu)} \Bigg[ \cos(\pi \mu) {\bigg(\frac{1+x}{1-x} \bigg)}^{\mu/2} \frac{1}{\Gamma[1-\mu]} F \big(\nu +1, -\nu; 1-\mu ; \tfrac{1}{2}(1-x) \big) \, , \\
    \nonumber & \hphantom{= \frac{\pi}{2 \sin(\pi \mu)} \Bigg[} - {\bigg(\frac{1-x}{1+x} \bigg)}^{\mu/2} \frac{\Gamma[\nu+\mu+1]}{\Gamma[\nu-\mu +1] \Gamma[1+\mu]} F \big(\nu +1, -\nu; 1+\mu ; \tfrac{1}{2}(1-x) \big) \Bigg] \, , \\
    \mathcal{Q}_{\nu}^{\mu}(x) &= \frac{\pi^{1/2} (x^2-1)^{\mu/2}}{2^{\nu+1}x^{\nu+\mu+1}} \frac{1}{\Gamma[\nu + 3/2]} F \big(\tfrac{1}{2} \nu + \tfrac{1}{2} \mu +1, \tfrac{1}{2} \nu + \tfrac{1}{2} \mu + \tfrac{1}{2}; \nu + \tfrac{3}{2} ; \tfrac{1}{x^2} \big) \, ,
\end{align}
with $F(a,b;c;x)$ the Gauss hypergeometric function.
\bibliographystyle{apsrev4-1}
\bibliography{Biblio-2021,bibliographyreview}

\begin{thebibliography}{64}%
\makeatletter
\providecommand \@ifxundefined [1]{%
 \@ifx{#1\undefined}
}%
\providecommand \@ifnum [1]{%
 \ifnum #1\expandafter \@firstoftwo
 \else \expandafter \@secondoftwo
 \fi
}%
\providecommand \@ifx [1]{%
 \ifx #1\expandafter \@firstoftwo
 \else \expandafter \@secondoftwo
 \fi
}%
\providecommand \natexlab [1]{#1}%
\providecommand \enquote  [1]{``#1''}%
\providecommand \bibnamefont  [1]{#1}%
\providecommand \bibfnamefont [1]{#1}%
\providecommand \citenamefont [1]{#1}%
\providecommand \href@noop [0]{\@secondoftwo}%
\providecommand \href [0]{\begingroup \@sanitize@url \@href}%
\providecommand \@href[1]{\@@startlink{#1}\@@href}%
\providecommand \@@href[1]{\endgroup#1\@@endlink}%
\providecommand \@sanitize@url [0]{\catcode `\\12\catcode `\$12\catcode
  `\&12\catcode `\#12\catcode `\^12\catcode `\_12\catcode `\%12\relax}%
\providecommand \@@startlink[1]{}%
\providecommand \@@endlink[0]{}%
\providecommand \url  [0]{\begingroup\@sanitize@url \@url }%
\providecommand \@url [1]{\endgroup\@href {#1}{\urlprefix }}%
\providecommand \urlprefix  [0]{URL }%
\providecommand \Eprint [0]{\href }%
\providecommand \doibase [0]{http://dx.doi.org/}%
\providecommand \selectlanguage [0]{\@gobble}%
\providecommand \bibinfo  [0]{\@secondoftwo}%
\providecommand \bibfield  [0]{\@secondoftwo}%
\providecommand \translation [1]{[#1]}%
\providecommand \BibitemOpen [0]{}%
\providecommand \bibitemStop [0]{}%
\providecommand \bibitemNoStop [0]{.\EOS\space}%
\providecommand \EOS [0]{\spacefactor3000\relax}%
\providecommand \BibitemShut  [1]{\csname bibitem#1\endcsname}%
\let\auto@bib@innerbib\@empty
\bibitem [{\citenamefont {Aghanim}\ \emph {et~al.}(2020)\citenamefont {Aghanim}
  \emph {et~al.}}]{Aghanim:2018eyx}%
  \BibitemOpen
  \bibfield  {author} {\bibinfo {author} {\bibfnamefont {N.}~\bibnamefont
  {Aghanim}} \emph {et~al.} (\bibinfo {collaboration} {Planck}),\ }\href
  {\doibase 10.1051/0004-6361/201833910} {\bibfield  {journal} {\bibinfo
  {journal} {Astron. Astrophys.}\ }\textbf {\bibinfo {volume} {641}},\ \bibinfo
  {pages} {A6} (\bibinfo {year} {2020})},\ \Eprint
  {http://arxiv.org/abs/1807.06209} {arXiv:1807.06209 [astro-ph.CO]}
  \BibitemShut {NoStop}%
\bibitem [{\citenamefont {Akrami}\ \emph {et~al.}(2018)\citenamefont {Akrami}
  \emph {et~al.}}]{Akrami:2018odb}%
  \BibitemOpen
  \bibfield  {author} {\bibinfo {author} {\bibfnamefont {Y.}~\bibnamefont
  {Akrami}} \emph {et~al.} (\bibinfo {collaboration} {Planck}),\ }\href@noop {}
  {\  (\bibinfo {year} {2018})},\ \Eprint {http://arxiv.org/abs/1807.06211}
  {arXiv:1807.06211 [astro-ph.CO]} \BibitemShut {NoStop}%
\bibitem [{\citenamefont {Caprini}\ and\ \citenamefont
  {Figueroa}(2018)}]{Caprini:2018mtu}%
  \BibitemOpen
  \bibfield  {author} {\bibinfo {author} {\bibfnamefont {C.}~\bibnamefont
  {Caprini}}\ and\ \bibinfo {author} {\bibfnamefont {D.~G.}\ \bibnamefont
  {Figueroa}},\ }\href {\doibase 10.1088/1361-6382/aac608} {\bibfield
  {journal} {\bibinfo  {journal} {Class. Quant. Grav.}\ }\textbf {\bibinfo
  {volume} {35}},\ \bibinfo {pages} {163001} (\bibinfo {year} {2018})},\
  \Eprint {http://arxiv.org/abs/1801.04268} {arXiv:1801.04268 [astro-ph.CO]}
  \BibitemShut {NoStop}%
\bibitem [{\citenamefont {Guzzetti}\ \emph {et~al.}(2016)\citenamefont
  {Guzzetti}, \citenamefont {Bartolo}, \citenamefont {Liguori},\ and\
  \citenamefont {Matarrese}}]{Guzzetti:2016mkm}%
  \BibitemOpen
  \bibfield  {author} {\bibinfo {author} {\bibfnamefont {M.~C.}\ \bibnamefont
  {Guzzetti}}, \bibinfo {author} {\bibfnamefont {N.}~\bibnamefont {Bartolo}},
  \bibinfo {author} {\bibfnamefont {M.}~\bibnamefont {Liguori}}, \ and\
  \bibinfo {author} {\bibfnamefont {S.}~\bibnamefont {Matarrese}},\ }\href
  {\doibase 10.1393/ncr/i2016-10127-1} {\bibfield  {journal} {\bibinfo
  {journal} {Riv. Nuovo Cim.}\ }\textbf {\bibinfo {volume} {39}},\ \bibinfo
  {pages} {399} (\bibinfo {year} {2016})},\ \Eprint
  {http://arxiv.org/abs/1605.01615} {arXiv:1605.01615 [astro-ph.CO]}
  \BibitemShut {NoStop}%
\bibitem [{\citenamefont {Tomita}(1967)}]{Tomita}%
  \BibitemOpen
  \bibfield  {author} {\bibinfo {author} {\bibfnamefont {K.}~\bibnamefont
  {Tomita}},\ }\href {\doibase 10.1143/PTP.37.831} {\bibfield  {journal}
  {\bibinfo  {journal} {Progress of Theoretical Physics}\ }\textbf {\bibinfo
  {volume} {37}},\ \bibinfo {pages} {831} (\bibinfo {year} {1967})},\ \Eprint
  {http://arxiv.org/abs/https://academic.oup.com/ptp/article-pdf/37/5/831/5234391/37-5-831.pdf}
  {https://academic.oup.com/ptp/article-pdf/37/5/831/5234391/37-5-831.pdf}
  \BibitemShut {NoStop}%
\bibitem [{\citenamefont {Matarrese}\ \emph {et~al.}(1993)\citenamefont
  {Matarrese}, \citenamefont {Pantano},\ and\ \citenamefont
  {Saez}}]{Matarrese:1992rp}%
  \BibitemOpen
  \bibfield  {author} {\bibinfo {author} {\bibfnamefont {S.}~\bibnamefont
  {Matarrese}}, \bibinfo {author} {\bibfnamefont {O.}~\bibnamefont {Pantano}},
  \ and\ \bibinfo {author} {\bibfnamefont {D.}~\bibnamefont {Saez}},\ }\href
  {\doibase 10.1103/PhysRevD.47.1311} {\bibfield  {journal} {\bibinfo
  {journal} {Phys. Rev. D}\ }\textbf {\bibinfo {volume} {47}},\ \bibinfo
  {pages} {1311} (\bibinfo {year} {1993})}\BibitemShut {NoStop}%
\bibitem [{\citenamefont {Matarrese}\ \emph {et~al.}(1994)\citenamefont
  {Matarrese}, \citenamefont {Pantano},\ and\ \citenamefont
  {Saez}}]{Matarrese:1993zf}%
  \BibitemOpen
  \bibfield  {author} {\bibinfo {author} {\bibfnamefont {S.}~\bibnamefont
  {Matarrese}}, \bibinfo {author} {\bibfnamefont {O.}~\bibnamefont {Pantano}},
  \ and\ \bibinfo {author} {\bibfnamefont {D.}~\bibnamefont {Saez}},\ }\href
  {\doibase 10.1103/PhysRevLett.72.320} {\bibfield  {journal} {\bibinfo
  {journal} {Phys. Rev. Lett.}\ }\textbf {\bibinfo {volume} {72}},\ \bibinfo
  {pages} {320} (\bibinfo {year} {1994})},\ \Eprint
  {http://arxiv.org/abs/astro-ph/9310036} {arXiv:astro-ph/9310036} \BibitemShut
  {NoStop}%
\bibitem [{\citenamefont {Ananda}\ \emph {et~al.}(2007)\citenamefont {Ananda},
  \citenamefont {Clarkson},\ and\ \citenamefont {Wands}}]{Ananda:2006af}%
  \BibitemOpen
  \bibfield  {author} {\bibinfo {author} {\bibfnamefont {K.~N.}\ \bibnamefont
  {Ananda}}, \bibinfo {author} {\bibfnamefont {C.}~\bibnamefont {Clarkson}}, \
  and\ \bibinfo {author} {\bibfnamefont {D.}~\bibnamefont {Wands}},\ }\href
  {\doibase 10.1103/PhysRevD.75.123518} {\bibfield  {journal} {\bibinfo
  {journal} {Phys. Rev. D}\ }\textbf {\bibinfo {volume} {75}},\ \bibinfo
  {pages} {123518} (\bibinfo {year} {2007})},\ \Eprint
  {http://arxiv.org/abs/gr-qc/0612013} {arXiv:gr-qc/0612013} \BibitemShut
  {NoStop}%
\bibitem [{\citenamefont {Baumann}\ \emph {et~al.}(2007)\citenamefont
  {Baumann}, \citenamefont {Steinhardt}, \citenamefont {Takahashi},\ and\
  \citenamefont {Ichiki}}]{Baumann:2007zm}%
  \BibitemOpen
  \bibfield  {author} {\bibinfo {author} {\bibfnamefont {D.}~\bibnamefont
  {Baumann}}, \bibinfo {author} {\bibfnamefont {P.~J.}\ \bibnamefont
  {Steinhardt}}, \bibinfo {author} {\bibfnamefont {K.}~\bibnamefont
  {Takahashi}}, \ and\ \bibinfo {author} {\bibfnamefont {K.}~\bibnamefont
  {Ichiki}},\ }\href {\doibase 10.1103/PhysRevD.76.084019} {\bibfield
  {journal} {\bibinfo  {journal} {Phys. Rev. D}\ }\textbf {\bibinfo {volume}
  {76}},\ \bibinfo {pages} {084019} (\bibinfo {year} {2007})},\ \Eprint
  {http://arxiv.org/abs/hep-th/0703290} {arXiv:hep-th/0703290} \BibitemShut
  {NoStop}%
\bibitem [{\citenamefont {Saito}\ and\ \citenamefont
  {Yokoyama}(2009)}]{Saito:2008jc}%
  \BibitemOpen
  \bibfield  {author} {\bibinfo {author} {\bibfnamefont {R.}~\bibnamefont
  {Saito}}\ and\ \bibinfo {author} {\bibfnamefont {J.}~\bibnamefont
  {Yokoyama}},\ }\href {\doibase 10.1103/PhysRevLett.102.161101} {\bibfield
  {journal} {\bibinfo  {journal} {Phys. Rev. Lett.}\ }\textbf {\bibinfo
  {volume} {102}},\ \bibinfo {pages} {161101} (\bibinfo {year} {2009})},\
  \bibinfo {note} {[Erratum: Phys.Rev.Lett. 107, 069901 (2011)]},\ \Eprint
  {http://arxiv.org/abs/0812.4339} {arXiv:0812.4339 [astro-ph]} \BibitemShut
  {NoStop}%
\bibitem [{\citenamefont {Saito}\ and\ \citenamefont
  {Yokoyama}(2010)}]{Saito:2009jt}%
  \BibitemOpen
  \bibfield  {author} {\bibinfo {author} {\bibfnamefont {R.}~\bibnamefont
  {Saito}}\ and\ \bibinfo {author} {\bibfnamefont {J.}~\bibnamefont
  {Yokoyama}},\ }\href {\doibase 10.1143/PTP.126.351} {\bibfield  {journal}
  {\bibinfo  {journal} {Prog. Theor. Phys.}\ }\textbf {\bibinfo {volume}
  {123}},\ \bibinfo {pages} {867} (\bibinfo {year} {2010})},\ \bibinfo {note}
  {[Erratum: Prog.Theor.Phys. 126, 351--352 (2011)]},\ \Eprint
  {http://arxiv.org/abs/0912.5317} {arXiv:0912.5317 [astro-ph.CO]} \BibitemShut
  {NoStop}%
\bibitem [{\citenamefont {Assadullahi}\ and\ \citenamefont
  {Wands}(2010)}]{Assadullahi:2009jc}%
  \BibitemOpen
  \bibfield  {author} {\bibinfo {author} {\bibfnamefont {H.}~\bibnamefont
  {Assadullahi}}\ and\ \bibinfo {author} {\bibfnamefont {D.}~\bibnamefont
  {Wands}},\ }\href {\doibase 10.1103/PhysRevD.81.023527} {\bibfield  {journal}
  {\bibinfo  {journal} {Phys. Rev. D}\ }\textbf {\bibinfo {volume} {81}},\
  \bibinfo {pages} {023527} (\bibinfo {year} {2010})},\ \Eprint
  {http://arxiv.org/abs/0907.4073} {arXiv:0907.4073 [astro-ph.CO]} \BibitemShut
  {NoStop}%
\bibitem [{\citenamefont {Assadullahi}\ and\ \citenamefont
  {Wands}(2009)}]{Assadullahi:2009nf}%
  \BibitemOpen
  \bibfield  {author} {\bibinfo {author} {\bibfnamefont {H.}~\bibnamefont
  {Assadullahi}}\ and\ \bibinfo {author} {\bibfnamefont {D.}~\bibnamefont
  {Wands}},\ }\href {\doibase 10.1103/PhysRevD.79.083511} {\bibfield  {journal}
  {\bibinfo  {journal} {Phys. Rev. D}\ }\textbf {\bibinfo {volume} {79}},\
  \bibinfo {pages} {083511} (\bibinfo {year} {2009})},\ \Eprint
  {http://arxiv.org/abs/0901.0989} {arXiv:0901.0989 [astro-ph.CO]} \BibitemShut
  {NoStop}%
\bibitem [{\citenamefont {Inomata}\ and\ \citenamefont
  {Nakama}(2019)}]{Inomata:2018epa}%
  \BibitemOpen
  \bibfield  {author} {\bibinfo {author} {\bibfnamefont {K.}~\bibnamefont
  {Inomata}}\ and\ \bibinfo {author} {\bibfnamefont {T.}~\bibnamefont
  {Nakama}},\ }\href {\doibase 10.1103/PhysRevD.99.043511} {\bibfield
  {journal} {\bibinfo  {journal} {Phys. Rev. D}\ }\textbf {\bibinfo {volume}
  {99}},\ \bibinfo {pages} {043511} (\bibinfo {year} {2019})},\ \Eprint
  {http://arxiv.org/abs/1812.00674} {arXiv:1812.00674 [astro-ph.CO]}
  \BibitemShut {NoStop}%
\bibitem [{\citenamefont {Cai}\ \emph {et~al.}(2019)\citenamefont {Cai},
  \citenamefont {Pi}, \citenamefont {Wang},\ and\ \citenamefont
  {Yang}}]{Cai:2019amo}%
  \BibitemOpen
  \bibfield  {author} {\bibinfo {author} {\bibfnamefont {R.-G.}\ \bibnamefont
  {Cai}}, \bibinfo {author} {\bibfnamefont {S.}~\bibnamefont {Pi}}, \bibinfo
  {author} {\bibfnamefont {S.-J.}\ \bibnamefont {Wang}}, \ and\ \bibinfo
  {author} {\bibfnamefont {X.-Y.}\ \bibnamefont {Yang}},\ }\href {\doibase
  10.1088/1475-7516/2019/05/013} {\bibfield  {journal} {\bibinfo  {journal}
  {JCAP}\ }\textbf {\bibinfo {volume} {05}},\ \bibinfo {pages} {013} (\bibinfo
  {year} {2019})},\ \Eprint {http://arxiv.org/abs/1901.10152} {arXiv:1901.10152
  [astro-ph.CO]} \BibitemShut {NoStop}%
\bibitem [{\citenamefont {Liu}\ \emph {et~al.}(2020)\citenamefont {Liu},
  \citenamefont {Guo},\ and\ \citenamefont {Cai}}]{Liu:2020oqe}%
  \BibitemOpen
  \bibfield  {author} {\bibinfo {author} {\bibfnamefont {J.}~\bibnamefont
  {Liu}}, \bibinfo {author} {\bibfnamefont {Z.-K.}\ \bibnamefont {Guo}}, \ and\
  \bibinfo {author} {\bibfnamefont {R.-G.}\ \bibnamefont {Cai}},\ }\href
  {\doibase 10.1103/PhysRevD.101.083535} {\bibfield  {journal} {\bibinfo
  {journal} {Phys. Rev. D}\ }\textbf {\bibinfo {volume} {101}},\ \bibinfo
  {pages} {083535} (\bibinfo {year} {2020})},\ \Eprint
  {http://arxiv.org/abs/2003.02075} {arXiv:2003.02075 [astro-ph.CO]}
  \BibitemShut {NoStop}%
\bibitem [{\citenamefont {Hajkarim}\ and\ \citenamefont
  {Schaffner-Bielich}(2020)}]{Hajkarim:2019nbx}%
  \BibitemOpen
  \bibfield  {author} {\bibinfo {author} {\bibfnamefont {F.}~\bibnamefont
  {Hajkarim}}\ and\ \bibinfo {author} {\bibfnamefont {J.}~\bibnamefont
  {Schaffner-Bielich}},\ }\href {\doibase 10.1103/PhysRevD.101.043522}
  {\bibfield  {journal} {\bibinfo  {journal} {Phys. Rev. D}\ }\textbf {\bibinfo
  {volume} {101}},\ \bibinfo {pages} {043522} (\bibinfo {year} {2020})},\
  \Eprint {http://arxiv.org/abs/1910.12357} {arXiv:1910.12357 [hep-ph]}
  \BibitemShut {NoStop}%
\bibitem [{\citenamefont {Bhattacharya}\ \emph {et~al.}(2020)\citenamefont
  {Bhattacharya}, \citenamefont {Mohanty},\ and\ \citenamefont
  {Parashari}}]{Bhattacharya:2019bvk}%
  \BibitemOpen
  \bibfield  {author} {\bibinfo {author} {\bibfnamefont {S.}~\bibnamefont
  {Bhattacharya}}, \bibinfo {author} {\bibfnamefont {S.}~\bibnamefont
  {Mohanty}}, \ and\ \bibinfo {author} {\bibfnamefont {P.}~\bibnamefont
  {Parashari}},\ }\href {\doibase 10.1103/PhysRevD.102.043522} {\bibfield
  {journal} {\bibinfo  {journal} {Phys. Rev. D}\ }\textbf {\bibinfo {volume}
  {102}},\ \bibinfo {pages} {043522} (\bibinfo {year} {2020})},\ \Eprint
  {http://arxiv.org/abs/1912.01653} {arXiv:1912.01653 [astro-ph.CO]}
  \BibitemShut {NoStop}%
\bibitem [{\citenamefont {Dom\`enech}(2020)}]{Domenech:2019quo}%
  \BibitemOpen
  \bibfield  {author} {\bibinfo {author} {\bibfnamefont {G.}~\bibnamefont
  {Dom\`enech}},\ }\href {\doibase 10.1142/S0218271820500285} {\bibfield
  {journal} {\bibinfo  {journal} {Int. J. Mod. Phys. D}\ }\textbf {\bibinfo
  {volume} {29}},\ \bibinfo {pages} {2050028} (\bibinfo {year} {2020})},\
  \Eprint {http://arxiv.org/abs/1912.05583} {arXiv:1912.05583 [gr-qc]}
  \BibitemShut {NoStop}%
\bibitem [{\citenamefont {Inomata}\ \emph
  {et~al.}(2019{\natexlab{a}})\citenamefont {Inomata}, \citenamefont {Kohri},
  \citenamefont {Nakama},\ and\ \citenamefont {Terada}}]{Inomata:2019zqy}%
  \BibitemOpen
  \bibfield  {author} {\bibinfo {author} {\bibfnamefont {K.}~\bibnamefont
  {Inomata}}, \bibinfo {author} {\bibfnamefont {K.}~\bibnamefont {Kohri}},
  \bibinfo {author} {\bibfnamefont {T.}~\bibnamefont {Nakama}}, \ and\ \bibinfo
  {author} {\bibfnamefont {T.}~\bibnamefont {Terada}},\ }\href {\doibase
  10.1088/1475-7516/2019/10/071} {\bibfield  {journal} {\bibinfo  {journal}
  {JCAP}\ }\textbf {\bibinfo {volume} {10}},\ \bibinfo {pages} {071} (\bibinfo
  {year} {2019}{\natexlab{a}})},\ \Eprint {http://arxiv.org/abs/1904.12878}
  {arXiv:1904.12878 [astro-ph.CO]} \BibitemShut {NoStop}%
\bibitem [{\citenamefont {Inomata}\ \emph
  {et~al.}(2019{\natexlab{b}})\citenamefont {Inomata}, \citenamefont {Kohri},
  \citenamefont {Nakama},\ and\ \citenamefont {Terada}}]{Inomata:2019ivs}%
  \BibitemOpen
  \bibfield  {author} {\bibinfo {author} {\bibfnamefont {K.}~\bibnamefont
  {Inomata}}, \bibinfo {author} {\bibfnamefont {K.}~\bibnamefont {Kohri}},
  \bibinfo {author} {\bibfnamefont {T.}~\bibnamefont {Nakama}}, \ and\ \bibinfo
  {author} {\bibfnamefont {T.}~\bibnamefont {Terada}},\ }\href {\doibase
  10.1103/PhysRevD.100.043532} {\bibfield  {journal} {\bibinfo  {journal}
  {Phys. Rev. D}\ }\textbf {\bibinfo {volume} {100}},\ \bibinfo {pages}
  {043532} (\bibinfo {year} {2019}{\natexlab{b}})},\ \Eprint
  {http://arxiv.org/abs/1904.12879} {arXiv:1904.12879 [astro-ph.CO]}
  \BibitemShut {NoStop}%
\bibitem [{\citenamefont {Gow}\ \emph {et~al.}(2021)\citenamefont {Gow},
  \citenamefont {Byrnes}, \citenamefont {Cole},\ and\ \citenamefont
  {Young}}]{Gow:2020bzo}%
  \BibitemOpen
  \bibfield  {author} {\bibinfo {author} {\bibfnamefont {A.~D.}\ \bibnamefont
  {Gow}}, \bibinfo {author} {\bibfnamefont {C.~T.}\ \bibnamefont {Byrnes}},
  \bibinfo {author} {\bibfnamefont {P.~S.}\ \bibnamefont {Cole}}, \ and\
  \bibinfo {author} {\bibfnamefont {S.}~\bibnamefont {Young}},\ }\href
  {\doibase 10.1088/1475-7516/2021/02/002} {\bibfield  {journal} {\bibinfo
  {journal} {JCAP}\ }\textbf {\bibinfo {volume} {02}},\ \bibinfo {pages} {002}
  (\bibinfo {year} {2021})},\ \Eprint {http://arxiv.org/abs/2008.03289}
  {arXiv:2008.03289 [astro-ph.CO]} \BibitemShut {NoStop}%
\bibitem [{\citenamefont {Dom\`enech}\ \emph {et~al.}(2020)\citenamefont
  {Dom\`enech}, \citenamefont {Pi},\ and\ \citenamefont
  {Sasaki}}]{Domenech:2020kqm}%
  \BibitemOpen
  \bibfield  {author} {\bibinfo {author} {\bibfnamefont {G.}~\bibnamefont
  {Dom\`enech}}, \bibinfo {author} {\bibfnamefont {S.}~\bibnamefont {Pi}}, \
  and\ \bibinfo {author} {\bibfnamefont {M.}~\bibnamefont {Sasaki}},\ }\href
  {\doibase 10.1088/1475-7516/2020/08/017} {\bibfield  {journal} {\bibinfo
  {journal} {JCAP}\ }\textbf {\bibinfo {volume} {08}},\ \bibinfo {pages} {017}
  (\bibinfo {year} {2020})},\ \Eprint {http://arxiv.org/abs/2005.12314}
  {arXiv:2005.12314 [gr-qc]} \BibitemShut {NoStop}%
\bibitem [{\citenamefont {Fumagalli}\ \emph
  {et~al.}(2021{\natexlab{a}})\citenamefont {Fumagalli}, \citenamefont
  {Renaux-Petel},\ and\ \citenamefont {Witkowski}}]{Fumagalli:2020nvq}%
  \BibitemOpen
  \bibfield  {author} {\bibinfo {author} {\bibfnamefont {J.}~\bibnamefont
  {Fumagalli}}, \bibinfo {author} {\bibfnamefont {S.}~\bibnamefont
  {Renaux-Petel}}, \ and\ \bibinfo {author} {\bibfnamefont {L.~T.}\
  \bibnamefont {Witkowski}},\ }\href {\doibase 10.1088/1475-7516/2021/08/030}
  {\bibfield  {journal} {\bibinfo  {journal} {JCAP}\ }\textbf {\bibinfo
  {volume} {08}},\ \bibinfo {pages} {030} (\bibinfo {year}
  {2021}{\natexlab{a}})},\ \Eprint {http://arxiv.org/abs/2012.02761}
  {arXiv:2012.02761 [astro-ph.CO]} \BibitemShut {NoStop}%
\bibitem [{\citenamefont {Dalianis}\ and\ \citenamefont
  {Kritos}(2021)}]{Dalianis:2020cla}%
  \BibitemOpen
  \bibfield  {author} {\bibinfo {author} {\bibfnamefont {I.}~\bibnamefont
  {Dalianis}}\ and\ \bibinfo {author} {\bibfnamefont {K.}~\bibnamefont
  {Kritos}},\ }\href {\doibase 10.1103/PhysRevD.103.023505} {\bibfield
  {journal} {\bibinfo  {journal} {Phys. Rev. D}\ }\textbf {\bibinfo {volume}
  {103}},\ \bibinfo {pages} {023505} (\bibinfo {year} {2021})},\ \Eprint
  {http://arxiv.org/abs/2007.07915} {arXiv:2007.07915 [astro-ph.CO]}
  \BibitemShut {NoStop}%
\bibitem [{\citenamefont {Abe}\ \emph {et~al.}(2021)\citenamefont {Abe},
  \citenamefont {Tada},\ and\ \citenamefont {Ueda}}]{Abe:2020sqb}%
  \BibitemOpen
  \bibfield  {author} {\bibinfo {author} {\bibfnamefont {K.~T.}\ \bibnamefont
  {Abe}}, \bibinfo {author} {\bibfnamefont {Y.}~\bibnamefont {Tada}}, \ and\
  \bibinfo {author} {\bibfnamefont {I.}~\bibnamefont {Ueda}},\ }\href {\doibase
  10.1088/1475-7516/2021/06/048} {\bibfield  {journal} {\bibinfo  {journal}
  {JCAP}\ }\textbf {\bibinfo {volume} {06}},\ \bibinfo {pages} {048} (\bibinfo
  {year} {2021})},\ \Eprint {http://arxiv.org/abs/2010.06193} {arXiv:2010.06193
  [astro-ph.CO]} \BibitemShut {NoStop}%
\bibitem [{\citenamefont {Braglia}\ \emph {et~al.}(2020)\citenamefont
  {Braglia}, \citenamefont {Hazra}, \citenamefont {Finelli}, \citenamefont
  {Smoot}, \citenamefont {Sriramkumar},\ and\ \citenamefont
  {Starobinsky}}]{Braglia:2020eai}%
  \BibitemOpen
  \bibfield  {author} {\bibinfo {author} {\bibfnamefont {M.}~\bibnamefont
  {Braglia}}, \bibinfo {author} {\bibfnamefont {D.~K.}\ \bibnamefont {Hazra}},
  \bibinfo {author} {\bibfnamefont {F.}~\bibnamefont {Finelli}}, \bibinfo
  {author} {\bibfnamefont {G.~F.}\ \bibnamefont {Smoot}}, \bibinfo {author}
  {\bibfnamefont {L.}~\bibnamefont {Sriramkumar}}, \ and\ \bibinfo {author}
  {\bibfnamefont {A.~A.}\ \bibnamefont {Starobinsky}},\ }\href {\doibase
  10.1088/1475-7516/2020/08/001} {\bibfield  {journal} {\bibinfo  {journal}
  {JCAP}\ }\textbf {\bibinfo {volume} {08}},\ \bibinfo {pages} {001} (\bibinfo
  {year} {2020})},\ \Eprint {http://arxiv.org/abs/2005.02895} {arXiv:2005.02895
  [astro-ph.CO]} \BibitemShut {NoStop}%
\bibitem [{\citenamefont {Braglia}\ \emph {et~al.}(2021)\citenamefont
  {Braglia}, \citenamefont {Chen},\ and\ \citenamefont
  {Hazra}}]{Braglia:2020taf}%
  \BibitemOpen
  \bibfield  {author} {\bibinfo {author} {\bibfnamefont {M.}~\bibnamefont
  {Braglia}}, \bibinfo {author} {\bibfnamefont {X.}~\bibnamefont {Chen}}, \
  and\ \bibinfo {author} {\bibfnamefont {D.~K.}\ \bibnamefont {Hazra}},\ }\href
  {\doibase 10.1088/1475-7516/2021/03/005} {\bibfield  {journal} {\bibinfo
  {journal} {JCAP}\ }\textbf {\bibinfo {volume} {03}},\ \bibinfo {pages} {005}
  (\bibinfo {year} {2021})},\ \Eprint {http://arxiv.org/abs/2012.05821}
  {arXiv:2012.05821 [astro-ph.CO]} \BibitemShut {NoStop}%
\bibitem [{\citenamefont {Atal}\ and\ \citenamefont
  {Dom\`enech}(2021)}]{Atal:2021jyo}%
  \BibitemOpen
  \bibfield  {author} {\bibinfo {author} {\bibfnamefont {V.}~\bibnamefont
  {Atal}}\ and\ \bibinfo {author} {\bibfnamefont {G.}~\bibnamefont
  {Dom\`enech}},\ }\href {\doibase 10.1088/1475-7516/2021/06/001} {\bibfield
  {journal} {\bibinfo  {journal} {JCAP}\ }\textbf {\bibinfo {volume} {06}},\
  \bibinfo {pages} {001} (\bibinfo {year} {2021})},\ \Eprint
  {http://arxiv.org/abs/2103.01056} {arXiv:2103.01056 [astro-ph.CO]}
  \BibitemShut {NoStop}%
\bibitem [{\citenamefont {Fumagalli}\ \emph
  {et~al.}(2021{\natexlab{b}})\citenamefont {Fumagalli}, \citenamefont
  {Renaux-Petel},\ and\ \citenamefont {Witkowski}}]{Fumagalli:2021cel}%
  \BibitemOpen
  \bibfield  {author} {\bibinfo {author} {\bibfnamefont {J.}~\bibnamefont
  {Fumagalli}}, \bibinfo {author} {\bibfnamefont {S.}~\bibnamefont
  {Renaux-Petel}}, \ and\ \bibinfo {author} {\bibfnamefont {L.~T.}\
  \bibnamefont {Witkowski}},\ }\href {\doibase 10.1088/1475-7516/2021/08/059}
  {\bibfield  {journal} {\bibinfo  {journal} {JCAP}\ }\textbf {\bibinfo
  {volume} {08}},\ \bibinfo {pages} {059} (\bibinfo {year}
  {2021}{\natexlab{b}})},\ \Eprint {http://arxiv.org/abs/2105.06481}
  {arXiv:2105.06481 [astro-ph.CO]} \BibitemShut {NoStop}%
\bibitem [{\citenamefont {Yuan}\ and\ \citenamefont
  {Huang}(2021)}]{Yuan:2021qgz}%
  \BibitemOpen
  \bibfield  {author} {\bibinfo {author} {\bibfnamefont {C.}~\bibnamefont
  {Yuan}}\ and\ \bibinfo {author} {\bibfnamefont {Q.-G.}\ \bibnamefont
  {Huang}},\ }\href@noop {} {\  (\bibinfo {year} {2021})},\ \Eprint
  {http://arxiv.org/abs/2103.04739} {arXiv:2103.04739 [astro-ph.GA]}
  \BibitemShut {NoStop}%
\bibitem [{\citenamefont {Dom\`enech}(2021)}]{Domenech:2021ztg}%
  \BibitemOpen
  \bibfield  {author} {\bibinfo {author} {\bibfnamefont {G.}~\bibnamefont
  {Dom\`enech}},\ }\href@noop {} {\  (\bibinfo {year} {2021})},\ \Eprint
  {http://arxiv.org/abs/2109.01398} {arXiv:2109.01398 [gr-qc]} \BibitemShut
  {NoStop}%
\bibitem [{\citenamefont {Allahverdi}\ \emph {et~al.}(2021)\citenamefont
  {Allahverdi} \emph {et~al.}}]{Allahverdi:2020bys}%
  \BibitemOpen
  \bibfield  {author} {\bibinfo {author} {\bibfnamefont {R.}~\bibnamefont
  {Allahverdi}} \emph {et~al.},\ }\href {\doibase 10.21105/astro.2006.16182}
  {\bibfield  {journal} {\bibinfo  {journal} {Open J. Astrophys.}\ }\textbf
  {\bibinfo {volume} {4}} (\bibinfo {year} {2021}),\
  10.21105/astro.2006.16182},\ \Eprint {http://arxiv.org/abs/2006.16182}
  {arXiv:2006.16182 [astro-ph.CO]} \BibitemShut {NoStop}%
\bibitem [{\citenamefont {Chen}(2010)}]{Chen:2010xka}%
  \BibitemOpen
  \bibfield  {author} {\bibinfo {author} {\bibfnamefont {X.}~\bibnamefont
  {Chen}},\ }\href {\doibase 10.1155/2010/638979} {\bibfield  {journal}
  {\bibinfo  {journal} {Adv. Astron.}\ }\textbf {\bibinfo {volume} {2010}},\
  \bibinfo {pages} {638979} (\bibinfo {year} {2010})},\ \Eprint
  {http://arxiv.org/abs/1002.1416} {arXiv:1002.1416 [astro-ph.CO]} \BibitemShut
  {NoStop}%
\bibitem [{\citenamefont {Chluba}\ \emph {et~al.}(2015)\citenamefont {Chluba},
  \citenamefont {Hamann},\ and\ \citenamefont {Patil}}]{Chluba:2015bqa}%
  \BibitemOpen
  \bibfield  {author} {\bibinfo {author} {\bibfnamefont {J.}~\bibnamefont
  {Chluba}}, \bibinfo {author} {\bibfnamefont {J.}~\bibnamefont {Hamann}}, \
  and\ \bibinfo {author} {\bibfnamefont {S.~P.}\ \bibnamefont {Patil}},\ }\href
  {\doibase 10.1142/S0218271815300232} {\bibfield  {journal} {\bibinfo
  {journal} {Int. J. Mod. Phys.}\ }\textbf {\bibinfo {volume} {D24}},\ \bibinfo
  {pages} {1530023} (\bibinfo {year} {2015})},\ \Eprint
  {http://arxiv.org/abs/1505.01834} {arXiv:1505.01834 [astro-ph.CO]}
  \BibitemShut {NoStop}%
\bibitem [{\citenamefont {Slosar}\ \emph {et~al.}(2019)\citenamefont {Slosar}
  \emph {et~al.}}]{Slosar:2019gvt}%
  \BibitemOpen
  \bibfield  {author} {\bibinfo {author} {\bibfnamefont {A.}~\bibnamefont
  {Slosar}} \emph {et~al.},\ }\href@noop {} {\  (\bibinfo {year} {2019})},\
  \Eprint {http://arxiv.org/abs/1903.09883} {arXiv:1903.09883 [astro-ph.CO]}
  \BibitemShut {NoStop}%
\bibitem [{\citenamefont {Zel'dovich}(1967)}]{Zeldovich:1967lct}%
  \BibitemOpen
  \bibfield  {author} {\bibinfo {author} {\bibfnamefont {I.~D.}\ \bibnamefont
  {Zel'dovich}, \bibfnamefont {Ya.B.;~Novikov}},\ }\href@noop {} {\bibfield
  {journal} {\bibinfo  {journal} {Soviet Astron. AJ (Engl. Transl. ),}\
  }\textbf {\bibinfo {volume} {10}},\ \bibinfo {pages} {602} (\bibinfo {year}
  {1967})}\BibitemShut {NoStop}%
\bibitem [{\citenamefont {Hawking}(1971)}]{Hawking:1971ei}%
  \BibitemOpen
  \bibfield  {author} {\bibinfo {author} {\bibfnamefont {S.}~\bibnamefont
  {Hawking}},\ }\href@noop {} {\bibfield  {journal} {\bibinfo  {journal} {Mon.
  Not. Roy. Astron. Soc.}\ }\textbf {\bibinfo {volume} {152}},\ \bibinfo
  {pages} {75} (\bibinfo {year} {1971})}\BibitemShut {NoStop}%
\bibitem [{\citenamefont {Carr}\ and\ \citenamefont
  {Hawking}(1974)}]{Carr:1974nx}%
  \BibitemOpen
  \bibfield  {author} {\bibinfo {author} {\bibfnamefont {B.~J.}\ \bibnamefont
  {Carr}}\ and\ \bibinfo {author} {\bibfnamefont {S.~W.}\ \bibnamefont
  {Hawking}},\ }\href@noop {} {\bibfield  {journal} {\bibinfo  {journal} {Mon.
  Not. Roy. Astron. Soc.}\ }\textbf {\bibinfo {volume} {168}},\ \bibinfo
  {pages} {399} (\bibinfo {year} {1974})}\BibitemShut {NoStop}%
\bibitem [{\citenamefont {Sasaki}\ \emph {et~al.}(2018)\citenamefont {Sasaki},
  \citenamefont {Suyama}, \citenamefont {Tanaka},\ and\ \citenamefont
  {Yokoyama}}]{Sasaki:2018dmp}%
  \BibitemOpen
  \bibfield  {author} {\bibinfo {author} {\bibfnamefont {M.}~\bibnamefont
  {Sasaki}}, \bibinfo {author} {\bibfnamefont {T.}~\bibnamefont {Suyama}},
  \bibinfo {author} {\bibfnamefont {T.}~\bibnamefont {Tanaka}}, \ and\ \bibinfo
  {author} {\bibfnamefont {S.}~\bibnamefont {Yokoyama}},\ }\href {\doibase
  10.1088/1361-6382/aaa7b4} {\bibfield  {journal} {\bibinfo  {journal} {Class.
  Quant. Grav.}\ }\textbf {\bibinfo {volume} {35}},\ \bibinfo {pages} {063001}
  (\bibinfo {year} {2018})},\ \Eprint {http://arxiv.org/abs/1801.05235}
  {arXiv:1801.05235 [astro-ph.CO]} \BibitemShut {NoStop}%
\bibitem [{\citenamefont {Carr}\ \emph {et~al.}(2020)\citenamefont {Carr},
  \citenamefont {Kohri}, \citenamefont {Sendouda},\ and\ \citenamefont
  {Yokoyama}}]{Carr:2020gox}%
  \BibitemOpen
  \bibfield  {author} {\bibinfo {author} {\bibfnamefont {B.}~\bibnamefont
  {Carr}}, \bibinfo {author} {\bibfnamefont {K.}~\bibnamefont {Kohri}},
  \bibinfo {author} {\bibfnamefont {Y.}~\bibnamefont {Sendouda}}, \ and\
  \bibinfo {author} {\bibfnamefont {J.}~\bibnamefont {Yokoyama}},\ }\href@noop
  {} {\  (\bibinfo {year} {2020})},\ \Eprint {http://arxiv.org/abs/2002.12778}
  {arXiv:2002.12778 [astro-ph.CO]} \BibitemShut {NoStop}%
\bibitem [{\citenamefont {Carr}\ and\ \citenamefont
  {Kuhnel}(2020)}]{Carr:2020xqk}%
  \BibitemOpen
  \bibfield  {author} {\bibinfo {author} {\bibfnamefont {B.}~\bibnamefont
  {Carr}}\ and\ \bibinfo {author} {\bibfnamefont {F.}~\bibnamefont {Kuhnel}},\
  }\href {\doibase 10.1146/annurev-nucl-050520-125911} {\bibfield  {journal}
  {\bibinfo  {journal} {Ann. Rev. Nucl. Part. Sci.}\ }\textbf {\bibinfo
  {volume} {70}},\ \bibinfo {pages} {355} (\bibinfo {year} {2020})},\ \Eprint
  {http://arxiv.org/abs/2006.02838} {arXiv:2006.02838 [astro-ph.CO]}
  \BibitemShut {NoStop}%
\bibitem [{\citenamefont {Green}\ and\ \citenamefont
  {Kavanagh}(2021)}]{Green:2020jor}%
  \BibitemOpen
  \bibfield  {author} {\bibinfo {author} {\bibfnamefont {A.~M.}\ \bibnamefont
  {Green}}\ and\ \bibinfo {author} {\bibfnamefont {B.~J.}\ \bibnamefont
  {Kavanagh}},\ }\href {\doibase 10.1088/1361-6471/abc534} {\bibfield
  {journal} {\bibinfo  {journal} {J. Phys. G}\ }\textbf {\bibinfo {volume}
  {48}},\ \bibinfo {pages} {4} (\bibinfo {year} {2021})},\ \Eprint
  {http://arxiv.org/abs/2007.10722} {arXiv:2007.10722 [astro-ph.CO]}
  \BibitemShut {NoStop}%
\bibitem [{\citenamefont {Inomata}\ \emph {et~al.}(2017)\citenamefont
  {Inomata}, \citenamefont {Kawasaki}, \citenamefont {Mukaida}, \citenamefont
  {Tada},\ and\ \citenamefont {Yanagida}}]{Inomata:2016rbd}%
  \BibitemOpen
  \bibfield  {author} {\bibinfo {author} {\bibfnamefont {K.}~\bibnamefont
  {Inomata}}, \bibinfo {author} {\bibfnamefont {M.}~\bibnamefont {Kawasaki}},
  \bibinfo {author} {\bibfnamefont {K.}~\bibnamefont {Mukaida}}, \bibinfo
  {author} {\bibfnamefont {Y.}~\bibnamefont {Tada}}, \ and\ \bibinfo {author}
  {\bibfnamefont {T.~T.}\ \bibnamefont {Yanagida}},\ }\href {\doibase
  10.1103/PhysRevD.95.123510} {\bibfield  {journal} {\bibinfo  {journal} {Phys.
  Rev. D}\ }\textbf {\bibinfo {volume} {95}},\ \bibinfo {pages} {123510}
  (\bibinfo {year} {2017})},\ \Eprint {http://arxiv.org/abs/1611.06130}
  {arXiv:1611.06130 [astro-ph.CO]} \BibitemShut {NoStop}%
\bibitem [{\citenamefont {Nakama}\ \emph {et~al.}(2017)\citenamefont {Nakama},
  \citenamefont {Silk},\ and\ \citenamefont {Kamionkowski}}]{Nakama:2016gzw}%
  \BibitemOpen
  \bibfield  {author} {\bibinfo {author} {\bibfnamefont {T.}~\bibnamefont
  {Nakama}}, \bibinfo {author} {\bibfnamefont {J.}~\bibnamefont {Silk}}, \ and\
  \bibinfo {author} {\bibfnamefont {M.}~\bibnamefont {Kamionkowski}},\ }\href
  {\doibase 10.1103/PhysRevD.95.043511} {\bibfield  {journal} {\bibinfo
  {journal} {Phys. Rev. D}\ }\textbf {\bibinfo {volume} {95}},\ \bibinfo
  {pages} {043511} (\bibinfo {year} {2017})},\ \Eprint
  {http://arxiv.org/abs/1612.06264} {arXiv:1612.06264 [astro-ph.CO]}
  \BibitemShut {NoStop}%
\bibitem [{\citenamefont {Ando}\ \emph {et~al.}(2018)\citenamefont {Ando},
  \citenamefont {Inomata}, \citenamefont {Kawasaki}, \citenamefont {Mukaida},\
  and\ \citenamefont {Yanagida}}]{Ando:2017veq}%
  \BibitemOpen
  \bibfield  {author} {\bibinfo {author} {\bibfnamefont {K.}~\bibnamefont
  {Ando}}, \bibinfo {author} {\bibfnamefont {K.}~\bibnamefont {Inomata}},
  \bibinfo {author} {\bibfnamefont {M.}~\bibnamefont {Kawasaki}}, \bibinfo
  {author} {\bibfnamefont {K.}~\bibnamefont {Mukaida}}, \ and\ \bibinfo
  {author} {\bibfnamefont {T.~T.}\ \bibnamefont {Yanagida}},\ }\href {\doibase
  10.1103/PhysRevD.97.123512} {\bibfield  {journal} {\bibinfo  {journal} {Phys.
  Rev. D}\ }\textbf {\bibinfo {volume} {97}},\ \bibinfo {pages} {123512}
  (\bibinfo {year} {2018})},\ \Eprint {http://arxiv.org/abs/1711.08956}
  {arXiv:1711.08956 [astro-ph.CO]} \BibitemShut {NoStop}%
\bibitem [{\citenamefont {Kohri}\ and\ \citenamefont
  {Terada}(2018{\natexlab{a}})}]{Kohri:2018qtx}%
  \BibitemOpen
  \bibfield  {author} {\bibinfo {author} {\bibfnamefont {K.}~\bibnamefont
  {Kohri}}\ and\ \bibinfo {author} {\bibfnamefont {T.}~\bibnamefont {Terada}},\
  }\href {\doibase 10.1088/1361-6382/aaea18} {\bibfield  {journal} {\bibinfo
  {journal} {Class. Quant. Grav.}\ }\textbf {\bibinfo {volume} {35}},\ \bibinfo
  {pages} {235017} (\bibinfo {year} {2018}{\natexlab{a}})},\ \Eprint
  {http://arxiv.org/abs/1802.06785} {arXiv:1802.06785 [astro-ph.CO]}
  \BibitemShut {NoStop}%
\bibitem [{\citenamefont {Franciolini}\ \emph {et~al.}(2021)\citenamefont
  {Franciolini}, \citenamefont {Baibhav}, \citenamefont {De~Luca},
  \citenamefont {Ng}, \citenamefont {Wong}, \citenamefont {Berti},
  \citenamefont {Pani}, \citenamefont {Riotto},\ and\ \citenamefont
  {Vitale}}]{Franciolini:2021tla}%
  \BibitemOpen
  \bibfield  {author} {\bibinfo {author} {\bibfnamefont {G.}~\bibnamefont
  {Franciolini}}, \bibinfo {author} {\bibfnamefont {V.}~\bibnamefont
  {Baibhav}}, \bibinfo {author} {\bibfnamefont {V.}~\bibnamefont {De~Luca}},
  \bibinfo {author} {\bibfnamefont {K.~K.~Y.}\ \bibnamefont {Ng}}, \bibinfo
  {author} {\bibfnamefont {K.~W.~K.}\ \bibnamefont {Wong}}, \bibinfo {author}
  {\bibfnamefont {E.}~\bibnamefont {Berti}}, \bibinfo {author} {\bibfnamefont
  {P.}~\bibnamefont {Pani}}, \bibinfo {author} {\bibfnamefont {A.}~\bibnamefont
  {Riotto}}, \ and\ \bibinfo {author} {\bibfnamefont {S.}~\bibnamefont
  {Vitale}},\ }\href@noop {} {\  (\bibinfo {year} {2021})},\ \Eprint
  {http://arxiv.org/abs/2105.03349} {arXiv:2105.03349 [gr-qc]} \BibitemShut
  {NoStop}%
\bibitem [{\citenamefont {Dalianis}\ \emph {et~al.}(2021)\citenamefont
  {Dalianis}, \citenamefont {Kodaxis}, \citenamefont {Stamou}, \citenamefont
  {Tetradis},\ and\ \citenamefont {Tsigkas-Kouvelis}}]{Dalianis:2021iig}%
  \BibitemOpen
  \bibfield  {author} {\bibinfo {author} {\bibfnamefont {I.}~\bibnamefont
  {Dalianis}}, \bibinfo {author} {\bibfnamefont {G.~P.}\ \bibnamefont
  {Kodaxis}}, \bibinfo {author} {\bibfnamefont {I.~D.}\ \bibnamefont {Stamou}},
  \bibinfo {author} {\bibfnamefont {N.}~\bibnamefont {Tetradis}}, \ and\
  \bibinfo {author} {\bibfnamefont {A.}~\bibnamefont {Tsigkas-Kouvelis}},\
  }\href@noop {} {\  (\bibinfo {year} {2021})},\ \Eprint
  {http://arxiv.org/abs/2106.02467} {arXiv:2106.02467 [astro-ph.CO]}
  \BibitemShut {NoStop}%
\bibitem [{\citenamefont {Spokoiny}(1993)}]{Spokoiny:1993kt}%
  \BibitemOpen
  \bibfield  {author} {\bibinfo {author} {\bibfnamefont {B.}~\bibnamefont
  {Spokoiny}},\ }\href {\doibase 10.1016/0370-2693(93)90155-B} {\bibfield
  {journal} {\bibinfo  {journal} {Phys. Lett. B}\ }\textbf {\bibinfo {volume}
  {315}},\ \bibinfo {pages} {40} (\bibinfo {year} {1993})},\ \Eprint
  {http://arxiv.org/abs/gr-qc/9306008} {arXiv:gr-qc/9306008} \BibitemShut
  {NoStop}%
\bibitem [{\citenamefont {Peebles}\ and\ \citenamefont
  {Vilenkin}(1999)}]{Peebles:1998qn}%
  \BibitemOpen
  \bibfield  {author} {\bibinfo {author} {\bibfnamefont {P.~J.~E.}\
  \bibnamefont {Peebles}}\ and\ \bibinfo {author} {\bibfnamefont
  {A.}~\bibnamefont {Vilenkin}},\ }\href {\doibase 10.1103/PhysRevD.59.063505}
  {\bibfield  {journal} {\bibinfo  {journal} {Phys. Rev. D}\ }\textbf {\bibinfo
  {volume} {59}},\ \bibinfo {pages} {063505} (\bibinfo {year} {1999})},\
  \Eprint {http://arxiv.org/abs/astro-ph/9810509} {arXiv:astro-ph/9810509}
  \BibitemShut {NoStop}%
\bibitem [{\citenamefont {Brax}\ and\ \citenamefont
  {Martin}(2005)}]{Brax:2005uf}%
  \BibitemOpen
  \bibfield  {author} {\bibinfo {author} {\bibfnamefont {P.}~\bibnamefont
  {Brax}}\ and\ \bibinfo {author} {\bibfnamefont {J.}~\bibnamefont {Martin}},\
  }\href {\doibase 10.1103/PhysRevD.71.063530} {\bibfield  {journal} {\bibinfo
  {journal} {Phys. Rev. D}\ }\textbf {\bibinfo {volume} {71}},\ \bibinfo
  {pages} {063530} (\bibinfo {year} {2005})},\ \Eprint
  {http://arxiv.org/abs/astro-ph/0502069} {arXiv:astro-ph/0502069} \BibitemShut
  {NoStop}%
\bibitem [{\citenamefont {Hossain}\ \emph {et~al.}(2014)\citenamefont
  {Hossain}, \citenamefont {Myrzakulov}, \citenamefont {Sami},\ and\
  \citenamefont {Saridakis}}]{Hossain:2014xha}%
  \BibitemOpen
  \bibfield  {author} {\bibinfo {author} {\bibfnamefont {M.~W.}\ \bibnamefont
  {Hossain}}, \bibinfo {author} {\bibfnamefont {R.}~\bibnamefont {Myrzakulov}},
  \bibinfo {author} {\bibfnamefont {M.}~\bibnamefont {Sami}}, \ and\ \bibinfo
  {author} {\bibfnamefont {E.~N.}\ \bibnamefont {Saridakis}},\ }\href {\doibase
  10.1103/PhysRevD.90.023512} {\bibfield  {journal} {\bibinfo  {journal} {Phys.
  Rev. D}\ }\textbf {\bibinfo {volume} {90}},\ \bibinfo {pages} {023512}
  (\bibinfo {year} {2014})},\ \Eprint {http://arxiv.org/abs/1402.6661}
  {arXiv:1402.6661 [gr-qc]} \BibitemShut {NoStop}%
\bibitem [{\citenamefont {Kawasaki}\ \emph {et~al.}(1999)\citenamefont
  {Kawasaki}, \citenamefont {Kohri},\ and\ \citenamefont
  {Sugiyama}}]{Kawasaki:1999na}%
  \BibitemOpen
  \bibfield  {author} {\bibinfo {author} {\bibfnamefont {M.}~\bibnamefont
  {Kawasaki}}, \bibinfo {author} {\bibfnamefont {K.}~\bibnamefont {Kohri}}, \
  and\ \bibinfo {author} {\bibfnamefont {N.}~\bibnamefont {Sugiyama}},\ }\href
  {\doibase 10.1103/PhysRevLett.82.4168} {\bibfield  {journal} {\bibinfo
  {journal} {Phys. Rev. Lett.}\ }\textbf {\bibinfo {volume} {82}},\ \bibinfo
  {pages} {4168} (\bibinfo {year} {1999})},\ \Eprint
  {http://arxiv.org/abs/astro-ph/9811437} {arXiv:astro-ph/9811437} \BibitemShut
  {NoStop}%
\bibitem [{\citenamefont {Kawasaki}\ \emph {et~al.}(2000)\citenamefont
  {Kawasaki}, \citenamefont {Kohri},\ and\ \citenamefont
  {Sugiyama}}]{Kawasaki:2000en}%
  \BibitemOpen
  \bibfield  {author} {\bibinfo {author} {\bibfnamefont {M.}~\bibnamefont
  {Kawasaki}}, \bibinfo {author} {\bibfnamefont {K.}~\bibnamefont {Kohri}}, \
  and\ \bibinfo {author} {\bibfnamefont {N.}~\bibnamefont {Sugiyama}},\ }\href
  {\doibase 10.1103/PhysRevD.62.023506} {\bibfield  {journal} {\bibinfo
  {journal} {Phys. Rev. D}\ }\textbf {\bibinfo {volume} {62}},\ \bibinfo
  {pages} {023506} (\bibinfo {year} {2000})},\ \Eprint
  {http://arxiv.org/abs/astro-ph/0002127} {arXiv:astro-ph/0002127} \BibitemShut
  {NoStop}%
\bibitem [{\citenamefont {Hannestad}(2004)}]{Hannestad:2004px}%
  \BibitemOpen
  \bibfield  {author} {\bibinfo {author} {\bibfnamefont {S.}~\bibnamefont
  {Hannestad}},\ }\href {\doibase 10.1103/PhysRevD.70.043506} {\bibfield
  {journal} {\bibinfo  {journal} {Phys. Rev. D}\ }\textbf {\bibinfo {volume}
  {70}},\ \bibinfo {pages} {043506} (\bibinfo {year} {2004})},\ \Eprint
  {http://arxiv.org/abs/astro-ph/0403291} {arXiv:astro-ph/0403291} \BibitemShut
  {NoStop}%
\bibitem [{\citenamefont {Hasegawa}\ \emph {et~al.}(2019)\citenamefont
  {Hasegawa}, \citenamefont {Hiroshima}, \citenamefont {Kohri}, \citenamefont
  {Hansen}, \citenamefont {Tram},\ and\ \citenamefont
  {Hannestad}}]{Hasegawa:2019jsa}%
  \BibitemOpen
  \bibfield  {author} {\bibinfo {author} {\bibfnamefont {T.}~\bibnamefont
  {Hasegawa}}, \bibinfo {author} {\bibfnamefont {N.}~\bibnamefont {Hiroshima}},
  \bibinfo {author} {\bibfnamefont {K.}~\bibnamefont {Kohri}}, \bibinfo
  {author} {\bibfnamefont {R.~S.~L.}\ \bibnamefont {Hansen}}, \bibinfo {author}
  {\bibfnamefont {T.}~\bibnamefont {Tram}}, \ and\ \bibinfo {author}
  {\bibfnamefont {S.}~\bibnamefont {Hannestad}},\ }\href {\doibase
  10.1088/1475-7516/2019/12/012} {\bibfield  {journal} {\bibinfo  {journal}
  {JCAP}\ }\textbf {\bibinfo {volume} {12}},\ \bibinfo {pages} {012} (\bibinfo
  {year} {2019})},\ \Eprint {http://arxiv.org/abs/1908.10189} {arXiv:1908.10189
  [hep-ph]} \BibitemShut {NoStop}%
\bibitem [{\citenamefont {Espinosa}\ \emph {et~al.}(2018)\citenamefont
  {Espinosa}, \citenamefont {Racco},\ and\ \citenamefont
  {Riotto}}]{Espinosa:2018eve}%
  \BibitemOpen
  \bibfield  {author} {\bibinfo {author} {\bibfnamefont {J.~R.}\ \bibnamefont
  {Espinosa}}, \bibinfo {author} {\bibfnamefont {D.}~\bibnamefont {Racco}}, \
  and\ \bibinfo {author} {\bibfnamefont {A.}~\bibnamefont {Riotto}},\ }\href
  {\doibase 10.1088/1475-7516/2018/09/012} {\bibfield  {journal} {\bibinfo
  {journal} {JCAP}\ }\textbf {\bibinfo {volume} {09}},\ \bibinfo {pages} {012}
  (\bibinfo {year} {2018})},\ \Eprint {http://arxiv.org/abs/1804.07732}
  {arXiv:1804.07732 [hep-ph]} \BibitemShut {NoStop}%
\bibitem [{\citenamefont {Kohri}\ and\ \citenamefont
  {Terada}(2018{\natexlab{b}})}]{Kohri:2018awv}%
  \BibitemOpen
  \bibfield  {author} {\bibinfo {author} {\bibfnamefont {K.}~\bibnamefont
  {Kohri}}\ and\ \bibinfo {author} {\bibfnamefont {T.}~\bibnamefont {Terada}},\
  }\href {\doibase 10.1103/PhysRevD.97.123532} {\bibfield  {journal} {\bibinfo
  {journal} {Phys. Rev. D}\ }\textbf {\bibinfo {volume} {97}},\ \bibinfo
  {pages} {123532} (\bibinfo {year} {2018}{\natexlab{b}})},\ \Eprint
  {http://arxiv.org/abs/1804.08577} {arXiv:1804.08577 [gr-qc]} \BibitemShut
  {NoStop}%
\bibitem [{\citenamefont {Pi}\ and\ \citenamefont {Sasaki}(2020)}]{Pi:2020otn}%
  \BibitemOpen
  \bibfield  {author} {\bibinfo {author} {\bibfnamefont {S.}~\bibnamefont
  {Pi}}\ and\ \bibinfo {author} {\bibfnamefont {M.}~\bibnamefont {Sasaki}},\
  }\href {\doibase 10.1088/1475-7516/2020/09/037} {\bibfield  {journal}
  {\bibinfo  {journal} {JCAP}\ }\textbf {\bibinfo {volume} {09}},\ \bibinfo
  {pages} {037} (\bibinfo {year} {2020})},\ \Eprint
  {http://arxiv.org/abs/2005.12306} {arXiv:2005.12306 [gr-qc]} \BibitemShut
  {NoStop}%
\bibitem [{\citenamefont {Fumagalli}\ \emph {et~al.}()\citenamefont
  {Fumagalli}, \citenamefont {Palma}, \citenamefont {Renaux-Petel},
  \citenamefont {Sypsas}, \citenamefont {Witkowski},\ and\ \citenamefont
  {Zenteno}}]{primordial-GWs}%
  \BibitemOpen
  \bibfield  {author} {\bibinfo {author} {\bibfnamefont {J.}~\bibnamefont
  {Fumagalli}}, \bibinfo {author} {\bibfnamefont {G.~A.}\ \bibnamefont
  {Palma}}, \bibinfo {author} {\bibfnamefont {S.}~\bibnamefont {Renaux-Petel}},
  \bibinfo {author} {\bibfnamefont {S.}~\bibnamefont {Sypsas}}, \bibinfo
  {author} {\bibfnamefont {L.~T.}\ \bibnamefont {Witkowski}}, \ and\ \bibinfo
  {author} {\bibfnamefont {C.}~\bibnamefont {Zenteno}},\ }\href@noop {}
  {\bibinfo  {journal} {in preparation}\ }\BibitemShut {NoStop}%
\bibitem [{\citenamefont {Unal}(2019)}]{Unal:2018yaa}%
  \BibitemOpen
\bibfield  {journal} {  }\bibfield  {author} {\bibinfo {author} {\bibfnamefont
  {C.}~\bibnamefont {Unal}},\ }\href {\doibase 10.1103/PhysRevD.99.041301}
  {\bibfield  {journal} {\bibinfo  {journal} {Phys. Rev. D}\ }\textbf {\bibinfo
  {volume} {99}},\ \bibinfo {pages} {041301} (\bibinfo {year} {2019})},\
  \Eprint {http://arxiv.org/abs/1811.09151} {arXiv:1811.09151 [astro-ph.CO]}
  \BibitemShut {NoStop}%
\bibitem [{\citenamefont {Adshead}\ \emph {et~al.}(2021)\citenamefont
  {Adshead}, \citenamefont {Lozanov},\ and\ \citenamefont
  {Weiner}}]{Adshead:2021hnm}%
  \BibitemOpen
  \bibfield  {author} {\bibinfo {author} {\bibfnamefont {P.}~\bibnamefont
  {Adshead}}, \bibinfo {author} {\bibfnamefont {K.~D.}\ \bibnamefont
  {Lozanov}}, \ and\ \bibinfo {author} {\bibfnamefont {Z.~J.}\ \bibnamefont
  {Weiner}},\ }\href@noop {} {\  (\bibinfo {year} {2021})},\ \Eprint
  {http://arxiv.org/abs/2105.01659} {arXiv:2105.01659 [astro-ph.CO]}
  \BibitemShut {NoStop}%
\bibitem [{{\relax DLMF}()}]{NIST:DLMF}%
  \BibitemOpen
  {\relax DLMF},\ \href {http://dlmf.nist.gov/} {\enquote {\bibinfo {title}
  {{\it NIST Digital Library of Mathematical Functions}},}\ }\bibinfo
  {howpublished} {http://dlmf.nist.gov/, Release 1.0.24 of 2019-09-15},\
  \bibinfo {note} {f.~W.~J. Olver, A.~B. {Olde Daalhuis}, D.~W. Lozier, B.~I.
  Schneider, R.~F. Boisvert, C.~W. Clark, B.~R. Miller, B.~V. Saunders, H.~S.
  Cohl, and M.~A. McClain, eds.}\BibitemShut {Stop}%
\end{thebibliography}%
\end{document}